\input harvmac.tex


\lref\AT{A.~Achucarro and P.~K.~Townsend,
``A Chern-Simons Action For Three-Dimensional Anti-De Sitter
Supergravity Theories,'' Phys.\ Lett.\ B {\bf 180}, 89 (1986).}

\lref\Wittengrav{E.~Witten,
``(2+1)-Dimensional Gravity As An Exactly Soluble System,''
Nucl.\ Phys.\ B {\bf 311}, 46 (1988).}

\lref\Matschull{H.-J.~Matschull, ``On the relation between $2+1$ Einstein
gravity and Chern Simons theory,'' Class.\ Quant.\ Grav. {\bf 16} (1999) 2599.}

\lref\Ezawa{K.~Ezawa,
``Classical and quantum evolutions of the de Sitter and
the anti-de Sitter universes in (2+1)-dimensions,''
Phys.\ Rev.\ D {\bf 49}, 5211 (1994)
[Addendum-ibid.\ D {\bf 50}, 2935 (1994)].}

\lref\WittenJones{
E.~Witten, ``Quantum Field Theory And The Jones Polynomial,''
Commun.\ Math.\ Phys.\  {\bf 121}, 351 (1989).}

\lref\Wittenamp{
E.~Witten, ``Topology Changing Amplitudes In (2+1)-Dimensional Gravity,''
Nucl.\ Phys.\ B {\bf 323}, 113 (1989).}

\lref\EMSS{S.~Elitzur, G.~W.~Moore, A.~Schwimmer and N.~Seiberg,
``Remarks On The Canonical Quantization Of The Chern-Simons-Witten Theory,''
Nucl.\ Phys.\ B {\bf 326}, 108 (1989).}

\lref\ADPW{S.~Axelrod, S.~Della Pietra and E.~Witten,
``Geometric Quantization Of Chern-Simons Gauge Theory,''
J.\ Diff.\ Geom.\  {\bf 33}, 787 (1991).}

\lref\Hitchin{N.~J.~Hitchin, ``Flat Connections And
Geometric Quantization,'' Commun.\ Math.\ Phys.\  {\bf 131}, 347 (1990).}

\lref\Weitsman{J.~Weitsman, ``Quantization via Real Polarization
of the Moduli Space of Flat Connections and Chern-Simons Gauge
Theory in Genus One,'' Commun.\ Math.\ Phys.\  {\bf 137}, 175 (1991);
``Real Polarization of the Moduli Space of Flat Connections on
a Riemann Surface,'' Commun.\ Math.\ Phys.\  {\bf 145}, 425 (1992).}

\lref\Woodhouse{N.M.J.~Woodhouse, ``Geometric Quantization,''
Oxford University Press, 1991.}

\lref\GQlect{S.~Bates, A.~Weinstein,
``Lectures on the Geometry of Quantization,'' Berkeley, 1997.}

\lref\Wittencpx{
E.~Witten, ``Quantization Of Chern-Simons Gauge Theory With
Complex Gauge Group,'' Commun.\ Math.\ Phys.\  {\bf 137}, 29 (1991).}

\lref\Hayashi{N.~Hayashi,
``Quantum Hilbert space of G(C) Chern-Simons-Witten theory and gravity,''
Prog.\ Theor.\ Phys.\ Suppl.\  {\bf 114}, 125 (1993).}

\lref\BNWitten{D.~Bar-Natan and E.~Witten,
``Perturbative expansion of Chern-Simons theory with noncompact
gauge group,'' Commun.\ Math.\ Phys.\  {\bf 141}, 423 (1991).}

\lref\BNair{M.~Bos and V.~P.~Nair,
``U(1) Chern-Simons Theory And C = 1 Conformal Blocks,''
Phys.\ Lett.\ B {\bf 223} (1989) 61;
``Coherent State Quantization Of Chern-Simons Theory,''
Int.\ J.\ Mod.\ Phys.\ A {\bf 5} (1990) 959.}

\lref\Manoliu{M.~Manoliu,
``Abelian Chern-Simons theory,'' J.\ Math.\ Phys.\ {\bf 39} (1998) 170;
``Quantization of symplectic tori in a real polarization,''
dg-ga/9609012.}

\lref\Jackiwsch{G.~V.~Dunne, R.~Jackiw and C.~A.~Trugenberger,
``Chern-Simons Theory In The Schrodinger Representation,''
Annals Phys.\  {\bf 194} (1989) 197.}

\lref\LLR{J.~M.~Labastida, P.~M.~Llatas and A.~V.~Ramallo,
``Knot Operators In Chern-Simons Gauge Theory,''
Nucl.\ Phys.\ B {\bf 348}, 651 (1991).}

\lref\NRZ{J.~E.~Nelson, T.~Regge and F.~Zertuche,
``Homotopy Groups And (2+1)-Dimensional Quantum De Sitter Gravity,''
Nucl.\ Phys.\ B {\bf 339}, 516 (1990).}

\lref\CarlipBH{S.~Carlip, ``The (2+1)-Dimensional Black Hole,''
Class.Quant.Grav. {\bf 12} (1995) 2853.}

\lref\Carliplect{S.~Carlip,
``Lectures on (2+1) dimensional gravity,''
J.\ Korean Phys.\ Soc.\  {\bf 28}, S447 (1995)
[arXiv:gr-qc/9503024].}

\lref\BTZ{M.~Banados, C.~Teitelboim, J.~Zanelli, ``The Black Hole in
Three Dimensional Space Time,'' Phys. Rev. Lett. {\bf 69} (1992) 1849;
M.~Banados, M.~Henneaux, C.~Teitelboim, J.~Zanelli, ``Geometry of
the 2+1 Black Hole,'' Phys.Rev. {\bf D48} (1993) 1506.}

\lref\CToffshell{
S.~Carlip and C.~Teitelboim, ``The Off-shell black hole,''
Class.\ Quant.\ Grav.\  {\bf 12}, 1699 (1995), [arXiv:gr-qc/9312002].}

\lref\CTaspects{S.~Carlip and C.~Teitelboim,
``Aspects of black hole quantum mechanics and thermodynamics
in (2+1)-dimensions,''
Phys.\ Rev.\ D {\bf 51}, 622 (1995), [arXiv:gr-qc/9405070].}


\lref\Roche{E.~Buffenoir, K.~Noui and P.~Roche,
``Hamiltonian quantization of Chern-Simons theory with SL(2,C) group,''
Class.\ Quant.\ Grav.\  {\bf 19}, 4953 (2002), [arXiv:hep-th/0202121].}

\lref\Krasnov{K.~Krasnov,
``Analytic continuation for asymptotically AdS 3D gravity,''
Class.\ Quant.\ Grav.\  {\bf 19}, 2399 (2002), [arXiv:gr-qc/0111049].}

\lref\KirillRiemann{K.~Krasnov,
``Holography and Riemann Surfaces,''
Adv. Theor. Math. Phys. {\bf 4} (2000) 929;
``On Holomorphic Factorization in Asymptotically AdS 3D Gravity,''
hep-th/0109198.}

\lref\FreidelK{L.~Freidel and K.~Krasnov,
``2D conformal field theories and holography,'' arXiv:hep-th/0205091.}


\lref\HartleH{J.~B.~Hartle and S.~W.~Hawking,
``Wave Function Of The Universe,''
Phys.\ Rev.\ D {\bf 28}, 2960 (1983).}

\lref\WDW{B.~S.~Dewitt,
Phys.\ Rev.\  {\bf 160}, 1113 (1967);
J.A.~Wheeler, in ``Battelle Rencontres'',
Benjamin, New York, 1968.}

\lref\GHartle{G.~W.~Gibbons and J.~B.~Hartle,
``Real Tunneling Geometries And The Large Scale Topology Of The Universe,''
Phys.\ Rev.\ D {\bf 42}, 2458 (1990).}

\lref\CarlipWDW{S.~Carlip,
``Notes on the (2+1)-dimensional Wheeler-DeWitt equation,''
Class.\ Quant.\ Grav.\  {\bf 11}, 31 (1994).}

\lref\Martinec{E.~J.~Martinec,
``Soluble Systems In Quantum Gravity,''
Phys.\ Rev.\ D {\bf 30}, 1198 (1984).}

\lref\Moncrief{V.~Moncrief,
``Reduction Of The Einstein Equations In (2+1)-Dimensions
To A Hamiltonian System Over Teichmuller Space,''
J.\ Math.\ Phys.\  {\bf 30}, 2907 (1989).}

\lref\Hosoya{A.~Hosoya and K.~i.~Nakao,
``(2+1)-Dimensional Pure Gravity For An Arbitrary Closed Initial Surface,''
Class.\ Quant.\ Grav.\  {\bf 7}, 163 (1990).}

\lref\Carlipobs{S.~Carlip,
``Observables, Gauge Invariance, And Time In (2+1)-Dimensional
Quantum Gravity,'' Phys.\ Rev.\ D {\bf 42}, 2647 (1990).}

\lref\STU{L.~Susskind, L.~Thorlacius and J.~Uglum,
``The Stretched horizon and black hole complementarity,''
Phys.\ Rev.\ D {\bf 48}, 3743 (1993).}

\lref\Thorne{K.~S.~Thorne, R.~H.~Price and D.~A.~Macdonald,
``Black Holes: The Membrane Paradigm,''
{\it  NEW HAVEN, USA: YALE UNIV. PR. (1986)}.}


\lref\DJT{S.~Deser, R.~Jackiw and G.~'t Hooft,
``Three-Dimensional Einstein Gravity: Dynamics Of Flat Space,''
Annals Phys.\  {\bf 152}, 220 (1984).}

\lref\DJ{S.~Deser and R.~Jackiw,
``Classical And Quantum Scattering On A Cone,''
Commun.\ Math.\ Phys.\  {\bf 118}, 495 (1988).}

\lref\tHooft{G.~'t Hooft,
``Nonperturbative Two Particle Scattering Amplitudes
In (2+1) Dimensional Quantum Gravity,''
Commun.\ Math.\ Phys.\  {\bf 117}, 685 (1988).}

\lref\Carlipscat{S.~Carlip,
``Exact Quantum Scattering In (2+1)-Dimensional Gravity,''
Nucl.\ Phys.\ B {\bf 324}, 106 (1989).}

\lref\Gerbert{P.~de Sousa Gerbert,
``On Spin And (Quantum) Gravity In (2+1)-Dimensions,''
Nucl.\ Phys.\ B {\bf 346}, 440 (1990).}

\lref\Bargman{V.~Bargman, Ann. Math. {\bf 48} (1947) 568.}

\lref\Naimark{M.A.~Naimark, ``Linear Representations of the Lorentz
Group,'' New York, 1964.}

\lref\Gelfand{I.M.~Gelfand, R.A.~Minlos, Z.Ya.~Shapiro,
``Representations of the Rotation and Lorentz Groups and
Their Applications,'' New York, 1963.}

\lref\KMVW{K.~Koehler, F.~Mansouri, C.~Vaz and L.~Witten,
``Wilson Loop Observables In (2+1)-Dimensional Chern-Simons Supergravity,''
Nucl.\ Phys.\ B {\bf 341}, 167 (1990).}

\lref\KMVWone{K.~Koehler, F.~Mansouri, C.~Vaz and L.~Witten,
``Two Particle Scattering In The Chern-Simons-Witten Theory Of Gravity
In (2+1)-Dimensions,'' Nucl.\ Phys.\ B {\bf 348}, 373 (1991).}

\lref\Ashtekar{A.~Ashtekar,
``New Variables For Classical And Quantum Gravity,''
Phys.\ Rev.\ Lett.\  {\bf 57}, 2244 (1986).}

\lref\RSknot{C.~Rovelli and L.~Smolin,
``Knot Theory And Quantum Gravity,''
Phys.\ Rev.\ Lett.\  {\bf 61}, 1155 (1988).}

\lref\RSloop{C.~Rovelli and L.~Smolin,
``Loop Space Representation Of Quantum General Relativity,''
Nucl.\ Phys.\ B {\bf 331}, 80 (1990).}

\lref\KirillLiouv{K.~Krasnov,
``3D gravity, point particles and Liouville theory,''
Class.\ Quant.\ Grav.\  {\bf 18} (2001) 1291.}


\lref\Jones{V.F.R.~Jones,
``A polynomial invariant for knots via von Newmann algebras,''
Bull. Amer. Math. Soc. {\bf 12} (1985) 103.}

\lref\Atiyah{M.F.~Atiyah, ``The Geometry and Physics of Knots,''
Cambridge Univ. Press, 1990.}

\lref\Thurston{W.~Thurston,
``Three-Dimensional Manifolds,
Kleinian Groups and Hyperbolic Geometry,''
Bull. Amer. Math. Soc. (N.S.) {\bf 6} (1982) 357--381.}

\lref\Thurstonbook{W. Thurston,
``The Geometry and Topology of Three-Manifolds,''
http://www.msri.org/publications/books/gt3m/}

\lref\DrorVas{D.~Bar-Natan, ``On the Vassiliev Knot Invariants,''
Topology {\bf 34} (1995) 423.}

\lref\Birman{J.S.~Birman, ``New Points of View in Knot Theory,''
Bull. Amer. Math. Soc. {\bf 28} (1993) 253;
J.S.~Birman and X.S.~Lin, ``Knot Polynomials and Vassiliev
Invariants,'' Invent. Math. {\bf 111} (1993) 225.}

\lref\LR{R.~Lawrence, L.~Rozansky, ``Witten-Reshetikhin-Turaev
Invariants of Seifert Manifolds,''
Commun. Math. Phys. {\bf 205} (1999) 287.}

\lref\Rozansky{L.~Rozansky,
``A Contribution To The Trivial Connection To Jones Polynomial
And Witten's Invariant Of 3-D Manifolds. 1,''
Commun.\ Math.\ Phys.\  {\bf 175} (1996) 275.}

\lref\CCGLS{D.~Cooper, M.~Culler, H.~Gillet, D.D.~Long, P.B. Shalen,
``Plane curves associated to character varieties of 3-manifolds,''
Invent. Math. {\bf 118} (1994) 47.}

\lref\CullerS{M.~Culler, P.B. Shalen, ``Bounding separating incompressible
surfaces in knot manifolds,'' Ann. Math. {\bf 117} (1983) 109.}

\lref\CLslope{D.~Cooper, D.~Long, ``An Undetected Slope in a Knot
Manifold,'' Topology '90, Walter de Gruyter (1992).}

\lref\CLrem{D.~Cooper, D.~Long, ``Remarks on the A-polynomial of a Knot,''
J. Knot Theory and Its Ramifications, {\bf 5} (1996) 609.}

\lref\CLrep{D.~Cooper, D.~Long, ``Representation Theory and the A-polynomial
of a Knot,'' Chaos, Solitons, and Fractals, {\bf 9} (1998) 749.}

\lref\Meyerhoff{R.~Meyerhoff, ``The Chern-Simons Invariant of
Hyperbolic 3-Manifolds,'' Thesis, Princeton University, 1981.}

\lref\Mostow{G.D.~Mostow, ``Quasi-conformal mappings in n-space
and the rigidity of hyperbolic space forms,'' 
Publ. IHES {\bf 34} (1968) 53.}

\lref\Gromov{M.~Gromov, ``Volume and bounded cohomology,''
Publ. Math. IHES {\bf 56} (1982) 5.}

\lref\Ruberman{D.~Ruberman, ``Mutation and volumes of knots in $S^3$,''
Invent. Math. {\bf 90} (1987) 189.}

\lref\NZagier{W.~Neumann, D.~Zagier, ``Volumes of Hyperbolic
Three-Manifolds,'' Topology {\bf 24} (1985) 307.}

\lref\Yoshida{T.~Yoshida, ``The $\eta$-invariant of hyperbolic
3-manifolds,'' Invent. Math. {\bf 81} (1985) 473.}

\lref\Hodgson{C.D.~Hodgson, ``Degeneration and regeneration of
geometric structures on three-manifolds,''
Thesis, Princeton University, 1986.}

\lref\Dunfield{N.~Dunfield, ``Cyclic surgery, degrees of maps of
character curves, and volume rigidity of hyperbolic manifolds,''
Invent. Math. {\bf 136} (1999) 623.}

\lref\HLMA{H.M.~Hilden, M.T.~Lozano, J.M.~Montesinos-Amilibia,
``On Volumes and Chern-Simons Invariants of Geometric 3-Manifolds,''
J. Math. Sci. Univ. Tokyo {\bf 3} (1996) 732.}

\lref\KKlassen{P.~Kirk, E.~Klassen, ``Chern-Simons Invariants of 3-Manifolds
Decomposed along Tori and the Circle Bundle over the Representation
Space of $T^2$,'' Commun. Math. Phys. {\bf 153} (1993) 521.}

\lref\Casson{A.~Casson, MSRI Lecture Notes, Berkeley (1985).}

\lref\Taubes{C.~Taubes, ``Casson's Invariant and Gauge Theory,
J. Diff. Geom. {\bf 31} (1990) 547.}

\lref\TV{V.G.~Turaev, O.Yu.~Viro, ``State-sum invariants of
3-manifolds and quantum 6j-symbols,'' Topology {\bf 31} (1992) 865.}

\lref\Turaev{V.G.~Turaev, C. R. Acad. Sci. Paris, t. 313,
S\'erie I (1991) 395; ``Topology of Shadow,'' preprint (1991).}

\lref\Walker{K.~Walker, ``On Witten's 3-Manifold Invariants,'' unpublished.}

\lref\RT{N.Y.~Reshetikhin, V.G.~Turaev,
``Invariants of 3-manifolds via link polynomials and quantum groups,''
Invent. Math. {\bf 103} (1991) 547.}

\lref\Beilinson{A.A.~Beilinson, V.G.~Drinfeld,
``Quantization of Hitchin's fibrations and Langlands' program,''
Math. Phys. Stud. {\bf 19}, Kluwer Acad. Publ. (1996) 3.}

\lref\Bonahon{F.~Bonahon, ``A Schlafli-type formula for convex cores
of hyperbolic 3-manifolds,'' J. Diff. Geom. {\bf 50} (1998) 24.}

\lref\Porti{J.~Porti, ``Torsion de Reidemeister pour les varietes
hyperboliques,''Mem. Amer. Math. Soc.  {\bf 128} (1997), no. 612.}


\lref\Kashaev{
R. M. Kashaev, ``Quantum Dilogarithm as a 6j-Symbol,''
Mod.\ Phys.\ Lett. {\bf A9} (1994) 3757;
R.M. Kashaev, ``A Link Invariant from Quantum Dilogarithm,''
q-alg/9504020;
R.M. Kashaev, ``The hyperbolic volume of knots from quantum dilogarithm,''
q-alg/9601025.}

\lref\MM{H.~Murakami, J.~Murakami,
``The colored Jones polynomials and the simplicial volume of a knot,''
math.GT/9905075.}

\lref\KT{R.M.~ Kashaev, O.~ Tirkkonen,
``Proof of the volume conjecture for torus knots,''
math.GT/9912210.}

\lref\MMOTY{H.~Murakami, J.~Murakami, M.~Okamoto, T.~Takata, Y.~Yokota,
``Kashaev's conjecture and the Chern-Simons invariants of knots and links,''
math.GT/0203119.}

\lref\Yokota{Y.~Yokota,
``On the volume conjecture for hyperbolic knots,'' math.QA/0009165.}

\lref\Hikami{K.~Hikami,
``Hyperbolic Structure Arising from a Knot Invariant,''
math-ph/0105039.}

\lref\BBqhi{S.~Baseilhac, R.~Benedetti,
``QHI, 3-manifolds scissors congruence classes and the volume conjecture,''
Geom. Topol. Monogr. {\bf 4} (2002) 13.}

\lref\SW{N.~Seiberg, E.~Witten, ``Monopole Condensation,
And Confinement In N=2 Supersymmetric Yang-Mills Theory,''
Nucl.Phys. {\bf B426} (1994) 19; Erratum-ibid. {\bf B430} (1994) 485.}

\lref\Wittencsstring{E.~Witten,
``Chern-Simons gauge theory as a string theory,''
Prog.\ Math.\  {\bf 133} (1995) 637, hep-th/9207094.}

\lref\GVonetwo{R.~Gopakumar, C.~Vafa,
``M-theory and topological strings. I - II,''
hep-th/9809187, hep-th/9812127.}

\lref\GopakumarV{R.~Gopakumar, C.~Vafa,
``Topological Gravity as Large N Topological Gauge Theory,''
Adv.\ Theor.\ Math.\ Phys. {\bf 2} (1998) 413.}

\lref\GopakumarVafa{R.~Gopakumar, C.~Vafa,
``On the Gauge Theory/Geometry Correspondence,''
Adv.\ Theor.\ Math.\ Phys. {\bf 3} (1999) 1415.}

\lref\OV{H.~Ooguri, C.~Vafa, ``Knot Invariants and Topological Strings,''
Nucl.Phys. {\bf B577} (2000) 419.}

\lref\LMV{J.M.F.~ Labastida, M.~ Marino, C.~ Vafa,
``Knots, links and branes at large N,'' JHEP {\bf 0011} (2000) 007.}

\lref\Curio{G.~Curio,
``Superpotentials for M-theory on a G2 holonomy manifold and
Triality symmetry,'' hep-th/0212211;
``Superpotential of the M-theory conifold and type IIA string theory,''
hep-th/0212233.}

\lref\Acharya{B.~Acharya, ``A Moduli Fixing Mechanism in M theory,''
hep-th/0212294.}

\lref\MurakamiMahler{H.~Murakami
``Mahler measure of the colored Jones polynomial
and the volume conjecture,'' math.GT/0206249.}

\lref\MurakamiOptimist{H.~Murakami, ``Optimistic calculations about
the Witten--Reshetikhin--Turaev invariants of closed three-manifolds
obtained from the figure-eight knot by integral Dehn surgeries,''
math.GT/0005289.}

\lref\Dylantalk{D.~Thurston,
``Hyperbolic Volume and the Jones Polynomial,''
talk presented at the Summer School in Grenoble, 1999.}

\lref\Benedetti{S.~Baseilhac and R.~Benedetti,
``Quantum Hyperbolic State Sum Invariants of 3-Manifolds,''
math.GT/0101234.}

\lref\Gelca{C.~Frohman, R.~Gelca, W.~Lofaro,
The A-polynomial from the noncommutative viewpoint,'' math.QA/9812048;
R.~Gelca, ``On the relation between the A-polynomial
and the Jones polynomial,'' math.QA/0004158.}

\lref\MMorton{P.~Melvin, H.~Morton, ``The Colored Jones Function,''
Commun. Math. Phys. {\bf 169} (1995) 501.}

\lref\DrorG{D.~Bar-Natan, S.~Garoufalidis, ``On the Melvin-Morton-Rozansky
Conjecture,'' Invent. Math. {\bf 125} (1996) 103.}

\lref\Roztriv{L.~Rozansky,
``The Trivial Connection Contribution to Witten's Invariant
and Finite Type Invariants of Rational Homology Spheres,'' q-alg/9503011.}

\lref\Roztwo{L.~Rozansky,
``A Contribution of the Trivial Connection to the Jones Polynomial
and Witten's Invariant of 3d Manifolds II,'' hep-th/9403021.}

\lref\Rozhigher{L.~Rozansky,
``Higher Order Terms in the Melvin-Morton Expansion of
the Colored Jones Polynomial,'' q-alg/9601009.}

\lref\RozBurau{L.~Rozansky,
``The Universal R-Matrix, Burau Representaion and the Melvin-Morton
Expansion of the Colored Jones Polynomial,'' q-alg/9604005.}

\lref\Milnortors{J.~Milnor, ``A Duality Theorem for Reidemeister Torsion,''
Ann. Math. {\bf 76} (1962) 137.}

\lref\Turaevtors{V.~Turaev, ``Reidemeister Torsion in Knot Theory,''
Russ. Math. Surveys {\bf 41} (1986) 97.}

\lref\FreedGompf{D.~Freed, R.~Gompf,
``Computer Calculation of Witten's 3-Manifold Invariants,''
Commun. Math. Phys. {\bf 141} (1991) 79.}

\lref\Lawrence{R.~Lawrence,
``Asymptotic Expansions of Witten-Reshetikhin-Turaev Invariants
of Some Simple 3-Manifolds,'' J. Math. Phys. {\bf 36} (1995) 6106.}

\lref\KaulG{R.~K.~Kaul and T.~R.~Govindarajan,
``Three-dimensional Chern-Simons theory as a theory of knots and links,''
Nucl.\ Phys.\ B {\bf 380}, 293 (1992), hep-th/9111063.}

\lref\Deguchi{Y.~Akutsu, T.~Deguchi, T.~Ohtsuki,
``Invariants of Colored Links,''
J. Knot Theory Ramif. {\bf 1} (1992) 161.}

\lref\Stavros{S.~Garoufalidis,
``Difference and differential equations for the colored Jones function,''
math.GT/0306229;
``On the characteristic and deformation varieties of a knot,''
math.GT/0306230.}


\let\includefigures=\iftrue
\newfam\black
\includefigures
\input epsf
\def\figin{\epsfcheck\figin}\def\figins{\epsfcheck\figins}
\def\epsfcheck{\ifx\epsfbox\UnDeFiNeD
\message{(NO epsf.tex, FIGURES WILL BE IGNORED)}
\gdef\figin##1{\vskip2in}\gdef\figins##1{\hskip.5in}
\else\message{(FIGURES WILL BE INCLUDED)}%
\gdef\figin##1{##1}\gdef\figins##1{##1}\fi}
\def\DefWarn#1{}
\def\figinsert{\goodbreak\midinsert}
\def\ifig#1#2#3{\DefWarn#1\xdef#1{fig.~\the\figno}
\writedef{#1\leftbracket fig.\noexpand~\the\figno}%
\figinsert\figin{\centerline{#3}}\medskip\centerline{\vbox{\baselineskip12pt
\advance\hsize by -1truein\noindent\footnotefont{\bf Fig.~\the\figno:} #2}}
\bigskip\endinsert\global\advance\figno by1}
\else
\def\ifig#1#2#3{\xdef#1{fig.~\the\figno}
\writedef{#1\leftbracket fig.\noexpand~\the\figno}%
\global\advance\figno by1}
\fi

\font\cmss=cmss10 \font\cmsss=cmss10 at 7pt

\def\IB{\relax\hbox{$\inbar\kern-.3em{\rm B}$}}
\def\IC{\relax\hbox{$\inbar\kern-.3em{\rm C}$}}
\def\IQ{\relax\hbox{$\inbar\kern-.3em{\rm Q}$}}
\def\ID{\relax\hbox{$\inbar\kern-.3em{\rm D}$}}
\def\IE{\relax\hbox{$\inbar\kern-.3em{\rm E}$}}
\def\IF{\relax\hbox{$\inbar\kern-.3em{\rm F}$}}
\def\IG{\relax\hbox{$\inbar\kern-.3em{\rm G}$}}
\def\IGa{\relax\hbox{${\rm I}\kern-.18em\Gamma$}}
\def\IH{\relax{\rm I\kern-.18em H}}
\def\IK{\relax{\rm I\kern-.18em K}}
\def\IL{\relax{\rm I\kern-.18em L}}
\def\IP{\relax{\rm I\kern-.18em P}}
\def\IR{\relax{\rm I\kern-.18em R}}
\def\Z{\relax\ifmmode\mathchoice
{\hbox{\cmss Z\kern-.4em Z}}{\hbox{\cmss Z\kern-.4em Z}}
{\lower.9pt\hbox{\cmsss Z\kern-.4em Z}}
{\lower1.2pt\hbox{\cmsss Z\kern-.4em Z}}\else{\cmss Z\kern-.4em
Z}\fi}

\def\II{\relax{\rm I\kern-.18em I}}

\def\S{{\bf S}}

\def\CP{{\bf CP}}

\def\CA {{\cal A}}
\def\CB {{\cal B}}

\def\CD {{\cal D}}

\def\CF {{\cal F}}

\def\CH {{\cal H}}

\def\CL {{\cal L}}
\def\CM {{\cal M}}
\def\CN {{\cal N}}

\def\CP {{\cal P}}


\def\p{\partial}

\def\tilde{\widetilde}
\def\hat{\widehat}
\def\bar{\overline}


\def\Hess{{\rm Hess}}

\def\Hom{{\rm Hom}}
\def\Tr{{\rm Tr}}

\def\Vol{{\rm Vol}}

\def\p{\partial}

\def\Det{{\rm Det}}

\def\Si{{\Sigma}}
\def\lieg{{\bf g}}

\def\clieg{{\bf g}_{\scriptscriptstyle{{\bf C}}}}

\def\CCG{G_{\scriptscriptstyle{\IC}}}

\def\inbar{\,\vrule height1.5ex width.4pt depth0pt}
\def\r{{\rm Re}}
\def\i{{\rm Im}}

\def\a{\alpha}

\def\g{\gamma}
\def\d{\delta}
\def\e{\epsilon}

\def\la{\lambda}
\def\th{\theta}
\def\s{\sigma}
\def\om{\omega}

\def\bar{\overline}

\def\example#1{\bgroup\narrower\footnotefont\baselineskip\footskip\bigbreak
\hrule\medskip\nobreak\noindent {\bf Example}. {\it #1\/}\par\nobreak}
\def\endexample{\medskip\nobreak\hrule\bigbreak\egroup}


\Title{\vbox{\baselineskip12pt\hbox{hep-th/0306165}
\hbox{HUTP-03/A003}
\hbox{ITEP-TH-50/02}
}} {\vbox{
\centerline{Three-Dimensional Quantum Gravity,}
\medskip
\centerline{Chern-Simons Theory, And}
\medskip
\centerline{The A-Polynomial}
}}
\centerline{Sergei Gukov}
\medskip
\vskip 8pt
\centerline{\it Jefferson Physical Laboratory, Harvard University,}
\centerline{\it Cambridge, MA 02138, USA}
\vskip 30pt
{\bf \centerline{Abstract}}
\noindent
We study three-dimensional Chern-Simons theory
with complex gauge group $SL(2,\IC)$,
which has many interesting connections with three-dimensional
quantum gravity and geometry of hyperbolic 3-manifolds.
We show that, in the presence of a single knotted Wilson loop
in an infinite-dimensional representation of the gauge group,
the classical and quantum properties of such theory are described
by an algebraic curve called the A-polynomial of a knot.
Using this approach,
we find some new and rather surprising relations between
the A-polynomial, the colored Jones polynomial,
and other invariants of hyperbolic 3-manifolds.
These relations generalize the volume conjecture and
the Melvin-Morton-Rozansky conjecture, and suggest
an intriguing connection between the $SL(2,\IC)$ partition
function and the colored Jones polynomial.

\smallskip
\Date{June 2003}


\newsec{Introduction and Motivation}

In this paper we study three-dimensional
Chern-Simons theory with complex gauge group.
%
%
Of particular interest is a Chern-Simons theory with gauge group
$\CCG=SL(2,\IC)$ (viewed as a complexification of $G=SU(2)$),
which has many interesting connections with three-dimensional
quantum gravity and geometry of hyperbolic three-manifolds.
In this introductory section we review some aspects of these
relations, formulate the problem, and describe various applications.

\subsec{Chern-Simons Theory}

Consider an oriented three-dimensional space $M$.
We wish to formulate a Chern-Simons gauge theory on $M$
with complex gauge group $\CCG$, whose real form we denote by $G$.
Let $\clieg$ and $\lieg$ be the corresponding Lie algebras.
In these notations, the gauge connection $\CA$ is
a one-form on $M$ valued in the complex Lie algebra $\clieg$.
Explicitly, we can write $\CA = \sum_a \CA^a \cdot T_a$
where $T_a$ denote the generators of $\lieg$,
which are assumed to be orthonormal, $\Tr (T_a T_b) = \d_{ab}$.
Then, the Chern-Simons action can be written
as a sum of the holomorphic and anti-holomorphic terms,
\eqn\iaction{\eqalign{
I = & {t \over 8 \pi} \int_M \Tr
\Big( \CA \wedge d \CA + {2 \over 3} \CA \wedge \CA \wedge \CA \Big) + \cr
& + {\bar t \over 8 \pi} \int_M \Tr
\Big( \bar \CA \wedge d \bar \CA
+ {2 \over 3} \bar \CA \wedge \bar \CA \wedge \bar \CA \Big)}}
where $t=k+is$ and $\bar t=k-is$ are the corresponding
coupling constants. Consistency of the quantum theory
requires the ``level'' $k$ to be an integer, $k \in \Z$.
The other parameter, $s$, is not quantized. However,
$s$ must obey certain constraints imposed by unitarity \Wittencpx.

In Euclidean space, unitarity implies that
the argument of the Feynman path integral
\eqn\zpathint{Z(M) = \int \CD \CA ~e^{iI}}
must be complex conjugated under a reversal of the orientation
on $M$. In the Chern-Simons theory defined by the action \iaction,
there are two possibilities to achieve this, corresponding to
either purely real or purely imaginary values of $s$.
In the first case, $\CA$ is invariant under the reversal of
the orientation and $\bar t$ is the usual complex conjugate of $t$. 
On the other hand, the second possibility, $s \in i\IR$,
is realized when the gauge connection transforms non-trivially
under the reversal of the orientation, $\CA \mapsto \bar \CA$.
In the rest of the paper we mainly consider the case of
imaginary $s$ and $\CCG=SL(2,\IC)$, which is related to
the Euclidean quantum gravity in three dimensions
(see discussion below).

\ifig\aknot{A knotted Wilson loop in the $\IR^3$.}
{\epsfxsize1.2in\epsfbox{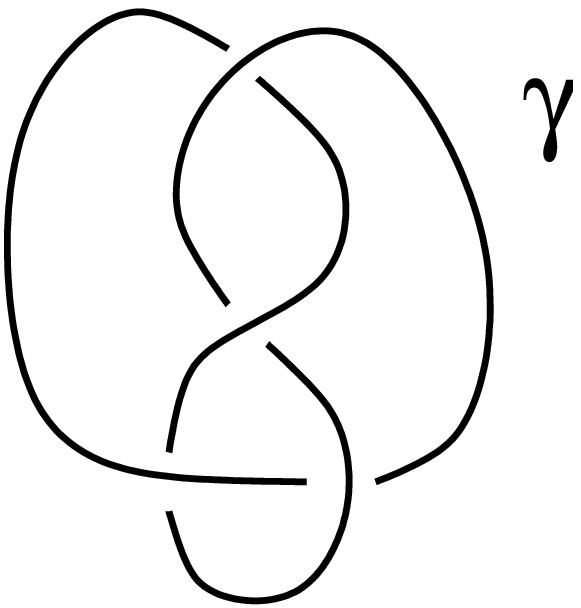}}

Now let us consider observables in this theory.
If $M$ is a manifold without boundary,
holonomies of the gauge connection provide a complete set
of observables, also known as ``Wilson lines'' in the context
of gauge theory or ``loop variables'' in the context
of gravity \RSloop.
Specifically, given a closed oriented curve $\g \subset M$
and a representation $R$ of $\CCG$, one can define a gauge
invariant observable as
\eqn\wloop{
W_{R} (\g) = \Tr_{R} ~{\rm Hol}_{\g} (\CA)
= \Tr_{R} \Big( P \exp \oint_{\g} \CA \Big)}

The Wilson loop observables $W_{R} (\g)$ are naturally associated
with knots in $M$. Indeed, even though intrinsically the curve $\g$
is simply a circle, its embedding in $M$ may be highly non-trivial,
represented by a knot, as in \aknot.
More generally, an embedding of a collection of circles
into $M$ is called a link,
and the image of each circle is called a component of the link.
Thus, given a link with (non-intersecting)
components $\g_i$, $i=1, \ldots, r$,
and a set of representations $R_i$ assigned to each component
of the link, one can study a natural generalization of
the Feynman path integral \zpathint:
\eqn\zwpathint{
Z(M;\g_i,R_i) = \int \CD \CA ~\exp({iI})~ \prod_{i=1}^r W_{R_i} (\g_i)}
which, following \WittenJones, we call the (unnormalized) expectation
value of the link.
By construction, it is a function of $t$ and $\bar t$,
which also depends on the topology of the three-manifold $M$,
on the choice of the Wilson lines $\g_i$,
and on the corresponding representations $R_i$.

A large class of representations $R_i$ can be naturally
obtained by complexification from the corresponding
representations of the real Lie algebra $\lieg$.
However, no new information can be gained by studying such
representations since the evaluation of \zwpathint\
essentially reduces to the Chern-Simons theory with
the real form of the gauge group, $G$, at least in perturbation
theory\foot{See {\it e.g.} exercise 6.32 in \DrorVas.}.
On the other hand, of particular interest are certain infinite
dimensional representations $R_i$ that we describe explicitly
in the next subsection, after explaining the connection with
three-dimensional quantum gravity.

\subsec{Three-Dimensional Quantum Gravity}

As we already mentioned earlier, there are several
intriguing connections between three-dimensional gravity and
Chern-Simons theory with complex gauge group $\CCG = SL(2, \IC)$
and imaginary values
of the parameter $s$ (for convenience,
in what follows we shall use a real parameter $\s = is$).
Thus, $SL(2,\IC)$ appears as the Poincare group in three-dimensional
Euclidean gravity with negative cosmological
constant\foot{More precisely,
${\rm Isom}_+ (\IH^3) = PSL(2,\IC) = SL(2,\IC)/\{\pm 1 \}$,
but according to \CullerS, a representation of $PSL(2,\IC)$
corresponding to a complete hyperbolic structure can always be
lifted to $SL(2,\IC)$, and it is $SL(2,\IC)$ with which we shall work.}.
Moreover, by writing the complex gauge field $\CA$
in terms of the real and imaginary components one can relate
the Chern-Simons action \iaction\ to the usual form of
the Einstein-Hilbert action of three-dimensional gravity
with negative cosmological constant \refs{\AT,\Wittengrav}.
Specifically, writing $\CA=w + ie$ and $\bar \CA=w - ie$
one finds
\eqn\icsgrav{\eqalign{
I = & {k \over 4 \pi} \int_M \Tr
\Big( w \wedge d w - e \wedge de
+ {2 \over 3} w \wedge w \wedge w
- 2 w \wedge e \wedge e \Big) + \cr
& + {i\s \over 2 \pi} \int_M \Tr
\Big( w \wedge de + w \wedge w \wedge e
- {1 \over 3} e \wedge e \wedge e \Big)
}}
The second term in this expression is indeed equivalent to
the Einstein-Hilbert action\foot{In our notations,
the length scale $\ell = 1$ and the Newton constant $G_N=1/(4\s)$.}
with negative cosmological constant, $\Lambda=-1$,
written in terms of the vielbein $e$ and the spin connection $w$.
We can also write it in the standard form:
\eqn\igrav{I_{{\rm grav}} = - {1 \over 4 \pi} \int_M d^3x
\sqrt{g} \Big( R + 2 \Big)}

As will be shown below, the first term in \icsgrav\
also has a nice interpretation. It is related to
the Chern-Simons invariant of the three-manifold $M$.
Therefore, it is convenient to denote this term as $I_{{\rm CS}}$.
In the new notations, we can write the original action \iaction\ as
\eqn\iks{I(k,\s) = k I_{{\rm CS}} + i \s I_{{\rm grav}}}

Summarizing, following Witten \Wittengrav, we conclude that
the real and imaginary components of the $SL(2, \IC)$ Chern-Simons
action \iaction\ have a nice physical interpretation.
In particular, a theory with $k=0$ represents,
at least (semi-)classically, a three-dimensional Euclidean
quantum gravity with negative cosmological
constant\foot{Chern-Simons theory with $SL(2,\IC)$ gauge group
and real values of the parameter $s$ is also related to
three-dimensional gravity, namely to de Sitter gravity
in 2+1 dimensions \refs{\AT,\Wittengrav}.
This theory can be treated similarly, and many of the arguments
below easily extend to this case.
For work on quantization of this theory
see \refs{\Wittencpx,\Ezawa,\Roche}.}.
However, this equivalence does not readily extend to quantum
theories due to a number of subtle issues, typically related
to degenerate vielbeins {\it etc.} (see {\it e.g.} \Matschull\
for a recent discussion of these questions).
For example,
in the Chern-Simons theory, it is natural
to expand around a trivial vacuum, $\CA=0$, which corresponds
to a very degenerate metric, $g_{ij}=0$. Also,
in the Chern-Simons path integral \zpathint\ one integrates
over all (equivalence classes of) gauge connections,
whereas in quantum gravity one takes only a subset
of those corresponding to positive-definite volume elements.
Nevertheless, one would hope that, for certain questions,
the relation to gravity can still be helpful
even beyond the classical limit.
Thus, in order to avoid the above problems
throughout the paper we shall mainly consider
the semi-classical expansion around an isolated critical point,
corresponding to a non-degenerate metric on $M$.
Then, the quantum fluctuations are small,
and both theories are expected to agree.

So far we discussed a relation between
$SL(2, \IC)$ Chern-Simons theory and pure gravity.
Now let us add sources representing point particles.
Assuming that particles don't have any internal
structure, they can be characterized by two
numbers: a mass and a spin.
As we will see later, it is natural to combine these
numbers into a single complex quantity, which labels
an infinite dimensional representation of $SL(2,\IC)$,
see \refs{\Gelfand,\Naimark}.
Interacting with gravity, matter particles produce
conical defects in the geometry of the space
manifold $M$ \refs{\DJT,\DJ,\tHooft}.
In particular, light-like particles correspond to cusps in $M$.
We shall say more about this aspect later, when we will be
talking about the relation to hyperbolic geometry.

The coupling of point-like sources to gravity can be
described by Wilson lines in the Feynman path integral \zwpathint,
see {\it e.g.} \refs{\Wittenamp,\Carlipscat,\Gerbert,\KMVW}.
For example, if we introduce extra variables $p_a$ and $x^a$,
which represent momentum and coordinate of a particle in space $M$,
then the Wilson line operator for a spinless particle can be
explicitly written as \Wittenamp:
\eqn\wsource{
W_R (\g) = \int \CD x^a(s) \CD p_a(s) \CD \la (s)
~\exp \left( i I_W \right) }
where $s$ is a parameter along the ``world-line'' $\g$,
and $I_W$ is the action
\eqn\iwline{I_W = \int_{\g} ds \Big( p_a {D x^a \over Ds}
- \la (\vec p^2 - m^2) \Big)}
In this action, $\la$ is a Lagrange multiplier that enforces
the on-shell condition $\vec p^2 = m^2$, and the coupling
to gravity is encoded in the covariant derivative
$$
{D x^a \over Ds} = {dx^a \over ds} + {{w_s}^a}_b x^b + i {e_s}^a
$$
where $(e_s, w_s)$ denotes the restriction of $(e,w)$ to $\g$.

Eqs. \wsource\ - \iwline\ define
an infinite dimensional representation $R$ of $SL(2,\IC)$
corresponding to a spinless particle in three dimensions.
In a similar way, one can define a general class of infinite
dimensional representations corresponding to massive spinning
particles \Gerbert\ and black holes,
which is precisely the class of representations we are going to
assign to Wilson lines $W_{R_i} (\g_i)$
in the Feynman path integral \zwpathint.
Below we shall give an equivalent geometric
description of such system.

\subsec{The Hartle-Hawking Wave Function in $2+1$ Dimensions}

There is yet another relation between $SL(2,\IC)$ Chern-Simons
theory with $t = - \bar t = \s$ and three dimensions gravity
with negative cosmological constant.
Namely, in the Hartle-Hawking no-boundary proposal \HartleH,
the ground state wave function of the 2+1 dimensional
{\it Lorentzian} universe is obtained from the {\it Euclidean} path
integral over metrics on a 3-manifold $M$ with boundary $\Si=\p M$,
\eqn\hhpsi{ \Psi_{\Si} (h) =
\sum_{M} \int \CD g ~\exp \left(- \s I_{{\rm grav}} \right) }
Here, the summation represents a sum over topologies,
and $I_{{\rm grav}}$ is the Euclidean gravity action \igrav.
Again, we consider pure gravity without coupling to matter.
The functional $\Psi_{\Si}(h)$, called the Hartle-Hawking wave function,
is a functional of the induced metric $h$ on the space-like surface $\Si$.
It should be interpreted as an amplitude of finding the universe
in a quantum state, characterized by the spatial geometry $\Si$
with metric $h$, as schematically shown on the figure below.
Furthermore, a ``smooth'' continuation from
the Riemannian to Lorentzian metrics across $\Si$
is possible only if we limit the sum \hhpsi\ to manifolds $M$
such that the boundary $\Si=\p M$ is totally geodesic \GHartle.

\ifig\hhawking{In the Hartle-Hawking no-boundary proposal,
the ``analytic continuation'' across a surface $\Si$
describes a real tunneling from ``nothing''
to a universe with ``space'' $\Si$.}
{\epsfxsize2.2in\epsfbox{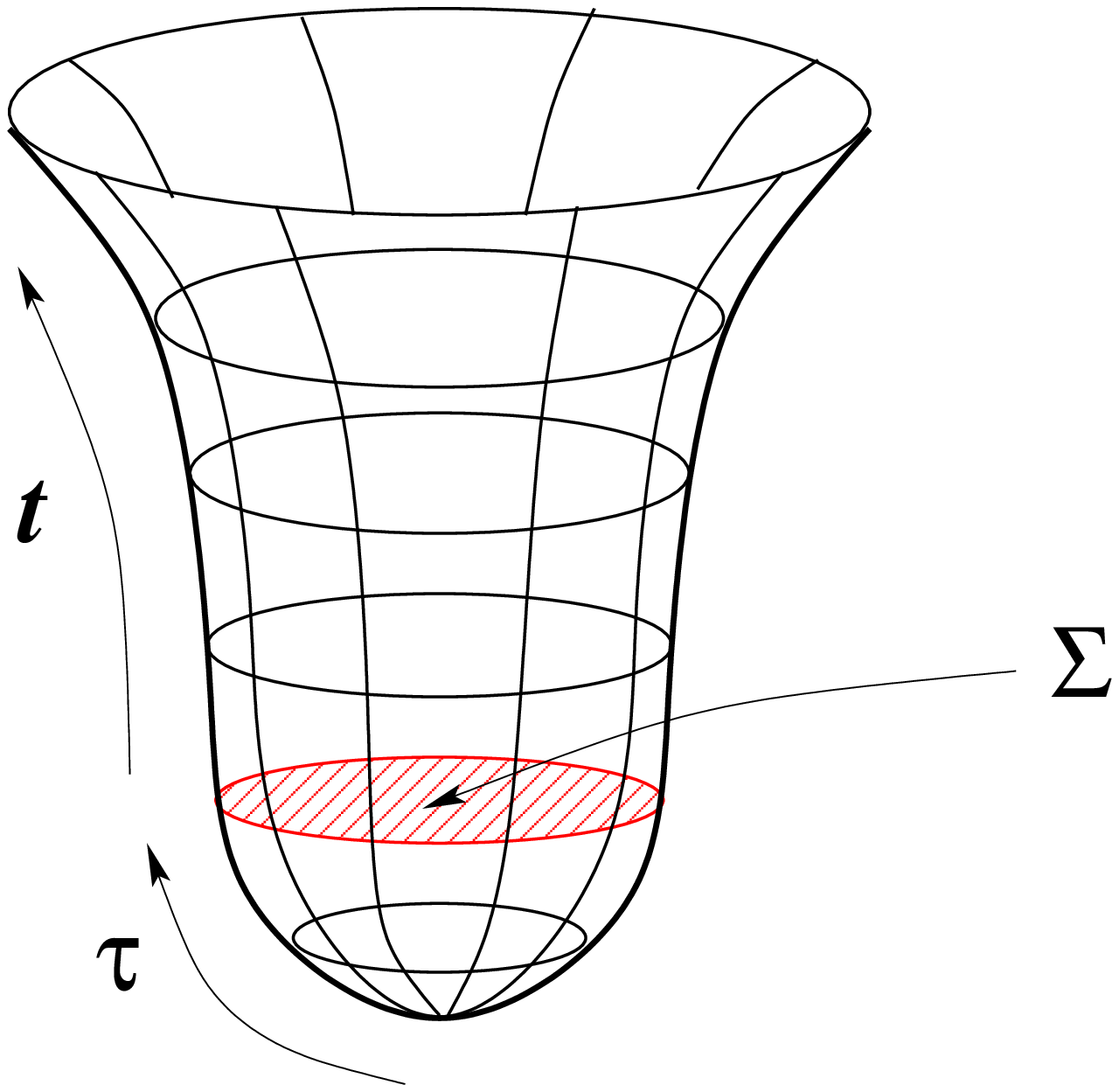}}

The wave function \hhpsi\ obeys the Wheeler-DeWitt equation \WDW,
which has the form of the Hamiltonian constraint
\eqn\wdweq{\hat H \Psi = 0}
where $\hat H$ is a second-order differential operator
determined by the topology of the surface $\Si$.
For instance, when $\Si$ is a torus, $\hat H$ essentially
reduces to the Laplace operator on the torus moduli space \CarlipWDW,
see also \refs{\Martinec, \Moncrief, \Hosoya, \Carlipobs, \Carliplect}.

The case when $\Si=T^2$ (or a collection of tori)
turns out to be closely related to the problem of computing
the Chern-Simons path integral \zwpathint\ with Wilson loops.
Indeed, modulo the subtleties related to the integration
domain in \hhpsi, one could replace the integral over
the metrics by the Chern-Simons path integral of the form \zpathint.
This would give a wave functional
\eqn\hhpsiz{ \Psi_{\Si} (h) = \sum_{M} Z(M;\Si)}
where each term $Z(M;\Si)$ depends on the topology of
the three-manifold $M$ and on the boundary conditions at $\Si=T^2$.
These boundary data can be identified
with the parameters of the Wilson line $W_R (\g)$
in the infinite dimensional representation $R$ of $SL(2,\IC)$,
so that one has \Rozansky:
\eqn\zmtvszk{Z(M;T^2) = Z(\S^3;\g,R)}
where $M = \S^3 \setminus \g$
is a compact three-manifold with a single torus boundary,
obtained by removing a Wilson line from the 3-sphere.
Indeed, one can split the path integral on the right-hand
side of \zmtvszk\ into three parts corresponding to:
$1)$ the integral over the connection $\CA$ inside a small
neighborhood of the Wilson line;
$2)$ the integral over its
complement, $M=\S^3 \setminus \g$, with certain
boundary conditions on the boundary, $\partial M = T^2$;
$3)$ and, finally, the integral over these boundary conditions.
Then, the first integral leads to a delta-function that
fixes the boundary conditions to certain values, so that
the entire path integral can be reduced to $Z(M;T^2)$ with
certain boundary conditions on the $T^2$.

\ifig\aknot{A compact 3-manifold $M$ with a single torus boundary
can be constructed by removing a small neighborhood of a knotted
Wilson loop from the 3-sphere.}
{\epsfxsize2in\epsfbox{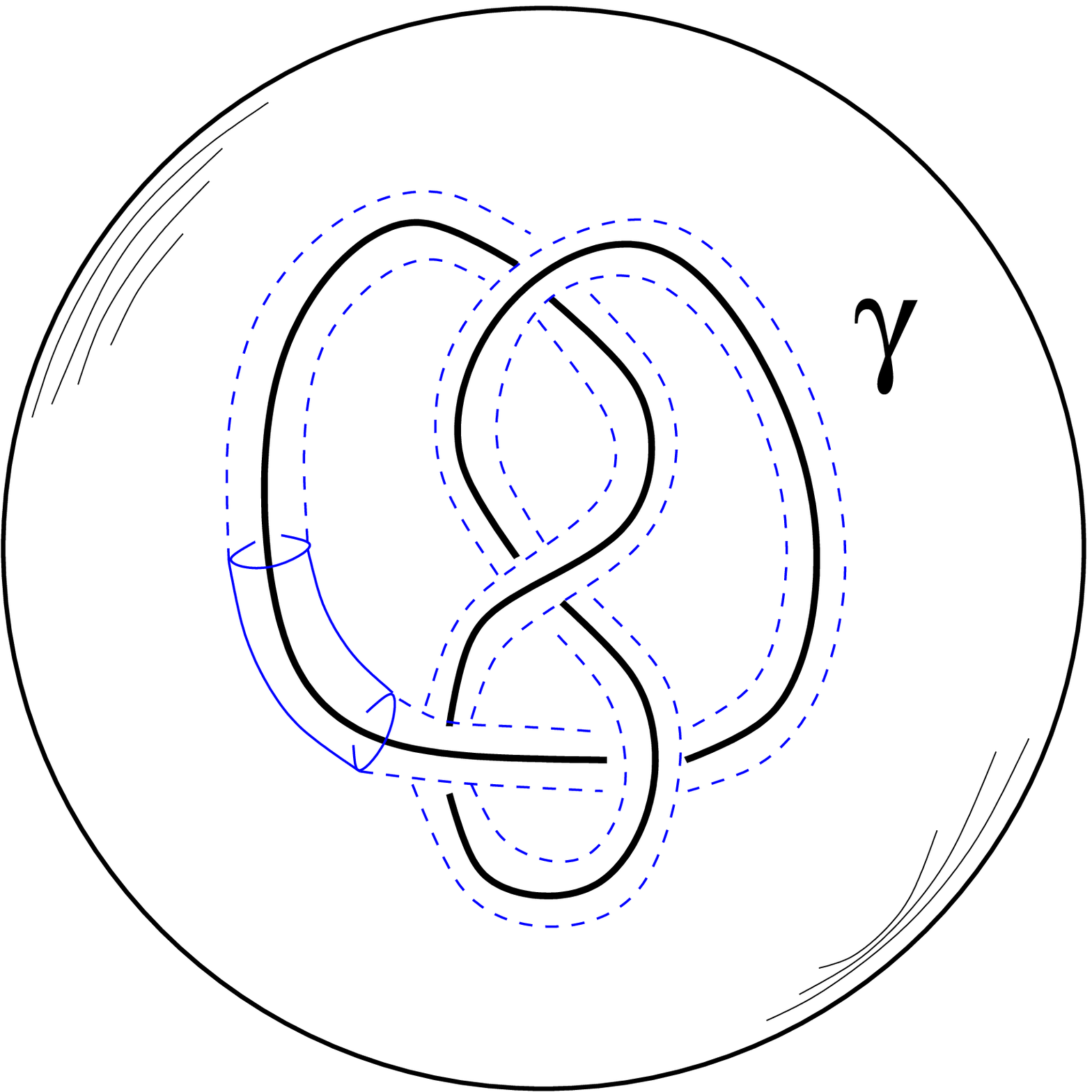}}

It is natural to expect that the wave function \hhpsiz\
obtained from the Chern-Simons path integral
satisfies an equation of the form \wdweq.
A more optimistic hope is that each term, $Z(M;\Si)$,
individually solves an analog of the Wheeler-DeWitt equation.
As we shall see below, this is indeed the case,
and the generalized partition function \zwpathint\
indeed obeys a functional equation of the form \wdweq.
Finally, we note that, even though in this work we mainly
focus on knot complements, the arguments can be extended
to generic 3-manifolds with arbitrary boundary $\Si$
(in particular, to 3-manifolds with infinite volume).
These manifolds also have a nice interpretation in
three-dimensional gravity \KirillRiemann, and will be
briefly discussed in section 4 and in Appendix A.

\subsec{Quantum Geometry of Hyperbolic Three-Manifolds}

Note, that the definition of the path integral \zwpathint\
does not involve a choice of metric on $M$.
Therefore, $Z(M;\g_i, R_i)$ is expected to be a topological
invariant of the three-manifold $M$ with a link, whose components
$\g_i$ are colored by (infinite-dimensional) representations, $R_i$,
of the complex gauge group $\CCG$. From the physical point of view,
this is not very surprising since we start with a Chern-Simons theory,
which has vanishing Hamiltonian and, therefore, is expected to be
a topological quantum field theory.

The quantum invariant $Z(M;\g_i, R_i)$ is particularly interesting
for the complex gauge group $\CCG=SL(2,\IC)$,
which is related to
three-dimensional Euclidean gravity with negative cosmological
constant. In particular, classical solutions in the gravity
theory \igrav\ correspond to manifolds with constant negative
curvature, and one can also interpret certain
Wilson line operators $W_R (\g)$
as operators creating cusps in $M$. Therefore, $Z(M;\g_i, R_i)$
is a natural invariant of hyperbolic three-manifolds with cusps.
If we remove the cusps from $M$, we can also interpret
$Z(M;\g_i, R_i)$ as a topological invariant of the complement
of the link, whose components are $\g_i$.

According to a famous theorem by Thurston, many 3-manifolds
are hyperbolic \Thurston.
For example, the complement of a knot in $\S^3$ admits a hyperbolic
structure unless it is a torus or satellite knot.
Moreover, after the Mostow Rigidity Theorem \Mostow, any geometric
invariant of a hyperbolic 3-manifold is a topological invariant.
Important geometric invariants, which can be defined even if
the hyperbolic manifold $M$ has cusps \Meyerhoff,
are the volume and the Chern-Simons invariant.
Thurston suggests to combine these two invariants into a single
complex invariant $Z(M)$ whose absolute value is $e^{2/\pi \Vol (M)}$
and whose argument is the Chern-Simons invariant of $M$ \Thurston:
\eqn\thurstinv{
Z(M) \sim \exp \left({2 \over \pi} \Vol (M) + 4 \pi i CS(M) \right)}
As one might expect from \iks, the path integral \zpathint\
reduces to an invariant like this in the semi-classical limit.

More surprisingly,
extending the well-known volume conjecture \refs{\Kashaev,\MM,\MMOTY},
we find that the $N$-colored Jones polynomial, $J_N (\g,e^{2\pi i/k})$,
also has a similar asymptotic behavior in the limit
$k,N \to \infty$, such that the ratio, $a=N/k$, is fixed.
Namely, depending on whether the parameter $a$ is rational
or not, the asymptotic behavior of $J_N (e^{2\pi i/k})$ is either
polynomial or exponential (the value $a=1$ is somewhat special).
According to a seminal work of Witten \WittenJones, the first case
is related to the $SU(2)$ Chern-Simons gauge theory
and rational conformal field theory in two dimensions.
On the other hand, as we show below, the exponential growth
of the colored Jones polynomial encodes a lot of interesting information
about the hyperbolic geometry of the knot complement and, therefore,
is more suggestive of the $SL(2,\IC)$ Chern-Simons theory.
Schematically, the whole picture can be summarized in the following table:

\vskip 0.8cm
\vbox{
\centerline{\vbox{
\hbox{\vbox{\offinterlineskip
\def\tablespace{height7pt&\omit&&\omit&&\omit&&\omit&\cr}
\def\tablerule{\tablespace\noalign{\hrule}\tablespace}

\hrule\halign{&\vrule#&\strut\hskip0.2cm\hfill #\hfill\hskip0.2cm\cr
\tablespace
& $a=N/k$   && asymptotic behavior && CFT && Chern-Simons  &\cr
&           && of $J_N (e^{2\pi i/k})$  && && theory & \cr
\tablerule
& Rational && Polynomial && Rational && $SU(2)$ & \cr
\tablerule
& Non-Rational && Exponential && Non-Rational && $SL(2,\IC)$ & \cr
\tablespace}\hrule}}}}
\centerline{\hbox{{\bf Table 1:}
{\it ~~Asymptotic behavior of the colored Jones polynomial
for hyperbolic knots.}}}
}
\vskip 0.5cm

\noindent
This heuristic picture agrees with the fact that, when $k=0$ and $\s \in \Z$,
the partition function \zpathint\ of the $SL(2,\IC)$ Chern-Simons theory
can be formally regarded as a product of two $SU(2)$ partition functions
by treating $\CA$ and $\bar \CA$ as independent $SU(2)$ gauge fields.
Moreover, it was argued in \Hayashi\ that at these values of
the coupling constants the Hilbert space of the $SL(2,\IC)$
Chern-Simons theory on a torus factorizes into
two copies of the Hilbert space of an $SU(2)$ theory.

The analytic continuation to non-integer values of $k$
in the $SU(2)$ Chern-Simons theory can be also motivated by
connection with topological string theory.
Just like ordinary quantum field theory can be embedded in
string theory, topological Chern-Simons theory can be
realized in topological open string theory \Wittencsstring,
where the string coupling constant is related to the level $k$.
In many cases, these theories admit a dual description
in terms of closed topological strings on a (non-compact)
Calabi-Yau manifold \GopakumarVafa.
Therefore, from the point of view of topological string theory,
it is also natural (and sometimes even necessary) to consider
non-integer values of $k$, which via duality is identified with
the complexified K\"ahler parameter of the Calabi-Yau manifold \GVonetwo.
Then, the results of this paper suggest that certain invariants
of hyperbolic 3-manifolds might emerge from topological closed
string theory in the ``zero radius limit''.

Finally, we note that Chern-Simons theory with $SL(2,\IC)$ gauge
group also has a number of interesting applications in string theory,
see {\it e.g.} \refs{\Curio,\Acharya} for some recent work.

\bigskip {\noindent {\it {Organization of the Paper}}}

The rest of the paper is organized as follows:
In section 2, we study classical aspects of Chern-Simons theory
with complex gauge group.
In particular, we introduce the A-polynomial of a knot and
identify it with the space of classical solutions in
the $SL(2,\IC)$ Chern-Simons theory on the knot complement.
Section 3 is devoted to quantization of this theory
(in a real polarization).
Starting with section 4, we discuss various extensions and
applications of this approach, including mathematical applications.
In section 5, we present some evidence for the picture
summarized in Table 1 and propose a generalization of the volume
conjecture to incomplete hyperbolic structures on knot complements.
Further aspects of the relation with $SU(2)$ Chern-Simons theory
and the colored Jones polynomial are discussed in section 6,
where we propose an analog of the Melvin-Morton-Rozansky conjecture
for a non-trivial (``hyperbolic'') flat connection.
Finally, in the appendices we illustrate the general ideas
using simple examples.
Thus, in Appendix A, we use the geometry of the BTZ black hole
to explain the origin and the interpretation of the A-polynomial
for the trivial knot.
In Appendix B, we discuss quantization of the $SL(2,\IC)$
Chern-Simons theory for torus knots, which leads to
a particularly simple (Gaussian) quantum mechanics.


\newsec{Classical Theory}

We wish to study quantum Chern-Simons theory \iaction\
with Wilson lines in the infinite dimensional
representations of the complex gauge group $\CCG$.
Like in any constrained system,
there are two ways of quantizing the theory:
one can either quantize the classical phase space
(that is the space of solutions of the classical
equations of motion);
or one can impose the constraints after quantization.
See {\it e.g.} \EMSS\ for a nice exposition of both methods.
Here, we will follow the first approach.
For this, we need to begin by studying
the classical solutions of the theory;
this will be the first place where we encounter the A-polynomial.

\subsec{The Moduli Space of Classical Solutions}

Away from the location of Wilson loops,
the classical Euler-Lagrange equations derived from
the Chern-Simons action \iaction\ look like
\eqn\feom{\CF = \bar \CF =0}
where $\CF = d\CA + \CA \wedge \CA$ is the field strength,
and $\bar \CF$ is its complex conjugate. Therefore,
the classical solutions in Chern-Simons theory are the so-called
flat connections, {\it i.e.} the gauge fields with zero curvature.

A flat connection on $M$ is determined
by its holonomies, that is by a homomorphism
\eqn\generalrho{ \rho \colon \pi_1(M) \to \CCG }
Hence, the moduli space of classical solutions is given by
the set of representations of the fundamental group, $\pi_1(M)$,
into the group $\CCG$
modulo gauge transformations, which act on $\rho$ by conjugation,
\eqn\generall{
L = {\rm Rep} \left( \pi_1 (M) \to \CCG \right) / {\rm conjugation}}
Similarly, if $M$ is a compact 3-manifold with boundary, $\p M = \Si$,
one can consider the moduli space of flat connections on $\Si$,
\eqn\generalp{
\CP = {\rm Rep} \left( \pi_1 (\Si) \to \CCG \right) / {\rm conjugation}}
There is a natural map from the representation variety $L$ to $\CP$
induced by restricting a flat connection on $M$ to $\Si$.
The image of $L$ under this map is a middle-dimensional
submanifold in $\CP$.
In the canonical quantization, that will be discussed in the next section,
$\CP$ is a classical phase space,
whereas the image of $L$ is associated with a semi-classical state.

\ifig\knots{Some simple knots in three-dimensional space.
We use the standard notation ${\bf k}_n$ to indicate
the $n$-th knot in the census of knots with $k$ crossings.}
{\epsfxsize5.0in\epsfbox{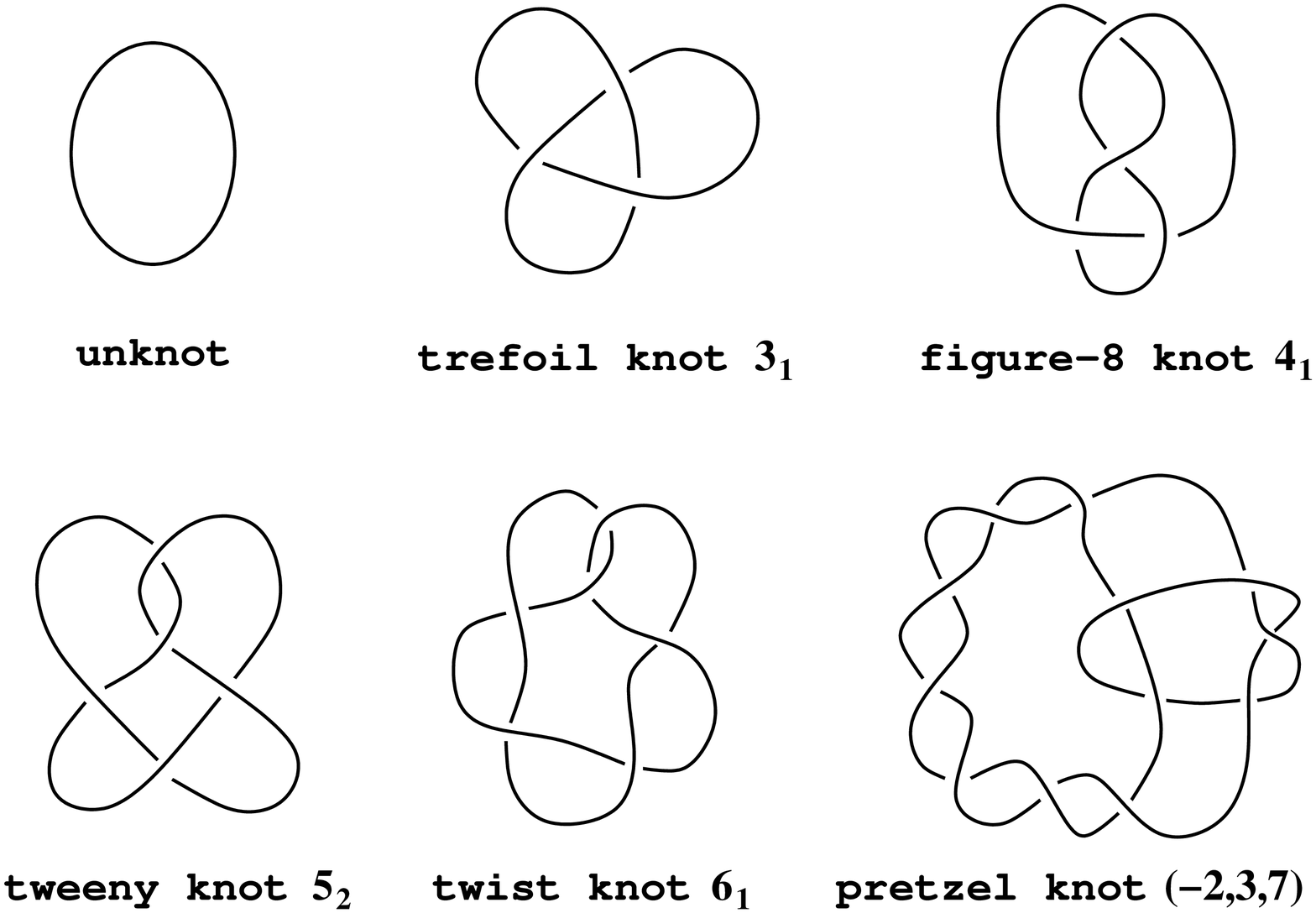}}

For simplicity, in what follows we focus on a large class of compact
oriented 3-manifolds with a single torus boundary component, $\Si = T^2$.
In particular, $M$ can be an exterior of a knotted Wilson loop,
which we denote by $\g=K$, in a rational homology sphere.
For example, see \knots\ for some simple knots in the $\S^3$.
Specifically, let $N(K)$ be a tubular neighborhood of the knot $K$;
it is homeomorphic to a solid torus, $N \cong {\bf D}^2 \times \S^1$.
Then, the knot complement (also known as a knot exterior),
\eqn\mkcompl{M=\S^3 \setminus N(K)}
is a compact 3-manifold whose boundary is a torus, $\Si = T^2$.
(This construction reminds the creation of the so-called ``stretched
horizon'' in the context of Euclidean gravity \refs{\Thorne,\STU}.)
The group $\pi_1 (M)$ is usually called the knot group,
and $\pi_1 (\Si)$ is called the peripheral subgroup of $M$.


\subsec{$SL(2,\IC)$ Chern-Simons Theory and the A-polynomial}

Of particular interest is a Chern-Simons theory
with complex gauge group $\CCG=SL(2,\IC)$
due to its relation to three-dimensional Euclidean gravity.
Indeed, as we already mentioned in the previous section,
the $SL(2,\IC)$ Chern-Simons action \iaction\ with
$t = - \bar t = \s$ is equivalent to the Einstein-Hilbert action,
$I_{{\rm grav}}$, written in the first order formalism.
Correspondingly, the classical field equations \feom\
in the Chern-Simons theory have the form of the usual Einstein
equations in three-dimensional general relativity
with negative cosmological constant, $\Lambda=-1$,
\eqn\einsteq{ R_{ij} = - 2 g_{ij} }
written in the first-order formalism.
In three dimensions, the full curvature tensor is completely
determined by the Ricci tensor\foot{One has the following identity:
$R_{ijkl} = g_{ik} R_{jl} + g_{jl} R_{ik} - g_{jk} R_{il} - g_{il} R_{jk}
- {1 \over 2} \left(g_{ik} g_{jl} - g_{il} g_{jk} \right) R$.}
and \einsteq\ implies that $g_{ij}$ is a metric with
constant negative curvature.

Therefore, hyperbolic metrics on a three-manifold $M$
can be naturally viewed as classical solutions to
the field equations \feom\ in the Chern-Simons theory.
Any hyperbolic manifold is locally isometric to the hyperbolic
three-space, $\IH^3$. We remind that $\IH^3$ can be defined
as the upper half-space with the standard hyperbolic metric
\eqn\hthreemetric{ds^2 = {1 \over z^2} (dx^2 + dy^2 + dz^2)
\quad , \quad z>0 }
If a hyperbolic space $M$ is geodesically complete,
it can be represented as a quotient space,
\eqn\mquotient{M = \IH^3 / \Gamma}
where a discrete, torsion-free subgroup $\Gamma \subset PSL(2,\IC)$
is a holonomy representation of the fundamental group $\pi_1 (M)$
into ${\rm Isom}_+ (\IH^3) = PSL(2,\IC)$.
Fortunately, every holonomy representation lifts to
a representation \generalrho\ into the matrix group $SL(2,\IC)$,
which is much easier to deal with \refs{\CullerS, \Thurston}.
In particular, the action of $\Gamma$ on $\IH^3$
can be conveniently expressed by identifying a point
$(x,y,z) \in \IH^3$ with a quaternion $q=x+y{\bf i}+z{\bf j}$
and writing
\eqn\gammaonh{q \mapsto (aq + b)/(cq + d)
\quad, \quad \pmatrix{a & b \cr c & d} \in SL(2,\IC) }

Since our main examples are three-manifolds \mkcompl\ with
a single torus boundary, which can be represented as knot
complements in a three-sphere, it is natural
to ask if such manifolds can admit a hyperbolic structure.
It turns out that many of them can. Namely, a famous
theorem of Thurston says that, unless $K$ is a torus or
a satellite knot, its complement admits a hyperbolic
metric \Thurston. Such knots are called hyperbolic.

\example{The Figure-eight Knot}
Among all hyperbolic knots, a complement of the figure-eight
knot 4$_1$ (see \knots)
has the least possible volume:
$\Vol ({\rm 4}_1) = 2.0298832128 \ldots$,
and admits a decomposition into two regular ideal
tetrahedra, $M = \Delta_0 \cup \Delta_0$.
The knot group, $\pi_1 (M)$, is generated by two elements,
$a$ and $b$, such that $a^{-1} b a b^{-1} ab = ba^{-1}ba$.
The corresponding representation into $SL(2,\IC)$ is given by
$$
\rho(a) = \pmatrix{1 & 1 \cr 0 & 1} \quad {\rm and} \quad
\rho(b) = \pmatrix{1 & 0 \cr {1 - \sqrt{-3} \over 2} & 1} 
$$
The complement of the figure-eight knot can be also
represented as a quotient space \mquotient, where the holonomy
group $\Gamma$ is generated by the above two matrices.

\endexample

Therefore, a complete hyperbolic structure on
the complement \mkcompl\ of a hyperbolic knot $K \subset \S^3$
can be naturally associated with a flat connection
in $SL(2,\IC)$ Chern-Simons theory with a Wilson loop $K$
or, equivalently, in a theory without the Wilson loop,
defined on its complement, $M = \S^3 \setminus K$.
However, it is important to stress here that $SL(2,\IC)$
Chern-Simons theory --- which is the main subject of this
paper --- makes sense even if the knot $K$ is not hyperbolic.
In particular, in Appendix B we will discuss torus knots.
Even though such examples are not related to hyperbolic geometry,
one can still study flat $SL(2,\IC)$ connections and their
moduli spaces, \generall\ and \generalp.
This is what we are going to do next for an arbitrary knot $K$.

\ifig\lmcycles{The longitude $\g_l$ and the meridian $\g_m$
furnish a basis of the peripheral subgroup
$\pi_1 (\Si) = \Z \times \Z$.}
{\epsfxsize3.5in\epsfbox{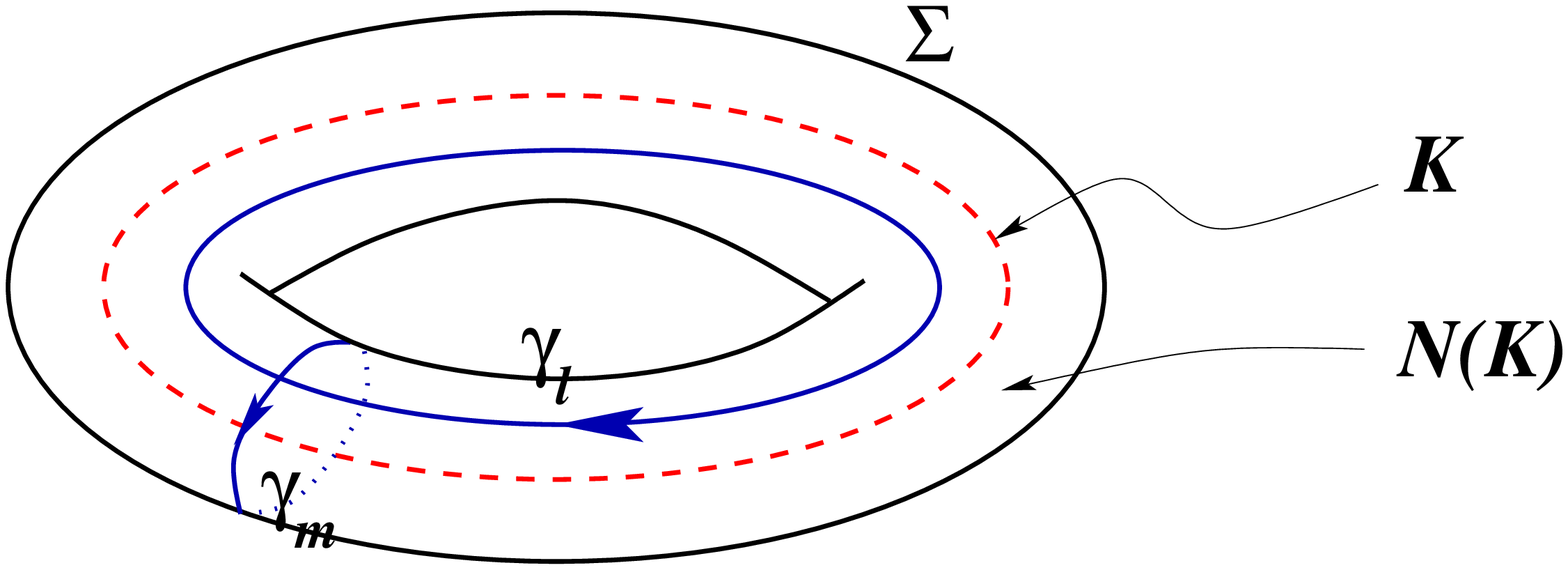}}

In the case of $SL(2,\IC)$ Chern-Simons theory defined on
a knot complement $M$ we can give a rather explicit description
of the representation variety, $L$, and the classical phase space, $\CP$.
Since the boundary of $M$ is a torus, $\Si=T^2$,
there are two simple closed curves, $\g_l$ and $\g_m$,
on $M$ called the longitude and meridian, which intersect
transversely in a single point, see \lmcycles.
These two curves generate the peripheral subgroup of $M$,
which is a free abelian group $\pi_1 (\Si) = \Z \times \Z$.
Therefore, a representation, $\rho$, of this group into $SL(2,\IC)$
can be conjugated to the upper triangular matrices
\eqn\lmholonomies{
\rho (\g_l) = \pmatrix{l & * \cr 0 & 1/l}
\quad , \quad
\rho (\g_m) = \pmatrix{m & * \cr 0 & 1/m} }
where $l$ and $m$ are complex numbers.
No further conjugacy of $\rho$
is possible\foot{Following \CCGLS, in our discussion
we suppress the quotient by the Weyl group.},
so that we can say that the pair of `eigenvalues' $(l,m)$
is the invariant data parametrizing the representation
of the peripheral subgroup into $SL(2,\IC)$.
This also gives an implicit description of the infinite dimensional
representations that we associate with Wilson loops $W_R (K)$
in the Chern-Simons path integral \zwpathint.
In order to emphasize the explicit dependence on $l$ and $m$,
in what follows we sometimes refer to such Wilson loops as $W_{(l,m)} (K)$.
Depending on the values of the holonomies,
such Wilson lines can be interpreted as either
black holes or massive particles\foot{
Namely, a Wilson line should be interpreted as a massive particle
when the holonomy $\rho (\gamma_m)$ is elliptic, and as a spinning
black hole otherwise, see {\it e.g.} \KirillLiouv.
I wish to thank K.~Krasnov for useful discussions
on various points related to this interpretation.
}
in three-dimensional gravity.
On the other hand, in the hyperbolic geometry of
the 3-manifold $M$, such Wilson lines correspond
to torus boundaries (cusps).
Roughly, the complex numbers $l$ and $m$ parameterize
the ``length'' and the ``torsion'' of the cusp.

Therefore, we find that for a compact oriented 3-manifold $M$
with a single torus boundary
the classical phase space $\CP$ is a two-dimensional
complex manifold
\eqn\pcstar{\CP = \IC^* \times \IC^*}
parametrized by the eigenvalues $(l,m)$ of the holonomies \lmholonomies.
Notice, that the phase space $\CP = \IC^* \times \IC^*$ can be regarded
as the total space of the cotangent bundle,
\eqn\pcot{\CP = T^* (\S^1 \times \S^1)}
where the two circles are parametrized by
$\arg (l)$ and $\arg (m)$,
and the directions along the fiber are parametrized by
$\log \vert l \vert$ and $\log \vert m \vert$.
Here, the base manifold,
\eqn\ssrep{\S^1 \times \S^1 = {\rm Hom} (\pi_1 (\Si) ; SU(2)) / SU(2) }
can be viewed as a representation space of
the peripheral subgroup, $\pi_1 (\Si)$,
into the compact part of the gauge group, $G=SU(2)$.

Our next task is to describe the representation variety,
\eqn\ldef{L = {\rm Hom} ( \pi_1 (M) ; SL(2,\IC)) / SL(2,\IC) }
Thurston showed that for a space $M$ with a single torus boundary,
the dimension of the numerator in this formula is equal to 4.
On the other hand, the Lie group $SL(2,\IC)$ has complex dimension 3.
Therefore, after we identify conjugate representations in \ldef\
we obtain a variety
of complex dimension one.
Furthermore, a basis $(\g_l, \g_m)$ for the peripheral subgroup of $M$
determines an embedding of $L$ into $\CP = \IC^* \times \IC^*$, and
using the standard techniques from algebraic geometry one
can show that the variety $L$ is the zero locus of a single
polynomial $A(l,m)$ in two variables, the so-called A-polynomial \CCGLS.
In a sense, in the $SL(2,\IC)$ Chern-Simons theory the A-polynomial
plays a role similar to the role of the Seiberg-Witten curve
in $\CN=2$ supersymmetric gauge theory \SW, which also describes
a moduli space of vacua. In fact, as we shall see below,
interesting physical quantities in the $SL(2,\IC)$ Chern-Simons
theory can be also expressed in terms of the period integrals
on the curve $L$ defined as the  zero locus of the A-polynomial.

Explicitly, we can write $L$ as:
\eqn\locusa{L = \{ (l,m) \in \IC^* \times \IC^* ~\vert~ A(l,m)=0 \}}
In three-dimensional Euclidean gravity, the A-polynomial
can be naturally interpreted as a mass shell condition
for a ``self-gravitating'' massive spinning particle
or a black hole
propagating along a knot $K$ in the $\IR^3$ (or in the 3-sphere).
For example, the A-polynomial of the unknot (trivial knot) is
\eqn\aunknot{A_{{\rm unknot}}(l,m)=l-1}
due to the abelian representations.
Essentially, this result appears in the study of the off-shell
BTZ black hole \refs{\CToffshell, \CTaspects, \CarlipBH, \Carliplect}.
In Appendix A, we review the geometry of the BTZ black hole
and reproduce the A-polynomial \aunknot.

Since $H_1(M) \cong \Z$ for any knot complement \mkcompl,
there is always a component of the character variety $L$ corresponding
to abelian representations. Usually, one ignores this component,
which contributes a factor of $(l-1)$ to the A-polynomial\foot{
However, in quantization of the theory it will be crucial
to include all the components of $L$.
I am grateful to G.~Moore and D.~Thurston
for emphasizing this aspect.}.
There is still some ambiguity in the definition of the A-polynomial.
Namely, the A-polynomial is defined up to scaling
and up to multiplication by powers of $l$ and $m$.
It turns out, however, that there is a natural normalization,
such that $A(l,m)$ is a polynomial with integer coefficients \CCGLS.
This is the normalization that will be used throughout this paper.

The explicit form of the A-polynomial depends on the knot $K$,
and it can be always systematically computed \CCGLS.
A-polynomials of some simple knots are listed below:

\vskip 0.8cm
\vbox{
\centerline{\vbox{
\hbox{\vbox{\offinterlineskip
\def\tablespace{height7pt&\omit&&\omit&&
 \omit&\cr}
\def\tablerule{\tablespace\noalign{\hrule}\tablespace}

\hrule\halign{&\vrule#&\strut\hskip0.2cm\hfill #\hfill\hskip0.2cm\cr
\tablespace
& Knot   && A-polynomial && Volume  &\cr
\tablerule
& $3_1$  && $lm^6 + 1$  && non-hyperbolic  &\cr
\tablespace
& $4_1$  && $-2+m^4+m^{-4}-m^2-m^{-2}-l-l^{-1}$  && $2.0298832 \ldots$  &\cr
\tablespace
& $5_1$  && $lm^{10} + 1$ && non-hyperbolic  &\cr
\tablespace
& $5_2$  && $1 + l(-1+2m^2+2m^4-m^8+m^{10}) + l^2 (m^4 - m^6 +$ &&
$2.8281220 \ldots$ &\cr
&  && $+ 2m^{10} + 2m^{12} - m^{14}) + l^3 m^{14}$ &&  &\cr
\tablespace
& $7_1$  && $m^{14} + l$  && non-hyperbolic  &\cr
\tablespace
& $(-2,3,7)$-pretzel &&
$-m^{110} + l m^{90} (m^2 - 1)^2 + l^2 (2 m^{74} + m^{72})-$
&& $2.8281221 \ldots$  &\cr
&  && $- l^4 (m^{38} + 2 m^{36}) - l^5 m^{16} (m^2 - 1)^2 + l^6$ &&  &\cr
}\hrule}}}}
\centerline{
\hbox{{\bf Table 2:}{\it ~~ A-polynomials of some knots
and the hyperbolic volumes of their complements.}}}
}
\vskip 0.5cm

When the knot $K$ is hyperbolic we can interpret
(a subset of) the zero locus of the A-polynomial as
the moduli space of hyperbolic structures on
the knot complement, $M = \S^3 \setminus K$.
Most of the points in this moduli space correspond
to incomplete hyperbolic metrics on $M$.
However, there are some special values of $l$ and $m$,
for which the metric on $M$ has extra nice properties.
For instance, at the point $(l,m)=(-1,1)$ we obtain
a complete hyperbolic manifold $M$, with a cusp along $K$.
The volumes of such manifolds evaluated at this
particular point in the moduli space are listed in Table 2.

Furthermore, if $l$ and $m$ satisfy an equation of the form
\eqn\lmdehn{l^p m^q=1}
for some co-prime integers $p$ and $q$, then
the knot complement $M$ can be completed into a compact
space $\overline M$ without boundary, which can be also
obtained by performing $(p,q)$-Dehn surgery on the knot $K$
in the 3-sphere:
\eqn\mmpq{\overline M = M(p,q)}
Specifically, let $N(K)$ be the tubular neighborhood of
the knot $K$, as on \lmcycles, and let $\g_l$ and $\g_m$
be the longitude and the meridian of the boundary torus.
Then, the $(p,q)$-Dehn surgery on $K$ is obtained
by replacing $N(K)$ with another solid torus,
such that the meridian in the solid torus matches up
with the curve $\g_l^p \g_m^q$ in $\p M$:
\eqn\surgery{M(p,q) = \left( \S^3 \setminus N(K) \right)
\cup_{T^2} {\bf D}^2 \times \S^1}

Except for a finite number of cases, the resulting
3-manifold $M(p,q)$ admits a hyperbolic metric.
The volume of the space $M(p,q)$ is always smaller
than the volume of the ``parent'' manifold $M_{cusped}$
with a cusp, and as $p^2+q^2 \to \infty$ the volume
of $M(p,q)$ approaches the volume of $M_{cusped}$,
as illustrated on the figure below:
\bigskip

\ifig\volumes{The volumes of the hyperbolic 3-manifolds
$M(p,q)$ obtained by Dehn surgery on the figure-eight knot
converge to the volume of the cusped manifold.}
{\epsfxsize3.5in\epsfbox{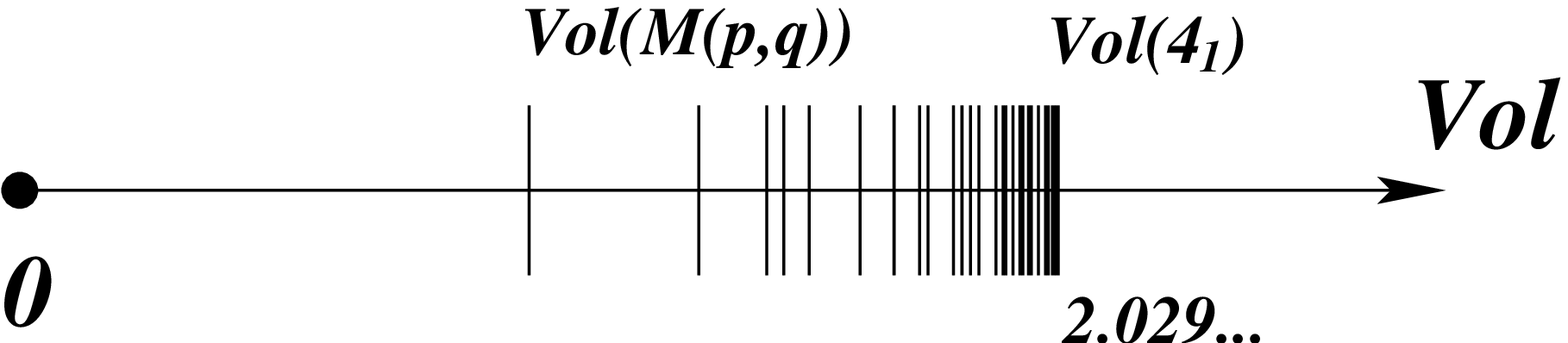}}

For example, for the figure-eight knot
the sequence $\Vol (M(p,q))$ looks like \NZagier:
\eqn\pqasympt{
\Vol \left( M(p,q) \right)
= \Vol ({\bf 4}_1) - {2 \sqrt{3} \pi^2 \over p^2 + 12q^2}
+ \ldots}
%

\subsec{Properties of the A-polynomial}

The A-polynomial of a knot has a number of interesting properties.
For example, as we already mentioned earlier, one can choose
a normalization such that $A(l,m)$ has integer coefficients.
For completeness, here we list some other basic properties of
the A-polynomial that will be useful to us in the discussion below
(however, one may skip this section in the first reading):

\item{$a)$}
If $K$ is a hyperbolic knot, then $A(l,m) \ne l-1$.
This already asserts that $A(l,m)$ is non-trivial
for a large class of knots.

\item{$b)$}
If $K$ is a knot in a homology sphere, then the A-polynomial
involves only even powers of $m$, {\it cf.} Table 2.

\item{$c)$}
Reversing the orientation of $K$ does not
change $A(l,m)$, but reversing the orientation of
the ambient space changes $A(l,m)$ to $A(l,1/m)$.
Note, in particular, that the A-polynomial is powerful
enough to distinguish mirror knots.

\item{$d)$}
An important property of the A-polynomial
is that under the change of basis
\eqn\basischange{
\pmatrix{\g_l \cr \g_m} \longrightarrow \pmatrix{a & b \cr c & d}
\pmatrix{\g_l \cr \g_m},
\quad \quad \pmatrix{a & b \cr c & d} \in SL(2,\Z)}
the A-polynomial transforms as
\eqn\abasischange{
A(l,m) \longrightarrow A(l^dm^{-b}, l^{-c}m^a).}

\item{$e)$}
The A-polynomial is reciprocal, {\it i.e.} $A(l,m)=\pm A(1/l,1/m)$
up to powers of $l$ and $m$.

\item{$f)$}
The A-polynomial is tempered, {\it i.e.} the faces
of the Newton polygon of $A(l,m)$ define cyclotomic
polynomials in one variable.

\item{$g)$}
If $K_1$ and $K_2$ are two knots and $K_1 \# K_2$
is their connected sum, then $A_{K_1 \# K_2}$ is divisible
by $A_{K_1} \cdot A_{K_2}/(l-1)$.

\item{$h)$}
Under some mild technical assumptions, the A-polynomial
of a knot complement has a property that
$A(l,\pm 1) = \pm (l-1)^{n_+} (l+1)^{n_-}$ for some
non-negative integers $n_+$ and $n_-$ \CCGLS.
The geometric interpretation of these integer numbers is not
known at present.

\item{$i)$}
Although it seems unlikely that the A-polynomial can be defined by
iterative skein relations as, for example, the Jones or Alexander polynomials,
there is a relation between the A-polynomial and the Alexander polynomial.
However, the A-polynomial is a strictly stronger invariant of knots
than the Alexander polynomial \CCGLS.

\item{$j)$} On the other hand, the A-polynomial is not
a complete invariant of knots: there are different knots
with the same A-polynomial, see {\it e.g.} \CLrem.
In particular, mutants seem to have the same A-polynomial.

\item{$k)$}
In the mathematical literature, the A-polynomial of a knot was used
to extract a great deal of subtle information about the knot complement.
For example, the slopes of the sides of the Newton polygon
of $A(l,m)$ are boundary slopes of incompressible
surfaces\foot{A proper embedding of a connected orientable
surface $F \to M$ is called incompressible if the induced
map $\pi_1 (F) \to \pi_1 (M)$ is injective.
Its boundary slope is defined as follows.
An incompressible surface $(F,\p F)$ gives rise to a collection
of parallel simple closed loops in $\p M$. Choose one such loop
and write its homology class as $\g_l^a \g_m^b$. Then, the boundary
slope of $(F,\p F)$ is defined as a rational number $a/b$.}
in $M$ which correspond to ideal points of $L$.

\noindent
Further properties of the A-polynomial can be found
in \refs{\CCGLS,\CLrem,\CLrep}.


\newsec{Quantization}

\subsec{A-polynomial as a Lagrangian Submanifold}

In the previous section we described the classical solutions in
a $SL(2,\IC)$ Chern-Simons theory with a single Wilson loop
$W_{(l,m)} (K)$ or, equivalently, the classical solutions in a theory
without the Wilson loop, defined on its complement, $M = \S^3 \setminus K$.
In particular, we found that the classical solutions
are parametrized by points on an algebraic curve $L$,
which is a zero locus of the A-polynomial, $A(l,m)$.
Since the knot complement $M$ is a compact 3-manifold with
a single torus boundary, $\Si=T^2$, there is a natural embedding
\eqn\linp{L \hookrightarrow \CP}
induced by restricting a flat connection from $M$ to $\Si$.
Here, $\CP = \IC^* \times \IC^*$ is the moduli space
of representations of $\pi_1(\Si)$ in $SL(2,\IC)$,
parametrized by two complex numbers, $l$ and $m$,
{\it cf.} \lmholonomies.

In the Hamiltonian approach \WittenJones, one regards $\CP$
as a classical phase space associated to a closed surface $\Si$.
Indeed, there is a symplectic structure $\om$ on $\CP$
derived from the classical Poisson bracket of gauge fields in
the Chern-Simons theory \iaction.
In order to see this explicitly, let us study the theory
on $\Si \times \IR^1$, {\it i.e.} near the boundary of $M$.
On $\Si \times \IR^1$ it is natural to choose the gauge $A_0=0$.
Then, the Chern-Simons Lagrangian \iaction,
which in this gauge looks like
\eqn\anotgauge{
I = {t \over 8 \pi} \int dt \int_{\Si} \e^{ij} \Tr \CA_i {d \over dt} \CA_j
+ {\bar t \over 8 \pi}
\int dt \int_{\Si} \e^{ij} \Tr \bar \CA_i {d \over dt} \bar \CA_j }
leads to the following Poisson brackets:
\eqn\cspb{\eqalign{
\{ \CA_i^a (x) , \CA_j^b (y) \}
&= {4 \pi \over t} \cdot \d^{ab} \e_{ij} \d^2(x-y) \cr
\{ \bar \CA_i^a (x) , \bar \CA_j^b (y) \}
&= {4 \pi \over \bar t} \cdot \d^{ab} \e_{ij} \d^2(x-y)
}}

These relations induce a Poisson structure on the classical
phase space $\CP$, which is parametrized by the $SL(2,\IC)$
holonomies \lmholonomies\ around the longitude $\g_l$ and
the meridian $\g_m$.
It is convenient to introduce new variables\foot{Note, that
our definition of $u$ and $v$ differs from \refs{\NZagier,\Yoshida,\HLMA}
by a factor of 2.} $u$ and $v$, such that
\eqn\uvdef{l = \exp (u) \quad, \quad m = \exp (v)}
Since the curves $\g_l$ and $\g_m$ intersect transversely in
a single point, from \cspb\ we find (see also \refs{\NRZ,\CarlipBH}):
\eqn\uvpb{\eqalign{
\{ u,v \} & = {4 \pi \over t}\cr
\{ \bar u, \bar v \} & = {4 \pi \over \bar t}
}}

These relations, together with $\{ u,\bar v \} = \{ \bar u , v \} =0$,
lead to the following non-trivial Poisson brackets
for the real and imaginary components of $u$ and $v$:
%
\eqn\uvkspb{\eqalign{
& \{\r (v) , \r (u) \} = \{ \i (u) , \i (v) \}
= - 2 \pi {k \over k^2 - \s^2} \cr
& \{ \r (u) , \i (v) \} = - \{ \r (v) , \i (u) \}
= i 2 \pi {\s \over k^2 - \s^2}
}}
Here we also used the explicit expression for
the parameters $t=k+\s$ and $\bar t = k - \s$,
both of which are real in the present discussion.

Writing the Poisson brackets \uvkspb\ in the form
$\{x^i , x^j\} = \om^{ij}$
and inverting the constant skew-symmetric matrix $\om^{ij}$,
we obtain the corresponding 2-form $\om = \om_{ij} dx^i \wedge dx^j$:
\eqn\totomega{
\om = {k \over \pi} \om_k + i {\s \over \pi} \om_{\s} }
where $\om_k$ and $\om_{\s}$ are real non-degenerate
2-forms on $\CP$, given by
\eqn\komega{
\om_k = d \r (v) \wedge d \r (u) + d \i (u) \wedge d \i (v) }
and
\eqn\somega{
\om_{\s} = - d \r (u) \wedge d \i (v) + d \r (v) \wedge d \i (u) }
The differential 2-form $\om$ is manifestly closed, $d \om =0$.
Also, notice that $\om$ has both real and imaginary components.
If its imaginary part vanishes (that is, if $\s \in i\IR$),
we can interpret \totomega\ as the usual symplectic
structure on the phase space $\CP$.
This system can be easily quantized
by regarding $u$ and $v$ as operators on a certain Hilbert
space, $\CH_{\Si}$, and by replacing the Poisson brackets
with commutators, $\{~,~\} \to i[~,~]$.
The resulting effective quantum mechanics
will be discussed in more detail below.

On the other hand, if $\s \in \IR$ (and, say, $k=0$),
the Poisson structure \uvkspb\ gives rise to an imaginary 2-form
$\om = i (\s / \pi) \om_{\s}$,
where $\om_{\s}$ is the natural symplectic structure on $\CP$.
Indeed, the phase space $\CP = \IC^* \times \IC^*$ can be regarded
as the total space of the cotangent bundle \pcot,
with the natural symplectic 2-form \somega.
However, since in \totomega\ we have an imaginary multiple
of this 2-form, the quantization of this system is
more naturally interpreted as a Euclidean quantum mechanics.
In general, if both $k$ and $\s$ are non-zero,
we deal with a mixed situation.
Notice, that this subtlety does not appear in a Chern-Simons
theory with real parameter $s$, which is related to
de Sitter gravity in 2+1 dimensions.
In this theory, one finds the Poisson brackets \uvkspb\
with purely real values on the right-hand side.
Therefore, a quantization of such system leads to
an ordinary quantum mechanics on a ``non-commutative'' 2-torus,
parametrized by $\i (v)$ and $\i (u)$.

To summarize, we found that the classical Poisson bracket in
the Chern-Simons theory gives $C^{\infty} (\CP)$ the structure
of a Lie algebra over $\IC$, and suggests to interpret
$(\CP,\om)$ as a classical Hamiltonian system.
To make this interpretation even more explicit, by a linear change
of variables we can always bring the 2-form $\om$ to the canonical form
\eqn\canomega{\om = \sum_i dp_i \wedge dq_i}
where $q_i$ and $p_i$, $i=1,2$, are the canonical coordinates
and momenta (given by linear combinations of the real and imaginary
components of $u$ and $v$).

In Hamiltonian mechanics, a (semi-)classical state of a system is
described by a Lagrangian submanifold in $\CP$, that is
a middle dimensional subvariety $L \subset \CP$ such that
the restriction of $\om$ to $L$ vanishes,
\eqn\lagrcond{\om \vert_L =0}
In our problem, a character variety $L$ that was defined
in \locusa\ as the zero locus of the A-polynomial
is a natural candidate for a classical state.
Indeed, it is a middle dimensional subvariety in $\CP$ which,
by definition, describes the classical field configurations
on the boundary $\Si = \p M$ that can be extended to a 3-manifold $M$.
On the other hand, $L$ is defined
as a holomorphic curve in $\CP = \IC^* \times \IC^*$,
with respect to its natural complex structure,
rather than a Lagrangian submanifold\foot{A similar
situation occurs in a different, though possibly
not unrelated context in string theory \LMV.}.
However, by performing a hyperK\"ahler rotation it is easy
to check that $L$ is indeed a Lagrangian submanifold,
with respect to the symplectic structures \komega\ and \somega\
(and, hence, with respect to any linear combination thereof).
Specifically, we can write the 2-form \komega\ as
\eqn\komegaviauv{\eqalign{
\omega_{k} & = d \r (v) \wedge d \r (u) + d \i (u) \wedge d \i (v) = \cr
& = {1 \over 4} (dv + d \bar v) \wedge (du + d \bar u)
- {1 \over 4} (du - d \bar u) \wedge (dv - d \bar v) = \cr
& = - {1 \over 2} (du \wedge dv + d \bar u \wedge d \bar v)
}}
Clearly, this 2-form vanishes
when restricted to the holomorphic curve $A(e^u,e^v)=0$.
Similarly, one can check that the restriction of
the 2-form \somega\ also vanishes
\eqn\somegaviauv{\eqalign{
\omega_{\s} & = - d \r (u) \wedge d \i (v) + d \r (v) \wedge d \i (u) = \cr
& = - {1 \over 4i} (du + d \bar u) \wedge (dv - d \bar v)
+ {1 \over 4i} (dv + d \bar v) \wedge (du - d \bar u) = \cr
& = {i \over 2} (du \wedge dv - d \bar u \wedge d \bar v)
}}
when restricted to $L$.
Hence, the same is true for any linear combination of
$\om_k$ and $\om_{\s}$.
In particular, the condition \lagrcond\ holds for
the complex 2-form \totomega.

We conclude that in the $SL(2,\IC)$ Chern-Simons theory
the zero locus $L$ of the A-polynomial is naturally regarded as
a Lagrangian submanifold in the phase space $(\CP,\om)$.
Hence, we are naturally led to the quantization of $(\CP,\om)$
in a real polarization\foot{For quantization of
Chern-Simons theory in complex polarizations
see \refs{\Wittencpx,\ADPW,\Hitchin,\Hayashi}.},
where elements of quantization are
associated with Lagrangian submanifolds, {\it cf.} \refs{\Weitsman}.
Then, it is natural to expect that
a quantization of this Hamiltonian system
gives the quantum Hilbert space, $\CH_{\Si}$,
of the $SL(2,\IC)$ Chern-Simons theory in genus one.
In particular, the semi-classical partition function \zpathint\
for a 3-manifold $M$ with a single torus boundary, $\Si = T^2$,
is given by the semi-classical wave function, $Z(M) \in \CH_{\Si}$,
supported on the Lagrangian submanifold $L$.
To be more precise, the partition function $Z(M)$
is a half-density\foot{Given an $n$-dimensional
manifold $X$, let $\CB X$ denote the frame bundle of $V=TX$; its
structure group is $GL(V)$. Then, an $r$-density on $X$ is
a smooth function $f \colon \CB X \to \IC$ which transforms under
the action of $GL(V)$ as $f \circ g = \vert \Det (g) \vert^r f$,
where $g \in GL(V)$ and $\Det(g)$ is the determinant of $g$.
For $r=1$ this definition gives the usual density on $X$,
and for $r=1/2$ it gives a half-density on $X$.} on $L$,
and the resulting quantum system can be naturally viewed
as a quantum mechanics on a ``non-commutative'' 2-torus.
Below, our goal will be to explain this prescription in more detail
and to verify that it leads to meaningful results.

\example{The Figure-eight Knot}
As in the previous example, let $K$ be the figure-eight knot,
and let $M=\S^3 \setminus K$ be its complement.
For a moment, let us also set $k=0$.
Then, the Poisson brackets \uvkspb\ suggest to define
the canonical variables on the four-dimensional
phase space $\CP$ as
\eqn\pqvarss{
\pmatrix{p_1 \cr p_2} = \pmatrix{- \r (u) \cr \r (v)}
\quad {\rm and} \quad
\pmatrix{q_1 \cr q_2} = \pmatrix{\i (v) \cr \i (u)}
}
In terms of these variables, the zero locus of the A-polynomial
of the figure-eight knot, with the factor $(l-1)$ included, is given by
\eqn\lpqfeight{
\Big( \cosh (4p_2 + 4iq_1) - \cosh (2p_2 + 2iq_1)
- \cosh (p_1 - iq_2) - 1 \Big) \left( e^{iq_2 - p_1} -1 \right) =0}
This complex equation defines two real constraints in a dynamical
system with trivial Hamiltonian,
whose quantization leads to an effective quantum mechanics
on a torus, parametrized by $q_1$ and $q_2$.
When the parameter $s$ in the Chern-Simons action \iaction\ is real,
one finds ordinary quantum mechanics, whereas for imaginary values
of $s$ ({\it i.e.} for $s=-i \s$) the resulting system
is best described as a Euclidean quantum mechanics.
In both cases, turning on the `level' $k$ leads to
a deformation of the Poisson structure \uvkspb,
which can be naturally interpreted as a non-commutativity
of the torus coordinates, $q_1$ and $q_2$
(as well as momenta, $p_1$ and $p_2$).

\endexample

Before we proceed, let us briefly point out that
the Chern-Simons action induces $SL(2,\IC)/SU(2)$ WZW theory
on the boundary manifold $\Si$.
However, the structure of the Hilbert space in
this theory is not completely clear at present.
One might hope to get further insights from
the present approach.

\subsec{Semi-Classical Approximation}

Following the general prescription formulated in the end of
the previous subsection, here we study the semi-classical
behavior of the partition function $Z(M)$ by quantizing
the Hamiltonian system associated with $(\CP,\om)$
and a Lagrangian submanifold $L$.
As in the standard classical mechanics, we introduce
a canonical 1-form (also known as a Liouville form),
which in the canonical variables $(p_i,q_j)$ can be written as
\eqn\canoneform{\th = \sum_i p_i d q_i}
We note that this 1-form may not be globally defined;
this happens, for example, when the phase space is compact.
However, locally we can always write $\om = d \th$, and because
$\om \vert_L =0$ holds by the definition of the Lagrangian submanifold $L$,
we find that the restriction of the canonical 1-form to $L$ is closed,
\eqn\oneformonl{d \th \vert_L = 0}
The cohomology class in $H^1(L,\IR)$ induced by the 1-form $\th$
is called the Liouville class of the Lagrangian immersion
$L \hookrightarrow \CP$. Lagrangian immersions for which
the Liouville class is trivial are called exact Lagrangian immersions.

Consistent quantization of $(\CP,\om)$ with a Lagrangian
submanifold $L$ requires the Liouville class of $L$ to obey
certain conditions. In order to explain these conditions,
let us proceed with the WKB quantization and define
a phase function $S$, usually called the action integral,
such that
\eqn\sdef{dS = \th \vert_L}
Then, the semi-classical approximation to the wave function
supported on the Lagrangian submanifold $L$ is given by
\eqn\semiclassb{Z \simeq \sum_{\a}
\psi \cdot e^{iS/ \hbar} e^{i \pi \eta /4}
+ O(\hbar) }
where the sum is over a discrete, finite set of the components of $L$,
$\eta$ is the famous Maslov correction to the phase of
the semi-classical wave function, and the amplitude $\psi$
is a half-density on $L$,
which obeys the transport equation \refs{\Woodhouse,\GQlect}:
\eqn\transport{\CL \psi = 0}
%

In order to understand the meaning of each term in \semiclassb,
it is instructive to consider a finite dimensional integral
\eqn\zex{Z = \int \prod_i dq_i ~e^{iS(q_i)/ \hbar}}
In the stationary phase approximation, this integral is
dominated by the contribution from the critical points,
\eqn\semiclassa{Z \simeq \sum_{\a}
{e^{iS/ \hbar} \over \sqrt{\det \big( -i \cdot \Hess (S) \big)}}
+ O(\hbar) }
where $\Hess (S)$ stands for the Hessian of the phase function $S$.
Here, a phase of the square root of the determinant requires extra care.
Namely, if $S$ is a real function,
$\Hess (S)$ has only real eigenvalues, $\la_i$.
Each positive eigenvalue contributes $(-i)^{-1/2} = e^{i \pi /4}$
to the phase of the above expression, whereas each negative eigenvalue
contributes $i^{-1/2} = e^{- i \pi /4}$.
Therefore, a refined version of the formula \semiclassa\ looks like:
\eqn\semiclassc{Z \simeq \sum_{\a}
e^{iS/ \hbar} e^{i \pi \eta /4}
~\vert \det \big( \Hess (S) \big) \vert^{-1/2}
+ O(\hbar) }
where the $\eta$-invariant
\eqn\etadef{\eta = \sum_i {{\rm sign~}} \la_i}
is defined as the (regularized) signature of the Hessian, $\Hess(S)$,
that is the total number of positive eigenvalues of $\Hess(S)$
minus the total number of negative eigenvalues of $\Hess(S)$.

Now, let us return to the semi-classical wave function \semiclassb\
associated with a Lagrangian submanifold $L$.
Notice, that when the restriction of $\th$ to
$L$ is an exact 1-form,
the phase function $S = \int \th$ is well defined.
However, if the Lioville class of $\th$ is non-trivial, the phase
function $S$ depends on the choice of the integration path in $L$.
In particular, a difference between two such choices
will change the action integral by a period of $\th$,
$$
\Delta S = \oint_C \th,
$$
where $C$ is a closed 1-cycle on $L$.
In order for the semi-classical expression \semiclassb\ to be
unambiguously defined, all the periods of $\th$ must be
integer multiples of $2 \pi \hbar$, for some value of $\hbar$:
\eqn\quantcond{
{1 \over 2 \pi \hbar} \oint_C \th \in \Z
\quad, \quad \forall ~C \in \pi_1 (L) }
A Lagrangian submanifolds $L \hookrightarrow \CP$
which obeys this condition is called quantizable,
and the corresponding values of $\hbar$ for which
\quantcond\ holds are called admissible for $L$.
Notice, that the set of quantizable Lagrangian submanifolds
with large first Betti number is usually rather limited.

\ifig\pdqint{The moduli space of the $SL(2,\IC)$ Chern-Simons theory
on a knot complement is described by the A-polynomial of the knot.
The action integral of the corresponding quantum mechanics can be
obtained by integrating $\theta = \sum_i p_i dq_i$ over
a path on the curve $A(l,m)=0$.}
{\epsfxsize4in\epsfbox{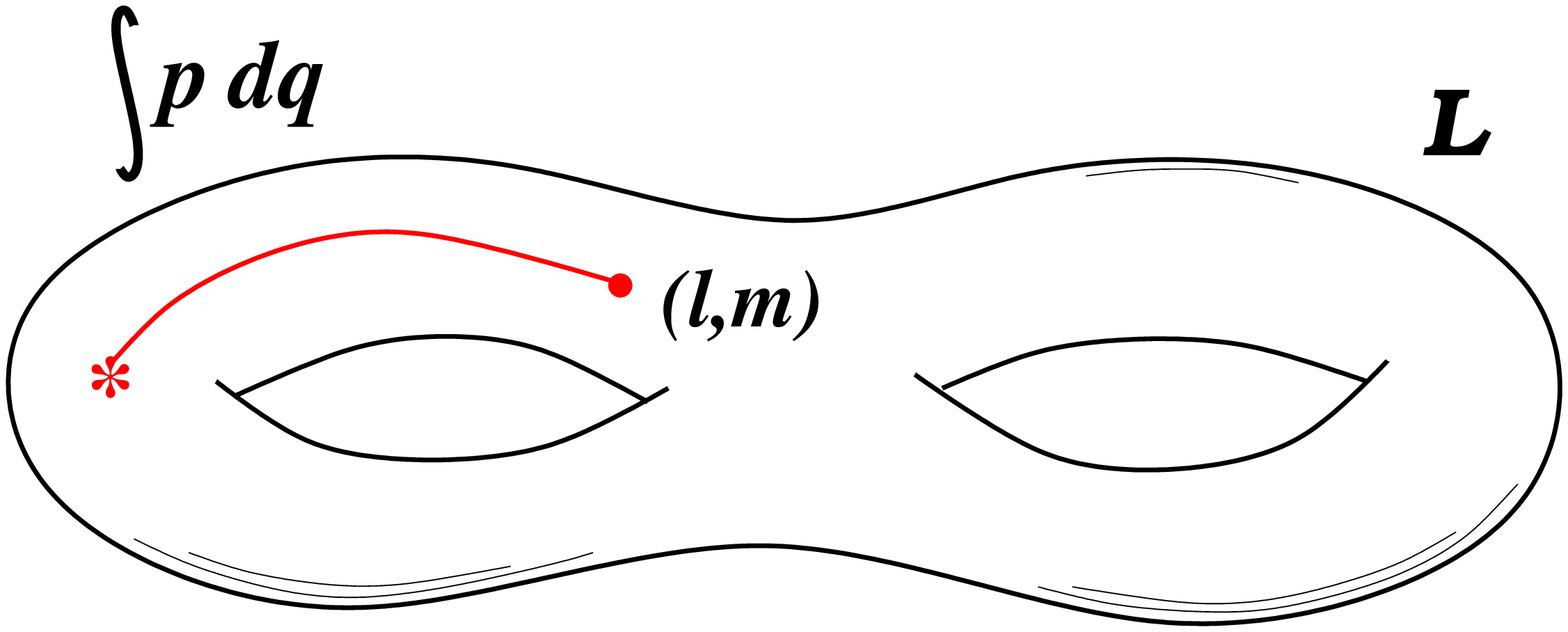}}

Now, let us apply this to our problem.
We set $\hbar=1$; the role of the Planck constant
is played by $k^{-1}$ and $\s^{-1}$, so that
the semi-classical limit corresponds to $k,\s \to \infty$.
Furthermore, from the defining equations \totomega\
and \canoneform\ we obtain a canonical 1-form,
\eqn\tottheta{
\th = {k \over \pi} \th_k + i {\s \over \pi} \th_{\s} }
where $\th_k$ and $\th_{\s}$ are real 1-forms,
\eqn\ktheta{
\th_k = \r (v) d \r (u) + \i (u) d \i (v) }
and
\eqn\stheta{
\th_{\s} = - \r (u) d \i (v) + \r (v) d \i (u) }

Since the 1-form $\th$ is complex-valued, the condition
for $L$ to be quantizable implies two independent sets of
constraints corresponding, respectively, to the imaginary
and real parts of the Bohr-Sommerfeld condition \quantcond:
\eqn\ksquantcond{
\underline{L {\rm ~quantizable~}:} \quad \quad \quad
\eqalign{& \oint_C \th_{\s} =0 \cr
{1 \over \pi^2} & \oint_C \th_k \in \IQ }}
Provided that these two conditions are satisfied for any
1-cycle $C \subset L$, the system can be consistently quantized
and one can write the semi-classical expression for
the partition function in the form \semiclassb.
The first condition in \ksquantcond\ asserts that
all the periods of $\th_{\s}$ must vanish.
Equivalently, $\th_{\s}$ should be an exact 1-form,
when restricted to $L$.

It turns out that, for a Lagrangian submanifold $L$
defined as a zero locus of the A-polynomial,
the restriction of $\th_{\s}$ to $L$ is always an exact 1-form.
Specifically, using a classical formula of Schl\"afli concerning
the volume of infinitesimally deformed polyhedra, one can show that
\eqn\volfla{\th_{\s} \vert_L
= {1 \over 2} d \Vol (M) }
where $\Vol (M)$ is the volume of the hyperbolic 3-manifold $M$,
which in the present discussion is a complement of a knot $K$.
This important result is due to Hodgson \Hodgson\
with improvements by Dunfield \Dunfield.
Similarly, the real part of the 1-form $\th$ is related
to the Chern-Simons invariant of the 3-manifold $M$,
\eqn\csfla{ \th_k \vert_L = - \pi^2 d CS (M) }
This formula was originally conjectured by Neumann and Zagier \NZagier,
and later proved by Yoshida \Yoshida, see also \refs{\KKlassen,\HLMA}.

{}From the equations \volfla\ and \csfla\ it follows that
imaginary and real components of the action integral $S$ are related,
respectively, to the volume and the Chern-Simons invariant
of the knot complement $M$:
\eqn\tots{S = {i\s \over 2 \pi} \Vol (M) - \pi k CS (M) }
This is indeed the expected semi-classical behavior of
the $SL(2,\IC)$ Chern-Simons action \iks.
We can also write it as
\eqn\totsviaf{dS = t dF + \bar t d \bar F}
where
$$
dF = {1 \over 4 \pi} \Big( v du - udv + d (u \bar v) \Big)
$$

Combining the formulas \semiclassb\ and \tots\ together, we obtain:
\eqn\semiclassz{Z(M) \simeq \sum_{\a}
\psi \cdot e^{i \pi \eta /4} \cdot
\exp \left(
- {\s \over 2 \pi} \Vol (M) - i \pi k CS (M)
\right) + \ldots }
This is the usual expression for the semi-classical wave
function in a quantum mechanics with the action integral \tots.
Let us now compare this result with a general form of
the (unnormalized) semi-classical partition function
in a Chern-Simons theory \WittenJones:
\eqn\semiclasscs{
Z(M) \simeq \sum_{\a}
{\det (\Delta) \over \sqrt{\vert \det (L_-) \vert} }
e^{i \pi \eta (\CA^{(\a)})/4} \cdot e^{i I(\CA^{(\a)})}
+ \ldots}
where $\Delta$ is the standard Laplacian and
$L_-$ denotes a restriction of the self-adjoint operator
$L=*D + D*$ to the space of odd forms on $M$
(not to be confused with the character variety $L$).
For a Chern-Simons theory with complex gauge group
one further has \refs{\Wittencpx,\BNWitten}:
\eqn\zeroeta{ \eta (\CA^{(\a)}) =0 }

It is clear that our result \semiclassz\ is very similar
to \semiclasscs.
In fact, one can easily see the terms in both expression
which have the same origin and, therefore, should be identified.
For example, each critical point in the quantum mechanics
problem corresponds to a flat $SL(2,\IC)$ connection, $\CA^{(\a)}$,
in the Chern-Simons theory.
Among these flat connections, there is a geometric one,
associated with the hyperbolic structure on the knot complement $M$.
The Chern-Simons action, $I(\CA^{(hyperb)})$, of this flat connection
is given by \tots, and it was already discussed in this section.
Moreover, $\eta (\CA^{(\a)})$ should be identified with
the Maslov correction to the phase of the semi-classical
expression \semiclassz, and \zeroeta\ implies that this
correction vanishes.
This should also follow directly from
the properties of the A-polynomial.
Finally, the absolute value of
the ratio of the determinants in \semiclasscs,
which is often called the Reidemeister-Ray-Singer torsion, $T(\CA^{(\a)})$,
\eqn\tdef{\sqrt{ T(\CA^{(\a)}) } =
{\det (\Delta) \over \sqrt{\vert \det (L_-) \vert} } }
is related to the amplitude, $\psi$, of the quantum
wave function \semiclassz.

Notice that, besides the critical point corresponding to
the hyperbolic $SL(2,\IC)$ connection, the sum \semiclassz\ also
contains terms corresponding to other flat connections.
Even though the Chern-Simons action, $I(\CA^{(\a)})$,
of these flat connections is given by \tots,
where the ``volume'' and the ``Chern-Simons invariant''
can be computed by integrating the 1-forms \ktheta\ and \stheta\
over different branches of the zero locus of the A-polynomial,
they no longer have a nice geometric interpretation.
For example, there are flat connections with negative ``volume''.
In particular, if $K$ is a hyperbolic knot, then the semi-classical behavior
of the partition function \semiclasscs\ is dominated by the flat
connection whose volume is equal to minus the hyperbolic volume of $M$,
\eqn\semiclasszm{Z(M) \simeq
\sqrt{ T(\CA^{(hyperb)}) }
\exp \left( {\s \over 2 \pi} \Vol (M) + i \pi k CS (M)
\right) + \ldots }
%

\subsec{Perturbation Theory}

Using effective quantum mechanics, we studied the semi-classical
limit of the partition function, $Z(M)$, in the $SL(2,\IC)$
Chern-Simons theory on a knot complement, $M = \S^3 \setminus K$.
Equivalently, we may also think of $Z(M)$ as
the (unnormalized) expectation value of a Wilson line, $W_{(l,m)}(K)$,
in an infinite-dimensional representation $R$ of $SL(2,\IC)$.
In either interpretation, it would be interesting to extend
this analysis beyond the leading order and to compute
the higher-order perturbative corrections to the partition
function \semiclassz.
General arguments suggest that the result should be in the form
\eqn\pertz{ Z \simeq
\sum_{\a} \sqrt{T(\CA_{\a})} ~e^{iI(\CA^{(\a)})}
\times \Big(1 + \sum_{m,n=1}^{\infty} {b_{m,n} (\a) \over k^m \s^n} \Big) }
where
the coefficients $b_{m,n} (\a)$ are expected to be related to
some perturbative invariants of a (decorated) knot $K$, analogous
to the Vassiliev invariants, {\it cf.} \refs{\DrorVas,\Birman}.

There are two important remarks that one should bear in mind.
First, a calculation of the coefficients $b_{m,n} (\a)$
requires a perturbative computation of the Feynman diagrams
in the background of a non-trivial flat connection $\CA^{(\a)}$.
The second remark has to do with the fact that $R$ is
an infinite-dimensional representation of $SL(2,\IC)$.
In fact, if $R$ was merely a complexification of
a finite-dimensional $SU(2)$ representation,
then the coefficients $b_{m,n} (\a)$ would be exactly
the same as in the $SU(2)$ Chern-Simons theory,
and nothing new could be gained \DrorVas.


\subsec{Non-perturbative Aspects}

We found the semi-classical expression for the partition function
$Z(M)$ and briefly discussed higher-order perturbative corrections.
However, using the effective quantum mechanics one
might hope to compute the entire series $Z(M)$,
say, using the technique of the geometric
quantization \refs{\ADPW,\Hitchin,\Weitsman,\Woodhouse,\GQlect}.
The semi-classical result \semiclassz\ is simply the leading order
approximation to the exact expression for $Z(M)$, which obeys
\eqn\azquantum{ \hat A Z =0 }
In quantum theory, this operator equation represents
the classical constraint $A(l,m)=0$, {\it cf. e.g.} \lpqfeight.
Namely, the operator $\hat A$ is obtained from the A-polynomial
by replacing $l$ and $m$ with the corresponding operators,
$\hat l$ and $\hat m$, whose commutation relations follow from \uvpb.
We also note that $Z(M)$ automatically satisfies
the Hamiltonian constraint, $\hat H Z = 0$,
which can be naturally interpreted as the Wheeler-DeWitt
equation \wdweq\ in three-dimensional Euclidean gravity.

%

As we explain in the later sections, the asymptotic behavior
of the colored Jones polynomial is very similar to
the perturbative expansion \semiclasszm\ -- \pertz.
This suggests a close connection between the exact,
non-perturbative partition function of the $SL(2,\IC)$ Chern-Simons
theory and a certain analytic continuation of the colored Jones polynomial,
\eqn\zmguess{
Z(M)
\longleftrightarrow^{\kern -11pt ?}~~
J_N (K,q) }
In particular, it is natural to speculate that the colored
Jones polynomial is related to the $SL(2,\IC)$ partition function
when the variable $q$ is not a root of unity, {\it cf.} Table 1.
A further evidence for this relation comes from the recent
work \Stavros, where it was shown that the colored Jones polynomial
for certain knots obeys the $q$-difference equation \azquantum.
We postpone further study of the expected relation \zmguess\ to future work.


\ifig\munion{The manifold $M$ is a connected sum of
the 3-manifolds $M_+$ and $M_-$, joined along their
common boundary $\Si$.}
{\epsfxsize3.3in\epsfbox{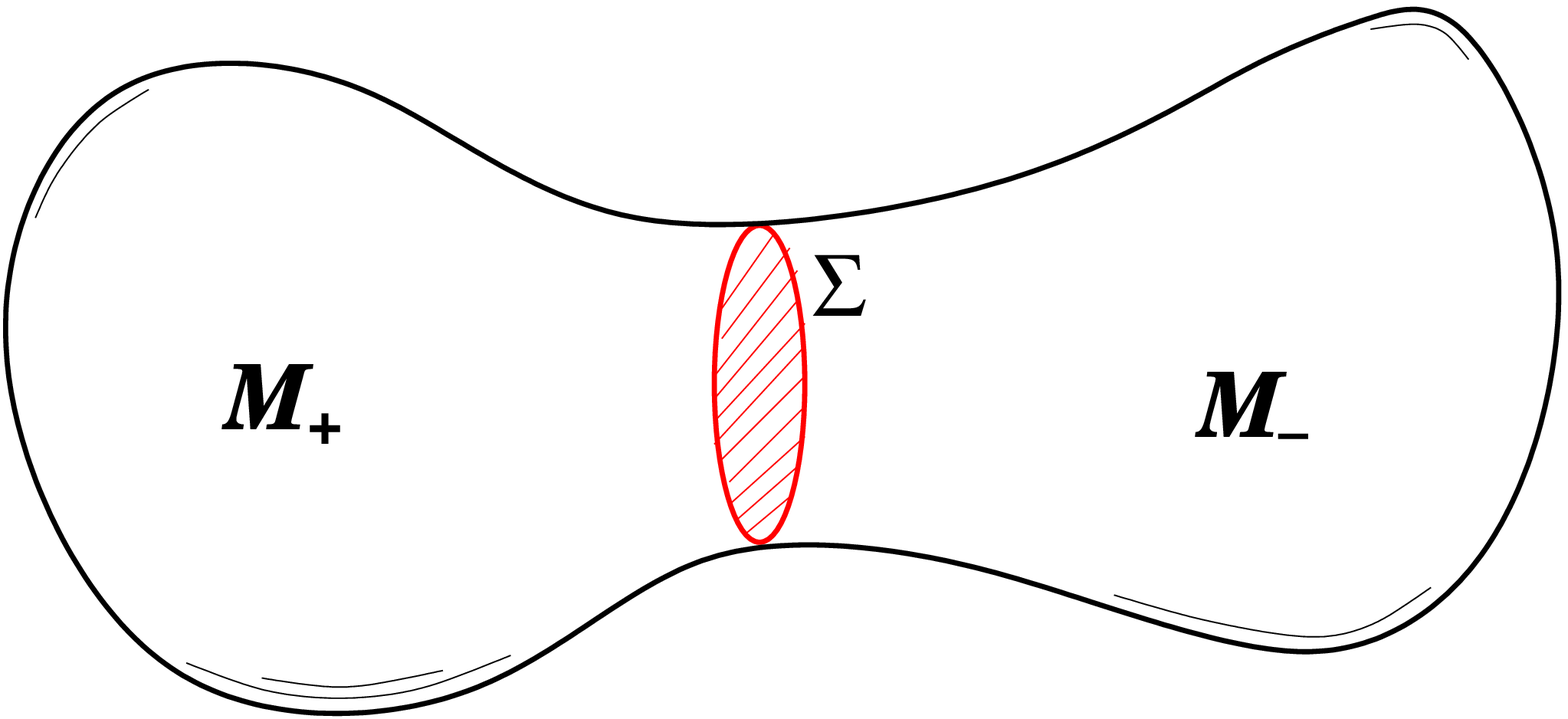}}

The non-perturbative partition function
$Z(M)$ has a number of nice properties
that follow directly from the path integral formulation \zwpathint.
Suppose, for example, that $M$ is a connected sum of
the 3-manifolds, $M_{\pm}$, joined along
the common sphere boundary $\Si=\S^2$ (see \munion):
\eqn\msum{ M = M_+ \cup_{\Si} M_- }
Then, the Chern-Simons partition function $Z(M)$ is given by
\eqn\zmsum{ Z(M) = {Z(M_+) \times Z(M_-) \over Z (\S^3)} }
where $Z (\S^3)$ denotes the partition function for a 3-sphere.

Similarly, one can compute $Z(M)$ for a homology 3-sphere $M$
represented as a union \msum\ of $M_+$ and $M_-$
joined along a common torus boundary, $\Si=T^2$.
Following the discussion in section 2,
to each $M_{\pm}$ we can associate a representation variety
(the zero locus of the corresponding A-polynomial):
$$
L_{\pm} = \Hom \left(\pi_1(M_{\pm}); SL(2,\IC)\right)/SL(2,\IC)
$$
Moreover, since $M_{\pm}$ are 3-manifolds with
a common torus boundary $\Si = T^2$,
the inclusions of $\Si$ into $M_{\pm}$ induce the embeddings:
$$
L_{\pm} \hookrightarrow \IC^* \times \IC^*
= \Hom \left(\pi_1(\Si); SL(2,\IC)\right)/SL(2,\IC)
$$
The intersection points of $L_+$ and $L_-$ are
in one-to-one correspondence with the conjugacy classes of
representations of the fundamental group $\pi_1(M)$ into $SL(2,\IC)$.
Therefore, we expect that the partition function
\eqn\zmtorussum{ Z(M) = \langle Z(M_+) \vert Z(M_-) \rangle }
is dominated by the contribution from the intersection
points of $L_+$ and $L_-$.

%
%
%

The above discussion is reminiscent of the definition of
the Casson invariant \refs{\Casson, \Taubes}.
Indeed, using the A-polynomial we can define (and in many cases
easily compute) the $SL(2,\IC)$ Casson invariant of $M$
as the weighted sum of the intersection points
of $L_+$ and $L_-$ inside $\IC^* \times \IC^*$,
\eqn\cassondef{ \la_{SL(2,{\bf C})} (M) = \# (L_+ \cap L_-) }

\example{$M=\S^3$}

Consider $M=\S^3$ with the standard Heegard splitting,
$M = M_+ \cup_{\Si} M_-$, where $\Si=T^2$ and each $M_{\pm}$
is a copy of the solid torus. Since the boundaries of $M_{\pm}$
are glued together with a relative $SL(2,\Z)$-transformation,
$$
\pmatrix{\g_l \cr \g_m} \longrightarrow \pmatrix{0 & -1 \cr 1 & 0}
\pmatrix{\g_l \cr \g_m}
$$
{}from \aunknot\ and \abasischange\ we get:
$$
L_+ = l-1 \quad , \quad L_- = m-1
$$
Clearly, the curves $L_+$ and $L_-$ intersect at a single point,
$m=l=1$, so that we find the $SL(2,\IC)$ Casson invariant
$\la_{SL(2,{\bf C})} (\S^3) = 1$.

\endexample

\noindent
It would be interesting to find a realization
of the $SL(2,\IC)$ Casson invariant in Chern-Simons
theory with fermionic symmetries, {\it cf.} \Wittenamp.

Finally, let us point out that
throughout the paper we mainly focus on the simplest case
of a single torus boundary, or a single Wilson loop $W_{(l,m)} (K)$.
It would be interesting to extend this discussion to links
with several components and three-manifolds with
arbitrary boundary $\Sigma$ (such generalizations
will be briefly discussed in the next section).
In a special case, when a link consists of $r$ unlinked
components $\g_k$, we have:
\eqn\zunlink{
{Z(\S^3;\g_1,R_1, \ldots, \g_r, R_r) \over Z(\S^3)}
= \prod_{k=1}^r {Z(\S^3;\g_k,R_k) \over Z(\S^3)} }
Suppose that $r=2$ and $R_1=R_2$,
{\it i.e.} we have two oriented unlinked components,
$\g_1$ and $\g_2$, carrying the same representation $R$.
Then, from the path integral formulation \zwpathint,
it follows that the partition function for the connected
sum $\g_1 \# \g_2$ is related to the product of the partition
functions for the individual link components,
\eqn\zconnlink{Z(\S^3;\g_1 \# \g_2, R)
= {Z(\S^3;\g_1,R) \times Z(\S^3;\g_2,R) \over Z(\S^3; {\rm unknot}, R)} }
This factorization formula also follows from the property $g)$
of the A-polynomial listed in section 2.3.


\newsec{Topological Quantum Field Theory and Invariants of Three-Manifolds}

As was already pointed out earlier,
the definition of the Chern-Simons path integral \zwpathint\ is
independent on the choice of metric on $M$, even in the quantum theory.
Hence, the Chern-Simons partition function
$Z(M;\g_i, R_i)$ must be a topological invariant.
This is not so obvious from the mathematical point of view
because Chern-Simons theory with complex gauge group operates on
an infinite-dimensional Hilbert space and, therefore,
does not fit into the standard definition of topological
quantum field theory of Atiyah and Segal \Atiyah.
Let us remind that, according to \Atiyah,
a topological quantum field theory is a functor $Z$,
such that: $(i)$ to each compact oriented 2-dimensional manifold $\Si$
without boundary one associates a finite dimensional complex vector
space $\CH_{\Si}$; and $(ii)$ a compact oriented 3-dimensional smooth
manifold $M$ with $\p M = \Si$ determines a vector $Z(M) \in \CH_{\Si}$.
Moreover, $Z$ must satisfy the following axioms:

\item{(1)} 
We denote by $-\Si$ the surface $\Si$ with the opposite orientation.
Then, we have $\CH_{-\Si} = \CH_{\Si}^*$ where $\CH_{\Si}^*$ is the dual
of $\CH_{\Si}$ as a complex vector space.

\item{(2)}
For a disjoint union $\Si_1 \sqcup \Si_2$ we have
$\CH_{\Si_1 \sqcup \Si_2} = \CH_{\Si_1} \otimes \CH_{\Si_2}$.

\item{(3)}
For the composition of cobordisms\foot{It follows from
the axioms (1) and (2) that a compact oriented
3-manifold $M$ with $\p M = (-\Si_1) \sqcup \Si_2$ determines
a linear map $Z(M) \in \Hom_{{\bf C}} (\CH_{\Si_1}, \CH_{\Si_2})$.
Such a manifold $M$ is called a cobordism between $\Si_1$ and $\Si_2$.}
$\p M_1 = (-\Si_1) \sqcup \Si_2$
and $\p M_2 = (-\Si_2) \sqcup \Si_3$, the relation
$Z (M_1 \cup M_2) = Z (M_2) \circ Z(M_1)$
holds, where the right-hand side stands for the composition
of linear maps $Z(M_1) \colon \CH_{\Si_1} \to \CH_{\Si_2}$
and $Z(M_2) \colon \CH_{\Si_2} \to \CH_{\Si_3}$.

\item{(4)}
For an empty set $\emptyset$ we have $Z (\emptyset) = \IC$.

\item{(5)}
Let $I$ denote the closed unit interval.
Then, $Z (\Si \times I)$ is the identity map as a linear
transformation of $\CH_{\Si}$.

It is easy to check that all of these conditions are satisfied
in a Chern-Simons theory with compact gauge group $G$.
Similarly, many of these properties extend
to a theory with the complexified gauge group $\CCG$,
essentially due to the path integral formulation \zpathint.
There is one important subtlety, however, related to the fact
that in the latter theory the phase space
\eqn\pmgen{
\CP = \Hom \left(\pi_1(\Si); \CCG \right)/ \CCG }
is no longer compact.
Indeed, the space $\CP$ is isomorphic to the total
space of the cotangent bundle, {\it cf.} \pcot:
\eqn\pmcot{\CP = T^* \CM}
where $\CM$ is a representation space of $\pi_1 (\Si)$
into the compact part of the gauge group, $G$.
Therefore, the corresponding Hilbert space $\CH_{\Si}$
is not finite dimensional.
To be more specific, by analogy with the $SL(2,\IC)$ case
studied in this paper, one might define the quantum Hilbert
space $\CH_{\Si}$ of a Chern-Simons theory with gauge group $\CCG$
as a space of half-densities on $\CP$.
Since this space is infinite-dimensional, one needs a refinement
of the above definition in order to prove that we deal with
a topological quantum field theory in a mathematical sense.
Physical considerations suggest, however, that there exists
a rigorous mathematical definition of the topological
invariant $Z(M;\g_i, R_i)$, which we postpone to future work.

%

\bigskip {\noindent {\it {Implications for Representation Varieties
and the A-polynomial}}} \medskip

Apart from defining a topological invariant of
three-manifolds with links colored by infinite-dimensional
representations, Chern-Simons theory can also shed some light
on the properties of representation varieties \generall,
in particular, on the properties of the A-polynomial.
Thus, a lot of interesting information about
$L = \Hom \left(\pi_1(M); \CCG \right)/ \CCG$
and about the 3-manifold $M$ itself can be obtained by
treating $L$ as a Lagrangian submanifold in $\CP$.
In particular, as we saw earlier,
the Lagrangian inclusion $L \hookrightarrow \CP$
is a natural starting point for quantization,
and can lead to new connections between
three-dimensional topology, symplectic geometry,
and perhaps even Langlands' program \Beilinson.

%

For instance, it directly follows from this new perspective
that the ``volume formula'' \volfla\ and
the ``Chern-Simons formula'' \csfla\ can be generalized
to hyperbolic 3-manifolds with arbitrary boundary $\Si$,
not necessarily connected.
There is a nice class of examples of such manifolds
called convex cores, which have finite volume \Bonahon\
and may be useful in this context.
Even though the explicit description of
the representation variety
\eqn\lmgen{
L = \Hom \left(\pi_1(M); SL(2,\IC) \right)/ SL(2,\IC)}
may be very complicated for a generic 
3-manifold $M$ with boundary $\Si$, the ambient space $\CP$
is always a symplectic space of the form \pmcot.
Specifically, in the case we are considering, the space $\CP$
is the total space of the cotangent bundle over
the moduli space, $\CM$, of flat $SU(2)$ connections on $\Si$.
Let $\om_{\s}$ denote the natural symplectic structure
on $\CP = T^* \CM$, and let $\th_{\s}$ be the corresponding
canonical 1-form, such that $\om_{\s} = d \th_{\s}$.
Then, the physical considerations suggest that
\eqn\genvolfla{\th_{\s} \vert_L = {1 \over 2} d \Vol (M)}
should hold for arbitrary $M$ (along with a similar
expression for the Chern-Simons invariant of $M$).
Furthermore, the 1-form $\th_{\s}$ should be exact, when
restricted to the Lagrangian submanifold $L \hookrightarrow \CP$,
$$
\oint_C \th_{\s} = 0 \quad, \quad \forall C \in \pi_1 (L)
$$
There is a similar set of constraints that follows from
the rationality of the periods of $\th_k$, {\it cf.} \ksquantcond.
Altogether, these conditions assert that $L$ is a quantizable
submanifold and impose severe restrictions on its geometry,
especially when $\pi_1 (L)$ is large.

For example,
these constraints lead to some non-trivial obstructions
to a polynomial arising as the A-polynomial of a knot.
Namely, if $\Si=T^2$ and $L$ is described by
the zero locus of the A-polynomial, from \ksquantcond\
we find that 
the integral of the 1-form $\th_{\s}$
around any closed loop $C$ on the curve $L$ must vanish,
\eqn\sobstruction{
\oint_C \log |l| d (\arg m) - \log |m| d (\arg l) =0 }
and, furthermore, that the integral of the 1-form $\th_k$
must be a rational number,
\eqn\kobstruction{ {1 \over \pi^2} \oint_C
\log |m| d \log |l| + (\arg l) d (\arg m) \in \IQ }
In particular, this gives the answer to a question
posed by Cooper and Long in \refs{\CLrem,\CLrep}:
``Which affine curves $L$ in $\IC^* \times \IC^*$ satisfy
the condition that $\th_{\s}$ is exact on $L$?''
Namely, the 1-form $\th_{\s}$ should be interpreted as
an ``imaginary part'' of the Lioville form \tottheta,
and then the condition \sobstruction\ is simply the condition
for $L$ to be a quantizable Lagrangian submanifold in $\IC^* \times \IC^*$.
This suggests a further relationship to symplectic geometry.
The first of the above conditions can be understood as
a consequence of the Sch\"afli-like formula \volfla,
and has a number of applications, see {\it e.g.} \CLrem.
To the best of our knowledge, the second condition
has not been discussed in the mathematical literature.

\example{}

Following \CLrem, let us demonstrate how the condition \sobstruction\
can be used to constrain the form of the A-polynomial.
For example, in Table 2 one finds the A-polynomial of
the figure-eight knot:
$$
A(l,m) = -2+m^4+m^{-4}-m^2-m^{-2}-l-l^{-1}
$$
A slight modification of this polynomial gives
$$
f(l,m) = -2+m^6+m^{-6}-m^2-m^{-2}-l-l^{-1}
$$
which is not the A-polynomial of any knot since the 1-form $\th_{\s}$
has non-vanishing periods on $f(l,m)=0$, therefore, violating \sobstruction.
However, $f(l,m)$ does exhibit every other property of the A-polynomial.
It would be interesting to invent similar examples which satisfy
the first condition \sobstruction, but fail \kobstruction.

\endexample

Notice, that the genus of the curve $A(l,m)=0$ grows very rapidly
with the complexity of the knot. (One can get a general impression,
say, by looking at the examples listed in Table 2.)
Therefore, the number of non-trivial constraints in
\sobstruction\ and \kobstruction\ also becomes very large,
especially for polynomials of large degree.
In fact, one might wonder if these constraints, together with
the ones mentioned in section 2.3, give a complete list:

{\bf Question:}
{\it Is every integral, reciprocal, and tempered polynomial $f(l,m)$
that satisfies both conditions \sobstruction\ and \kobstruction\
is the A-polynomial of some knot?}


\newsec{A Generalization of the Volume Conjecture}

In this section\foot{The work presented in this
section originated from discussions with K.~Krasnov.}
we show how the above approach
can be used to extend the volume conjecture to
incomplete hyperbolic structures on knot
complements\foot{See also
\refs{\Dylantalk,\MurakamiOptimist,\Benedetti}
for a previous work in this direction.}.
The generalized volume conjecture, then, can be interpreted
as a relation between the A-polynomial and
the the colored Jones polynomial.
Furthermore, in the next section we discuss a similar generalization
of the Melvin-Morton-Rozansky conjecture, also inspired by physics.

\subsec{A Brief Review of the Volume Conjecture}

A well-known invariant of knots is
the Jones polynomial $J (K,q)$ \Jones.
Here we consider a more general invariant, the so-called
$N$-colored Jones polynomial $J_N (K,q)$, associated with
an $N$-dimensional irreducible representation of $SU(2)$.
Thus, the usual Jones polynomial appears as a special case,
$J (K,q) = J_2 (K,q)$.
Like the ordinary Jones polynomial, $J_N (K,q)$ can be defined by
skein relations and is a Laurent polynomial in the variable $q^{1/2}$.
Remarkably, there is a relation between the colored Jones
polynomial of a knot and the volume of its complement.

The first observation along these lines was made by Kashaev, who
introduced a link invariant associated with quantum dilogarithm \Kashaev.
The Kashaev's invariant assocated with a knot $K$ --- usually denoted
by $\langle K \rangle_N$ --- is based on the theory of quantum
dilogarithms at the $N$-th root of unity, $q = \exp (2\pi i/N)$.
Moreover, it was noticed in \Kashaev\ that for certain knots
the asymptotic behavior of this invariant is related
to the volume of the knot complement
\eqn\kashaevconj{\lim_{N \to \infty}
{\log \vert \langle K \rangle_N \vert \over N}
= {1 \over 2 \pi} \Vol (M) }
A generalization of this relation to all knots
is known as the volume conjecture \Kashaev.

Later, it was realized in \MM\ that the Kashaev's invariant
is related to the colored Jones polynomial
evaluated at the special value of $q$:
\eqn\qroot{ q = e^{2 \pi i / N} }
Therefore, the volume conjecture can be reformulated as
a relation between the volume of the knot complement and
a special limit of the colored Jones polynomial:

{\bf The Volume Conjecture:}
\eqn\volumeconj{\lim_{N \to \infty}
{\log \vert J_N (K, e^{2 \pi i / N} ) \vert \over N}
= {1 \over 2 \pi} \Vol (M)
}
In this form, the volume conjecture has been verified in a number
of examples. In particular, Kashaev and Tirkkonen \KT\ proved
that it is true for all torus knots, in a sense that
the limit \volumeconj\ is zero.

The volume conjecture was extended further in \MMOTY,
where it was shown that for a large class of knots
one can remove the absolute value in \volumeconj,
so that the following limit holds\foot{Notice, that
our normalization of the Chern-Simons invariant
agrees with \refs{\NZagier, \Yoshida}, but differs from
the normalization used in \MMOTY\ by a factor of $2\pi^2$.}:
\eqn\csvolumeconj{\lim_{N \to \infty}
{\log J_N (K, e^{2 \pi i / N} ) \over N}
= {1 \over 2 \pi} \left( \Vol (M) + i 2\pi^2 CS (M) \right)
}
This version of the volume conjecture ---
which relates the asymptotic behavior of $J_N (K,q)$ to
the volume and Chern-Simons invariant of the knot complement
--- is the one we are going to use here.
For more work on the volume conjecture see {\it e.g.}
\refs{\Benedetti,\BBqhi,\Yokota,\Hikami}.

\subsec{The A-Polynomial and the Generalized Volume Conjecture}

Incomplete hyperbolic structures on
knot complements come in continuous families.
For example, if $K$ is a hyperbolic knot in the 3-sphere,
then the moduli space of hyperbolic metrics on $M = \S^3 \setminus K$
has at least one component of complex dimension one.
Apart from a special point representing the cusped 3-manifold,
a generic point in this moduli space corresponds to
an incomplete hyperbolic structure on $M$ with
a conical singularity or some other kind of degeneration
along the knot $K$, see {\it e.g.} \Hodgson\ for more details.

A convenient way to describe the space of incomplete hyperbolic
structures on $M = \S^3 \setminus K$ is to identify
a hyperbolic structure with a $SL(2,\IC)$ structure.
Then, the moduli space of $SL(2,\IC)$ structures is described
by the character variety which, in turn, is given by
the zero locus of the A-polynomial \CCGLS,
\eqn\azerolocus{ A(l,m) = 0 }
where $l$ and $m$ are the complex-valued `eigenvalues' of
the holonomies \lmholonomies\
around the longitude and the meridian of a knot.
In these variables, the point in the moduli space representing
the complete metric on $M$ is located at $(l,m)=(-1,1)$.
The change of the volume and the Chern-Simons invariant is described,
respectively, by the differential 1-forms \volfla\ and \csfla\
on the curve \azerolocus:
\eqn\csvoldiff{\eqalign{
& d ~\Vol (M)
= 2 \Big( - \log |l| d (\arg m) + \log |m| d (\arg l) \Big) \cr
& d ~CS(M)
= - {1 \over \pi^2} \Big( \log |m| d \log |l| + (\arg l) d (\arg m) \Big)
}}

Therefore, if $K$ is a hyperbolic knot, the right-hand side of
the volume conjecture \csvolumeconj\ has a natural generalization.
For a given knot $K$, we can consider a family
of (incomplete) hyperbolic metrics on $M$
and regard $\Vol(M)$ and $CS(M)$ as functions
on the curve $A(l,m)=0$ obtained by integrating \csvoldiff,
{\it cf.} \pdqint,
\eqn\csvolint{\eqalign{
& \Vol (l,m) = \Vol (K)
+ 2 \int \Big( - \log |l| d (\arg m) + \log |m| d (\arg l) \Big) \cr
& CS(l,m) = CS(K)
- {1 \over \pi^2} \int \Big( \log |m| d \log |l| + (\arg l) d (\arg m) \Big)
}}
Here, $\Vol (K)$ and $CS(K)$ refer to the volume and
the Chern-Simons invariant of the complete hyperbolic
metric on the knot complement
(more generally, the Gromov norm of $K$).

In order to find a suitable generalization of the left-hand
side of the volume conjecture \csvolumeconj, let us look at
the path integral definition of the colored Jones polynomial.
In quantum field theory, the colored Jones polynomial
evaluated at the $k$-th root of unity
appears as the normalized\foot{Here, the word ``normalized'' refers to
a factor of $Z_{SU(2)} (\S^3)$ in the denominator of the path integral.
In the following section we will introduce another
version of the colored Jones polynomial, often used in
the mathematical literature, where $J_N (K,q)$
is further normalized relative to $J_N ({\rm unknot},q)$.}
expectation value of a Wilson line
in the $SU(2)$ Chern-Simons theory \WittenJones,
\eqn\njoneshol{
J_N (K, e^{2\pi i /k})
= \langle \Tr_{R_j} ~{\rm P}\exp \oint_K A \rangle
}
where $k$ stands for the (renormalized) value of the level,
and $R_j$ is an irreducible spin-$j$ representation of $SU(2)$
of dimension $N=2j+1$.
Therefore, the left-hand side of the volume conjecture \csvolumeconj\
can be interpreted as a classical limit of the Chern-Simons-Witten
invariant \njoneshol, such that
\eqn\kashaevlimit{
k \to \infty
\quad , \quad
N \to \infty
\quad , \quad
{N \over k} = 1
}

In physics,
Wilson lines represent trajectories of massive point-like particles.
Interacting with gravity such particles produce conical defects
in the geometry of space-time, so that the deficit angle
is proportional to the mass of the particle \DJT.
These heuristic arguments suggest that, in the present context,
a Wilson line \njoneshol\ should be associated with a conical
singularity along the knot $K$, with a deficit angle $\sim N/k$.
On the other hand, deformations of the hyperbolic structure on $M$
with a conical singularity along the knot $K$ are parametrized
by the curve \azerolocus, with $|m|=1$. Therefore, one might expect
\eqn\expectedm{
\log (m) \sim 2\pi i \left( 1 - {N \over k} \right) }

These considerations suggest that the proper modification of
the left-hand side of the volume conjecture \csvolumeconj, that includes
deformations of the hyperbolic structure on $M$, should be obtained by
replacing \kashaevlimit\ with a more general double-scaling limit,
\eqn\thelimit{
k \to \infty
\quad , \quad
N \to \infty
\quad , \quad
a \equiv {N \over k} = {\rm ~fixed}
}

The ratio, $a=N/k$, which is kept fixed in this limit,
does not need to be a rational number.
In fact, the relation \expectedm\ suggests that,
in order to compare with the volume and the Chern-Simons invariant
computed from the A-polynomial, the parameter $a$ must be treated
as a continuous complex variable.
Moreover, rational values of $a$ (except for $a=1$) are special,
in a sense that the asymptotic
behavior of the colored Jones polynomial ``jumps'' in such cases.
Again, this suggests to consider generic values of $a$.
Since the colored Jones polynomial is defined for all values of $q$,
not only for the roots of unity,
a natural way to realize this is to keep $N$ integer,
and take $k = N/a$ to be non-integer\foot{Alternatively,
one could consider an analytic continuation of $J_N (K,q)$
to non-integer values of the color, $N$.
Even though eventually this possibility may play an important
r\'ole in the connection with the $SL(2,\IC)$ Chern-Simons theory,
it seems less obvious at present. See however \Deguchi,
which may be relevant here.
I wish to thank D.~Thurston for pointing out this reference
and for helpful discussions on related topics.}.
Then, by analogy with \csvolumeconj,
we expect the following conjecture to be true:

{\bf The Generalized Volume Conjecture:}
{\it In the limit \thelimit,
the $N$-colred Jones polynomial has the following asymptotic behavior:
\eqn\generalconj{\lim_{N,k \to \infty}
{\log J_{N} (K, e^{2 \pi i / k} ) \over k}
= {1 \over 2 \pi} \left( \Vol (l,m) + i 2\pi^2 CS (l,m) \right)
}
where $\Vol (l,m)$ and $CS (l,m)$ are the functions \csvolint\
on the zero locus of the A-polynomial, evaluated at the point}
\eqn\mvalue{m = - \exp ({i \pi a}) }
This identification of the parameters agrees
with the expected relation \expectedm,
where we fixed the exact numerical factor
by considering specific examples, see below.

Note, that the generalized volume conjecture \generalconj\ presents
a 1-parameter family of relations, which include the well-known
volume conjecture \csvolumeconj\ as a special case, $m=1$.
Below, we demonstrate that \generalconj\
is true for the figure-eight knot.


\subsec{Proof of the Generalized Volume Conjecture
for the Figure-Eight Knot ${\bf 4}_1$}

The $N$-colored Jones polynomial of the figure-eight knot
has the following form, see {\it e.g.} \KaulG:
\eqn\njoneseight{J_N ( {\bf 4}_1 , q)
= \sum_{i=0}^{N-1} \prod_{j=1}^i
\left( q^{(N+j)/2} - q^{-(N+j)/2} \right)
\left( q^{(N-j)/2} - q^{-(N-j)/2} \right)
}
Notice, that at the $N$-th root of unity, $q = \exp (2\pi i/N)$,
it is indeed equal to the Kashaev's invariant of
the figure-eight knot \Kashaev:
$$
\langle {\bf 4}_1 \rangle_N = \sum_{i=0}^{N-1} \prod_{j=1}^i 
\left( 1- q^j \right) \left( 1- \bar q^j \right)
$$

In the limit \thelimit,
the asymptotic behavior of the colored Jones polynomial \njoneseight\
can be obtained using the saddle-point approximation.
The result has the following form, {\it cf.} \MurakamiMahler:
\eqn\limiteight{
\lim_{k,N \to \infty}
{\log J_N ( {\bf 4}_1 , e^{2\pi i/k} ) \over k} = {1 \over \pi} V(a) }
where the function
\eqn\vfunction{
V(a) = \Lambda (a \pi + \th (a)/2) - \Lambda (a \pi - \th (a)/2) }
is defined via
$\th (a) = \arccos \left( \cos (2\pi a) - 1/2 \right)$
and the Lobachevsky function,
\eqn\Lobachevsky{
\Lambda (z) = - \int_0^z \log |2 \sin x| dx
= {1 \over 2} \sum_{n=1}^{\infty} {\sin (2n z) \over n^2}}
The graph of $V(a)$ is shown on the figure below:

\ifig\volumefncn{The graph of the volume function $V(a)$ near $a=1$.}
{\epsfxsize3in\epsfbox{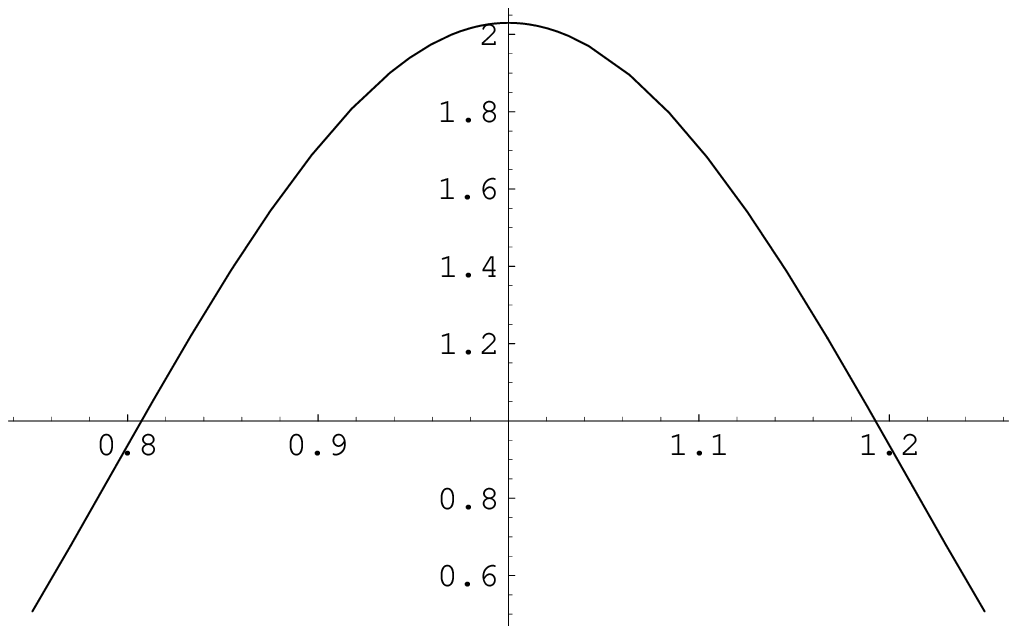}}

Notice, that at the special value $a=1$ corresponding
to the complete hyperbolic structure on the knot complement,
we have
\eqn\kashlimiteight{
2\pi \lim_{N \to \infty}
{\log J_N ( {\bf 4}_1 , e^{2\pi i/N}) \over N}
= 2 \Big( \Lambda \left( 7 \pi/ 6 \right)
- \Lambda \left( 5 \pi / 6 \right) \Big) }
Using the $\pi$-periodicity of the Lobachevsky function
and the identity
$$
\Lambda (nz)
= n \sum_{j~({\rm mod}~n)} \Lambda \left( z + {\pi j \over n} \right)
$$
one can check that \kashlimiteight\ is indeed equal to
the volume of the figure-eight knot complement,
\eqn\volumefigeight{
\Vol ({\bf 4}_1) = 6 \Lambda (\pi/3) \approx 2.0298832 \ldots }

Now, let us study the limit \limiteight\ as a function of
the parameter $a$, near the point $a=1$.
For reasons that will become clear in a moment, it is convenient
to introduce a new parameter $y$, such that
\eqn\yparameter{a = 1 + {y \over \pi}}
According to the relation \mvalue, we have
\eqn\mviay{m = \exp ({iy}) }
It is clear that $y$ is a good expansion parameter near
$m=1$ ({\it i.e.} near $y=0$).

Since we already know that for $y=0$ the volume conjecture is true,
we need to compare only the $y$-dependent terms on both
sides in \generalconj, say, by differentiating with
respect to $y$. On the left-hand side, this leads to
a derivative of the function \vfunction, representing
the limiting behavior of the colored Jones polynomial.
Explicitly, we find
\eqn\dvdy{\eqalign{
{dV \over dy}
= & \left(1 - {\sin(2y) \over
\sqrt{1 - (\cos (2y) -1/2)^2}} \right)
\log | 2 \sin \Big(y - {1 \over 2}
\arccos \left( \cos (2y) -1/2 \right) \Big) | - \cr
& -\left(1 + {\sin(2y) \over
\sqrt{1 - (\cos (2y) -1/2)^2}} \right)
\log | 2 \sin \Big(y + {1 \over 2}
\arccos \left( \cos (2y) -1/2 \right) \Big) |
}}
It is easy to see that this is an odd function of $y$,
which has the following power series expansion,
\eqn\dvdyeightseries{ {dV \over dy}
= -2 \sqrt{3} y + {8 \over \sqrt{3}} y^3 - {184 \over 15\sqrt{3}} y^5
+ {1424 \over 45 \sqrt{3}} y^7 - {86248 \over 945 \sqrt{3}} y^9 + \ldots }
The graph of the function $dV/dy$ is shown on the figure below:

\ifig\dervgraph{The graph of the function $dV/dy$.}
{\epsfxsize3in\epsfbox{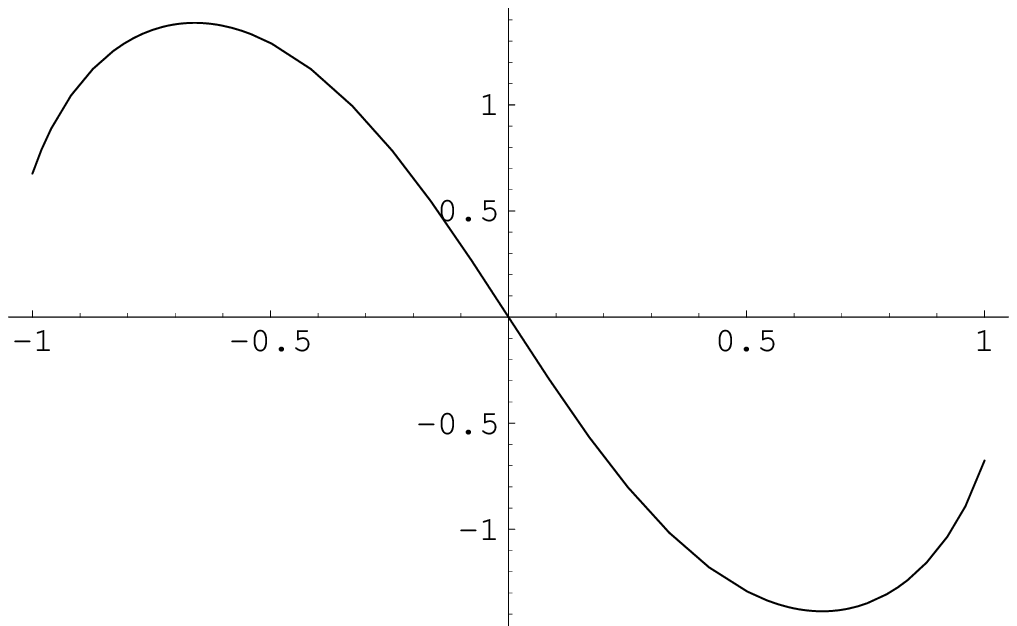}}

Now, let us look at the right-hand side of the proposed
relation \generalconj. It involves the volume function, $\Vol (l,m)$,
and the Chern-Simons function, $CS(l,m)$, defined by the integrals
\csvolint\  on the zero-locus of the A-polynomial.
For the figure-eight knot, the A-polynomial has
the following form \CCGLS:
\eqn\apolyeight{
A(l,m) = -2+m^4+m^{-4}-m^2-m^{-2}-l-l^{-1}}

For simplicity, let us take $y$ to be real.
Then, using the fact that $CS({\bf 4}_1)=0$, we can write
the integrals \csvolint\ in the following simple form,
\eqn\voleightint{\Vol (m) = \Vol ({\bf 4}_1)
- 2 \int_0^y \log |l| d (\arg m) }
and
\eqn\cseightint{CS(m) =
- {1 \over \pi^2} \int_0^y (\arg l) d (\arg m) }
where $\log |l|$ and $\arg l$ should be determined from
the equation $A (l,e^{iy})=0$.
Since the A-polynomial \apolyeight\ is quadratic in $l$,
it is easy to solve this equation. One finds
\eqn\leightsoln{
l = \cos (4y) - \cos (2y) -1 \pm \sqrt{(\cos (4y) - \cos (2y)-1)^2-1} }
It is the second root (with the ``$-$'' sign) that corresponds
to the geometric branch of the moduli space of flat $SL(2,\IC)$
connections. Choosing this root, we find that $l$ is real
for real values of $y$ (sufficiently close to $y = 1$).
Hence, the Chern-Simons integral \cseightint\ is identically zero.
This agrees with the fact that the function $V$, that describes
the limiting behavior of the colored Jones polynomial
in \limiteight\ -- \vfunction\ is also real.
Therefore, the imaginary part of the proposed relation \generalconj\
is true --- both sides vanish identically --- and we only need
to verify the real part.

Fortunately, we do not need to evaluate the integral \voleightint\
explicitly in order to check the real part
of the generalized volume conjecture \generalconj.
Since we are interested only in $y$-dependent terms,
we only have to compare \dvdy\ with the derivative of
the volume function \voleightint,
\eqn\dvdycompare{
- \log |l| = - \log \Big[
\cos (4y) - \cos (2y) -1 - \sqrt{(\cos (4y) - \cos (2y)-1)^2-1} \Big] }
It is amusing to verify that the functions \dvdy\ and \dvdycompare\
are indeed identical.


\subsec{A Relation Between The Colored Jones Polynomial And The A-Polynomial}

A generalization of the volume conjecture discussed above
could be interpreted as a purely algebraic relation between
the A-polynomial and the colored Jones polynomial.
Specifically, given a colored Jones polynomial $J_N (K,q)$
of a knot $K$ one can consider the limit \thelimit,
$$
k \to \infty
\quad , \quad
N \to \infty
\quad , \quad
a \equiv {N \over k}
\quad \quad a \in \IC \setminus \IQ
$$
and define a function $l(a)$, such that $l(1)=-1$ and
\eqn\lasalimit{
\log l = - {d \over da} \lim_{{N,k \to \infty \atop N/k=a}}
{\log J_N (K, e^{2 \pi i / k} ) \over k}
}
{}From the definition, it might seem that $l(a)$ is a rather
complicated function of $a$. However, the generalized volume
conjecture \generalconj\ implies that $l$ and $m=-e^{ia}$ satisfy
a simple polynomial relation \azerolocus\ with integer coefficients.
Namely, they belong to the zero locus of the A-polynomial.

\noindent
{\bf Remarks:}

\item{1)}
Among other things, the generalized volume conjecture implies that
the A-polynomial is an invariant of knots not stronger than the colored
Jones polynomial (regarded as a function of the color as well).
This seems to be consistent with the existing observations.
For example, both the A-polynomial and the colored Jones
polynomial can not distinguish mutants, {\it etc}.

\item{2)}
Given that the colored Jones polynomial and the A-polynomial
have a very different nature, it might be useful to understand
a relation between them better. For instance, the colored
Jones polynomial can be defined by skein relations, whereas
no such definition is known for the A-polynomial.

\item{3)}
It would be interesting to understand a relation to the work \Gelca,
where a similar connection between the A-polynomial and the Jones
polynomial was proposed from the the non-commutative point of view.


\newsec{Non-Trivial Flat Connections and
the Melvin-Morton-Rozansky Conjecture}

The volume conjecture and its generalization discussed in
the previous section imply that, for generic values of $a$,
the asymptotic behavior of the colored Jones polynomial
is dominated by a flat $SL(2,\IC)$ connection.
This observation forms a very nice and complete picture
once considered along with the Melvin-Morton-Rozansky conjecture
regarding the asymptotic behavior of the colored Jones polynomial
for integer values of $k$.
Using path integral interpretation, in this section we will try
to explain this general picture, which will allow us to formulate
an analog of the Melvin-Morton-Rozansky conjecture
for the contribution of a non-trivial flat connection.

\subsec{A Brief Review of the Melvin-Morton-Rozansky Conjecture}

Let $J_N (K,q)$ be the $N$-colored Jones polynomial of a knot $K$.
We define a {\it reduced} Jones polynomial as
\eqn\vnjones{
V_N (K,q) = {J_N (K,q) \over J_N ({\rm unknot},q)},
\quad \quad V_N \in \Z [q,q^{-1}] }
where $q=\exp (2\pi i/k)$ and
\eqn\njonesunknot{J_N ({\rm unknot},q) = [N]
= {q^{N/2} - q^{-N/2} \over q^{1/2} - q^{-1/2} }  }
is the $N$-colored Jones polynomial for the trivial knot.

Following \Roztriv, let us denote by $J^{(tr)}_N (K,q)$
(resp. $V^{(tr)}_N (K,q)$) the trivial connection contribution
to the colored Jones polynomial.
Consider the Melvin-Morton expansion of the colored
Jones polynomial in powers of $N$ and $h=q-1$ \MMorton,
\eqn\mmexpansion{V^{(tr)}_N (K,q)
= \sum_{m,n \ge 0} D_{m,n} (K) N^{2m} h^n  }
Here, the rational numbers $D_{m,n} (K)$ are Vassiliev invariants
of order $n$ \refs{\DrorVas,\Birman},
and we assume that both $N$ and $k$ are integer.
The extension to non-rational values of $a=N/k$ is related
to the generalized volume conjecture and will be discussed
further below.

It was conjectured by Melvin and Morton \MMorton\
and later proved by Rozansky \Rozansky,
and by Bar-Natan and Garoufalidis \DrorG,
that the coefficients $D_{m,n} (K)$ in the expansion \mmexpansion\
have the following properties\foot{To avoid cluttering,
in what follows we suppress the dependence of $D_{m,n}(K)$
on the knot $K$.},
\eqn\mmconj{\eqalign{
& D_{m,n} = 0 \quad {\rm for} \quad m > n/2 \cr
& \sum_{m \ge 0} D_{m,2m}  a^{2m}
= {1 \over \nabla_A (K,e^{i\pi a} - e^{-i\pi a})}
}}
where $\nabla_A (K,z)$ is the Alexander polynomial of the knot $K$,
normalized such that $\nabla_A ({\rm unknot},z)=1$.
We remind, that the Alexander polynomial can be defined by
the skein relation,
\eqn\skeinalex{\nabla_A (L_+,z) - \nabla_A (L_-,z) = z \nabla_A (L_0,z)}
corresponding to the link diagrams shown on the figure below.

\ifig\dervgraph{Link diagrams connected by the skein relation.}
{\epsfxsize3.5in\epsfbox{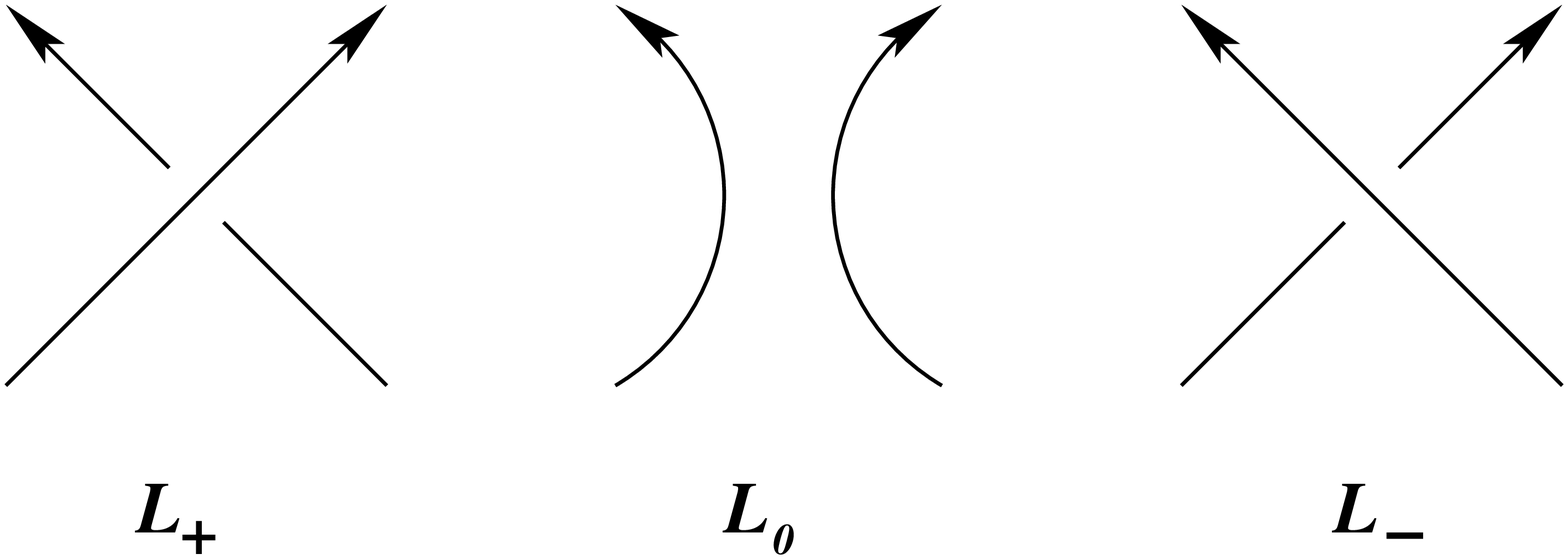}}

The bound on the powers of $N$ allows to rewrite \mmexpansion\
in the form
\eqn\mmnhexpansion{
V^{(tr)}_N (K,q)
= \sum_{n \ge 0} h^n \sum_{m \ge 0} D_{m,n+2m} (Nh)^{2m} }
which is reminiscent of the t' Hooft expansion in a $U(N)$ gauge theory.
Let us rewrite this expansion in yet another form,
replacing $(Nh)$ with a new parameter
$$
z = q^{N/2} - q^{-N/2}=2i \sin (\pi a),
$$
such that \Rozhigher:
\eqn\nhviaz{
Nh = 2 \log \left( \sqrt{1 + z^2/4} + z/2 \right)
{h \over \log (1+h) } = z + \ldots  }
Then, the Melvin-Morton expansion \mmnhexpansion\ of the colored
Jones polynomial can be written in the form
\eqn\mmzexpansion{
V^{(tr)}_N (K,q)
= \sum_{n \ge 0} V^{(n)} (K,z) h^n
= \sum_{n \ge 0} h^n \sum_{m \ge 0} d_m^{(n)} z^{2m}  }
In terms of the new variable $z$, the Melvin-Morton
conjecture \mmconj\ turns into a statement about the first
line in the expansion \mmzexpansion,
\eqn\mmconjviaz{ V^{(0)} (K,z) = {1 \over \nabla_A (K,z)} }
which, among other things, implies that the coefficients $d_m^{(0)}$
are all integer numbers.
Rozansky conjectured \Rozhigher\ and proved later \RozBurau,
that all the coefficients $d_m^{(n)}$ are integer numbers
and that the $n$-th line in the expansion \mmzexpansion\
is a rational function of $z$:
\eqn\rozanskyconj{
V^{(n)} (K,z) = {P^{(n)} (K,z) \over \nabla_A^{2n+1} (K,z)}
\quad ,\quad P^{(n)} (K,z) \in \Z [z^2] }

In order to understand a connection between the Melvin-Morton-Rozansky
and the volume conjectures,
it is helpful to look at the path integral interpretation.


\subsec{Path Integral Interpretation}

As we already pointed out earlier, the colored Jones
polynomial can be viewed as the expectation value of
the Wilson loop observable \wloop\
in the $SU(2)$ Chern-Simons theory \WittenJones,
\eqn\njonesviaw{
J_N (K, e^{{2\pi i \over k_0 + 2}}) = \langle W_{R_j} (K) \rangle
= {1 \over Z_{SU(2)} (\S^3)}
\int \CD A ~W_{R_j} (K)~ e^{{i k_0 \over 4\pi} S_{CS} (A)}
}
where $k_0$ is the tree-level value of the coupling
constant, $R_j$ is an irreducible $N=(2j+1)$ dimensional
representation of $SU(2)$, and $S_{CS} (A)$ is the Chern-Simons
functional of the $su(2)$-values gauge connection $A$,
\eqn\csaction{
S_{CS} = \Tr \int_{\S^3 } (A \wedge dA + {2 \over 3} A \wedge A \wedge A) }

The Wilson loop expectation value \njonesviaw\ is
normalized by the Chern-Simons path integral on $\S^3$,
\eqn\zforsphere{
Z_{SU(2)} (\S^3) = \sqrt{2 \over k} \sin \left( {\pi \over k} \right)
}
where $k=k_0+2$ is the {\it renormalized} value of the level.

In the semi-classical limit, $k \to \infty$, the path
integral \njonesviaw\ can be represented as a sum
over flat connections,
\eqn\zsumc{Z_{SU(2)} (W_{R_j},k) = \sum_{\a} Z_{SU(2)}^{(\a)} (W_{R_j},k)}
where each term has the form of a perturbation series,
\eqn\zcseries{
Z_{SU(2)}^{(\a)} (W_{R_j},k) = \exp {i k \over 4 \pi} \left(
S_{CS}^{(\a)} + \sum_{n=1}^{\infty} (2\pi/k)^n S_n^{(\a)}  \right)  }
Therefore, the leading contribution to the path integral
\njonesviaw\ comes from the classical value of the Chern-Simons
functional \csaction\ evaluated on the flat connection $A^{(\a)}$.
Then, the next, ``1-loop'' term, $S_1^{(\a)}$, has the form,
see {\it e.g.} \refs{\WittenJones,\FreedGompf,\Lawrence},
\eqn\sconeloop{
\exp \left( {i \over 2} S_1^{(\a)} \right)
= {\sqrt{T (A^{(\a)})} \over \Vol (H_{\a}) }
\left( {4 \pi^2 \over k} \right)^{(h^0 (A^{(\a)}) - h^1 (A^{(\a)}) )/2}
\exp \left( - {i\pi \over 4} N_{ph} \right)
}
where $h^k (A^{(\a)})$ is the dimension of the $k$-th cohomology
of $M = \S^3 \setminus K$ with coefficients twisted by $A^{(\a)}$,
and $T (A^{(\a)})$ is the $SU(2)$ Ray-Singer torsion of $M$
twisted by $A^{(\a)}$.
The isotropy group, $H_{\a}$, is a subgroup of $SU(2)$
that commutes with the holonomies of the connection $A^{\a}$.
Its tangent space, $TH_{\a}$, can be identified with
the 0-th cohomology space of $M$ twisted by $A^{(\a)}$.
Therefore, we have
\eqn\hzerofla{h^0 (A^{(\a)}) = \dim (H_{\a})}
On the other hand, the elements of the 1-st twisted cohomology
can be identified with infinitesimal deformations of
the flat connection $A^{\a}$. However, since these deformations
may be obstructed there is no simple formula for $h^1 (A^{(\a)})$,
analogous to \hzerofla.
Finally, $N_{ph}$ is given by
\eqn\nphformula{
N_{ph} = 2 SF^{(\a)}
+ h^0 (A^{(\a)}) + h^1 (A^{(\a)}) + (1 + b^1) \dim G }
where $SF^{(\a)}$ is a spectral flow of the self-adjoint operator
$L=*D + D*$ restricted to the space of odd forms.

Now, following Rozansky \Rozansky,
let us evaluate \sconeloop\ in the case of a trivial connection.


\subsec{A Contribution of the Trivial Connection}

For the trivial connection we have $S_{CS}^{(tr)}=0$. Therefore,
the leading contribution to the colored Jones polynomial
comes from the 1-loop term. Since a reducible trivial connection
with fixed boundary conditions on $M$ has no moduli, we have
\eqn\htrivial{h^0 =1 \quad , \quad h^1=0}
The isotropy group is $H_{\a} = U(1)$, and $\Vol (H_{\a}) = \sqrt{8} \pi$.
Moreover, from the results of Milnor \Milnortors\ and Turaev \Turaevtors\
it follows that in the present case
the Ray-Singer torsion is related to the Alexander polynomial,
\eqn\milnorturaev{
\sqrt{T(a)} = {2 \sin (\pi a) \over \nabla_A (K,e^{2\pi ia})} }
where $a=N/k$ is the $U(1)$ holonomy around the Wilson line $W_{R_j} (K)$.
Therefore, substituting \htrivial\ and \milnorturaev\ into
\sconeloop, we find that the leading contribution
of the trivial connection is given by \Rozansky:
\eqn\zrtivial{
Z_{SU(2)}^{(tr)} (W_{R_j},k) \simeq \sqrt{{2 \over k}}
{\sin (\pi a) \over \nabla_A (K,e^{2\pi ia})}
}
Normalizing by \zforsphere, we find that in the limit $k \to \infty$
the contribution of the trivial connection to the colored Jones
polynomial looks like
\eqn\trivjones{
J_N^{(tr)} (K,e^{2\pi i/k})
\simeq {k \sin (\pi a) \over \pi \nabla_A (K,e^{2\pi ia})}
+ \ldots }
This implies the following asymptotic behavior of the reduced Jones
polynomial $V_N (K,q)$,
\eqn\trivvnjones{
V_N^{(tr)} (K,e^{2\pi i/k})
\simeq {1 \over \nabla_A (K,e^{2\pi ia})} + \ldots}
which, in turn, implies the Melvin-Morton conjecture \mmconj.


\subsec{A Contribution of the ``Hyperbolic'' Flat Connection}

Now let us explain the relation to the volume conjecture and
its generalization \generalconj\ discussed in the previous section.
In the path integral interpretation, the exponential growth
of the colored Jones polynomial means that its asymptotic
behavior is dominated by the non-trivial $SL(2,\IC)$ flat connection,
associated with the hyperbolic structure on the knot complement, $M$.
We shall denote this connection by $A^{(hyperb)}$.
Comparing \generalconj\ with \zcseries, we conclude that
the Chern-Simons action for this complex-valued connection
is given by
\eqn\ahyperbact{
S_{CS} ( A^{(hyperb)} ) = - 2i \Big( \Vol (m) + i 2\pi^2 CS(m) \Big) }
For hyperbolic knots, the imaginary part of this expression
does not vanish and, hence, leads to the exponential
growth of $J_N (K,q)$ in the limit \thelimit.
Notice, that in order to see this exponential growth it is crucial
to allow the parameter $a=N/k$ to take generic (non-rational) values,
which is more natural in the $SL(2,\IC)$ Chern-Simons theory,
rather than in the $SU(2)$ theory, {\it cf.} Table 1.


This means that we have to find a suitable generalization
of the Melvin-Morton conjecture \mmconj\ for non-rational
values of $a$. Such a generalization can be found using
path integral arguments, similar to the above.
Indeed, let us look at the loop expansion \zcseries\
of the Chern-Simons-Witten invariant
around the flat connection $A^{(hyperb)}$.
The leading term in this expansion is given by the classical
action \ahyperbact. In the next, 1-loop term we have
\eqn\hhzero{h^0 (A^{(hyperb)}) =0 \quad , \quad h^1 (A^{(hyperb)}) =0}
This follows from eq.\hzerofla\ and the fact that the flat
connection $A^{(hyperb)}$, associated with the hyperbolic
structure on the knot complement, is irreducible and rigid,
so that $\dim (H_{hyperb})=0$.
(In fact, the same relations describe the contribution of
a non-trivial flat connection to the Reshetikhin-Turaev-Witten
invariant for torus knots \Rozansky.)
Therefore, from \zcseries\ we get,
\eqn\leadingz{ Z_{SU(2)}^{(hyperb)} (W_{R_j},k) = 
\sqrt{T (A^{(hyperb)})}~
e^{ {i k \over 4 \pi} \left(
S_{CS}^{(hyperb)} + \sum_{n>1} (2\pi/k)^n S_n^{(hyperb)}  \right) }
}
where we ignore a constant phase.
In this expression, $T (A^{(hyperb)})$ denotes
the $SL(2,\IC)$ Ray-Singer torsion of the knot complement
twisted by $A^{(hyperb)}$.
Normalizing by $Z_{SU(2)} (\S^3)$, we find the asymptotic
behavior of the $N$-colored Jones polynomial,
$$
J_N (K,q) = {\sqrt{k} \over \sqrt{2} \sin (\pi/k)} \sqrt{T}
\exp \left( {k \over 2\pi} (\Vol (m) + i 2\pi^2 CS(m))
+ {i \over 2} \sum_{n=1}^{\infty} (2\pi/k)^n S_{n+1} \right)
$$
It is convenient to write this expression in the logarithmic form:
\eqn\lognjones{
\log J_N (K,q) =
{k \over 2\pi} (\Vol (m) + i 2\pi^2 CS(m))
+ \log \left( {\sqrt{k T } \over \sqrt{2} \sin (\pi/k)} \right)
+ {i \over 2} \sum_{n=1}^{\infty} (2\pi/k)^n S_{n+1}  }
Notice, that this expansion is very similar to
the perturbative expansion of the $SL(2,\IC)$
partition function, {\it cf.} \semiclasszm\ -- \pertz.
It would be very interesting to find a better understanding
of this relation.

Now, by analogy with \vnjones, let us define a properly normalized
version of the colored Jones polynomial that would remain finite
in the limit \thelimit.
In eq.\vnjones\ this was achieved by dividing by the colored
Jones polynomial of the unknot, which automatically removed
the linear $k$-dependence from \trivjones, as well as the universal
factor of $\sin (\pi a)$.
By considering specific examples (see below), we find that
the Ray-Singer torsion, $T (A^{(hyperb)})$, does not have
the universal factor $\sin (\pi a)$ in the present case.
Moreover, as can be easily seen from \lognjones,
the polynomial growth of $J_N (K,q)$ comes from the normalization
of the Chern-Simons partition function \leadingz\ by $Z_{SU(2)} (\S^3)$.
Therefore, in the present case, it is natural to define
the reduced Jones polynomial $\tilde V_N (K,q)$
by restoring the original normalization in \leadingz,
\eqn\newvnjones{
\tilde V_N (K,q) = J_N (K,q) \cdot Z_{SU(2)} (\S^3) \cdot
\exp \left( - {k \over 2\pi} (\Vol (m) + i 2\pi^2 CS(m)) \right)
}
where we also explicitly eliminated the exponential growth using
the volume and the Chern-Simons functions introduced in \csvolint.
It follows from the loop expansion \leadingz, that
the resulting Jones polynomial has the following asymptotic behavior,
$$
\tilde V_N (K,q) = \sqrt{T (A^{(hyperb)})} \cdot
\exp \left({i \over 2} \sum_{n=1}^{\infty} (2\pi/k)^n S_{n+1} \right)
$$
In particular, this expression remains finite in the limit \thelimit.
Hence, as in the case of the trivial flat connection,
we can consider the Melvin-Morton expansion of $\tilde V_N (K,q)$,
\eqn\newmmexpansion{\tilde V_N (K,q)
= \sum_{m,n \ge 0} \tilde D_{m,n} N^{m} h^n  }
The perturbative expansion \zcseries\ of the Chern-Simons path
integral implies that the coefficients $\tilde D_{m,n}$ should
vanish unless $m \le n$.
Moreover, the following analog of the Melvin-Morton conjecture
should hold in the limit \thelimit, with non-rational $a$,

{\bf The Generalized Melvin-Morton Conjecture:}
\eqn\newmmconj{\eqalign{
& \tilde D_{m,n} = 0 \quad {\rm for} \quad m > n \cr
& \sum_{m \ge 0} \tilde D_{m,m}  a^{m}
= \sqrt{T (A^{(hyperb)}) }
}}

To get further insights into geometric information encoded
in the power series \newmmexpansion, following \Rozhigher,
let us write it in terms of the variable
$z = q^{N/2} - q^{-N/2} = 2i \sin (\pi a)$,
\eqn\newmmzexpansion{
\tilde V_N (K,q)
= \sum_{n \ge 0} \tilde V^{(n)} (K,z) h^n
= \sum_{n \ge 0} h^n \sum_{m \ge 0} \tilde d_m^{(n)} z^{m}  }
In contrast to the case of the reducible connection,
the numbers $\tilde d_m^{(n)}$ do not appear to be integer.
However, it is plausible that $\tilde d_m^{(n)}$ take
values in $\IQ(\sqrt{-1})$, or in some other number field,
which might be related to the arithmetic properties of
the knot complement.

\noindent
{\bf Questions:}

\item{$1)$}
{\it What is the geometric interpretation of
the numbers $\tilde D_{m,n}$ and $\tilde d_m^{(n)}$ ?}

\item{$2)$}
{\it What is the relation, if any, between $\tilde d_m^{(n)}$
and the coefficients $b_{m,n}$ in the perturbative
expansion \pertz\ of the $SL(2,\IC)$ partition function?}

%
%


\example{The Figure-eight Knot}

As we explained above, in order to verify the generalized
version of the Melvin-Morton conjecture \newmmconj,
it is enough to check that the colored Jones polynomial
has asymptotic expansion of the form \lognjones.
For the figure-eight knot we find the following asymptotic
behavior of the colored Jones polynomial,
\eqn\lognjoneseight{
\log J_N ({\bf 4}_1, e^{2\pi i/k}) =
{k \over 2\pi} \Vol ({\bf 4}_1)
+ {3 \over 2} \log k
+ {i \over 2} S_1 (a)
+ {i \over 2} S_2 (a) \left( {2\pi \over k} \right)
+ \ldots}
where the dominant linear term was identified earlier
in \kashlimiteight, and the coefficient of the logarithmic term
agrees with the expected cohomology \hhzero\
and with the general formula \lognjones.
The subleading terms, $S_n(a)$, contain the information
about the coefficients $\tilde D_{m,n}$ and $\tilde d_m^{(n)}$.
Numerically, we find

\eqn\dtildeeight{\eqalign{
\tilde d_0^{(0)} & = 3.3755 \ldots \cr
\tilde d_0^{(1)} & = - i 0.301 \ldots \cr
\tilde d_0^{(2)} & = 0.06 \ldots + i 0.151 \ldots
}}
%


\endexample


\vskip 30pt

\centerline{\bf Acknowledgments}

It is a pleasure to thank D.~Bar-Natan, R.~Dijkgraaf, N.~Dunfield,
S.~Garoufalidis, R.~Gopakumar,
G.~Horowitz, D.~Long, M.~Mari\~no, S.~Minwalla,
H.~Ooguri, F.~Rodriguez-Villegas, L.~Rozansky,
C.~Vafa, E.~Witten, S.-T.~Yau,
and especially
K.~Krasnov, G.~Moore, A.~Strominger, and D.~Thurston
for valuable and stimulating discussions.
This research was conducted during the period S.G.
served as a Clay Mathematics Institute Long-Term Prize Fellow.
This work is also supported in part by RFBR grant 01-01-00549
and RFBR grant for Young Scientists 02-01-06322.
I would also like to thank
the University of California at Santa Barbara,
Stanford University,
California Institute of Technology,
and Rutgers University
for kind hospitality while this work was in progress.

\vskip 30pt


\appendix{A}{The BTZ Black Hole}

The metric of the Lorentzian BTZ black hole is
described by the line element\foot{We remind that,
in our notations, $\ell =1$.} \BTZ:
\eqn\btzlor{ds^2 = - \Big( r^2 - M \Big) dt^2
+ \Big( r^2 - M \Big)^{-1} dr^2 + r^2 d \phi^2 }
Analytic continuation leads to the Euclidean metric
\eqn\btzeucl{ds^2 = \Big( r^2 - M \Big) dt^2
+ \Big( r^2 - M \Big)^{-1} dr^2 + r^2 d \phi^2 }
This is a special case of a more general metric
corresponding to a rotating BTZ black hole.
Namely, a solution corresponding to a black hole with
mass $M$ and angular momentum $J$ assumes the following form:
\eqn\btzrot{\eqalign{
ds^2 & = N^2 d\tau^2
+ N^{-2} dr^2 + r^2 \Big(d \phi^2 + N^{\phi} d\tau \Big)^2 \cr
N & = \sqrt{r^2 - M - {J^2 \over 4r^2}}
\quad , \quad
N^{\phi} = - {J \over 2r^2} \cr
r^2_{\pm} &
= {M \over 2} \Big[ 1 \pm \sqrt{1 + \Big( {J \over M} \Big)^2} \Big]
}}
This geometry can be represented as a quotient \mquotient\
of the hyperbolic 3-space,
$$
M = \IH^3 / \Gamma
$$
by a discrete group $\Gamma$.
The explicit change of variables, which brings the Euclidean
metric \btzrot\ to the standard hyperbolic metric \hthreemetric\
on $\IH^3$, is given by
\eqn\btzxyz{\eqalign{
x & = \sqrt{{r^2 - r_+^2 \over r^2 - r_-^2}}
\cdot \cos \Big(r_+ \tau + \vert r_- \vert \phi \Big)
\cdot e^{r_+ \phi - \vert r_- \vert \tau} \cr
y & = \sqrt{{r^2 - r_+^2 \over r^2 - r_-^2}}
\cdot \sin \Big(r_+ \tau + \vert r_- \vert \phi\Big)
\cdot e^{r_+ \phi - \vert r_- \vert \tau} \cr
z & = \sqrt{{r_+^2 - r_-^2 \over r^2 - r_-^2}}
\cdot e^{r_+ \phi - \vert r_- \vert \tau}
}}

In order to identify the quotient group $\Gamma$,
it is convenient to write $(x,y,z)$ in terms of
the spherical coordianates,
\eqn\xyzvia{\eqalign{
x & = R \cos \th \cos \chi \cr
y & = R \sin \th \cos \chi \cr
z & = R \sin \chi
}}
In these variables, the black hole metric has the form
\eqn\rchimetric{
ds^2 = {1 \over \sin^2 \chi} \left( {dR \over R} \right)^2
+ \cot^2 \chi ~d \th^2 + {1 \over \sin^2 \chi} d \chi^2 }
and the identifications on $\IH^3$
that generate the holonomy group $\Gamma$ are
\eqn\btzident{\eqalign{
(a): \quad & (R,\th,\chi)
\sim (R \cdot e^{2 \pi r_+},\th + 2\pi \vert r_- \vert ,\chi) \cr
(b): \quad & (R,\th,\chi) \sim (R,\th + 2 \pi, \chi)
}}

\ifig\euclbtzfig{The Euclidean BTZ black hole has a geometry of
the solid torus, $M \cong {\bf D}^2 \times \S^1$. We can also
view $M$ as a complement of the unknot (the trivial knot) in the 3-sphere.}
{\epsfxsize3.5in\epsfbox{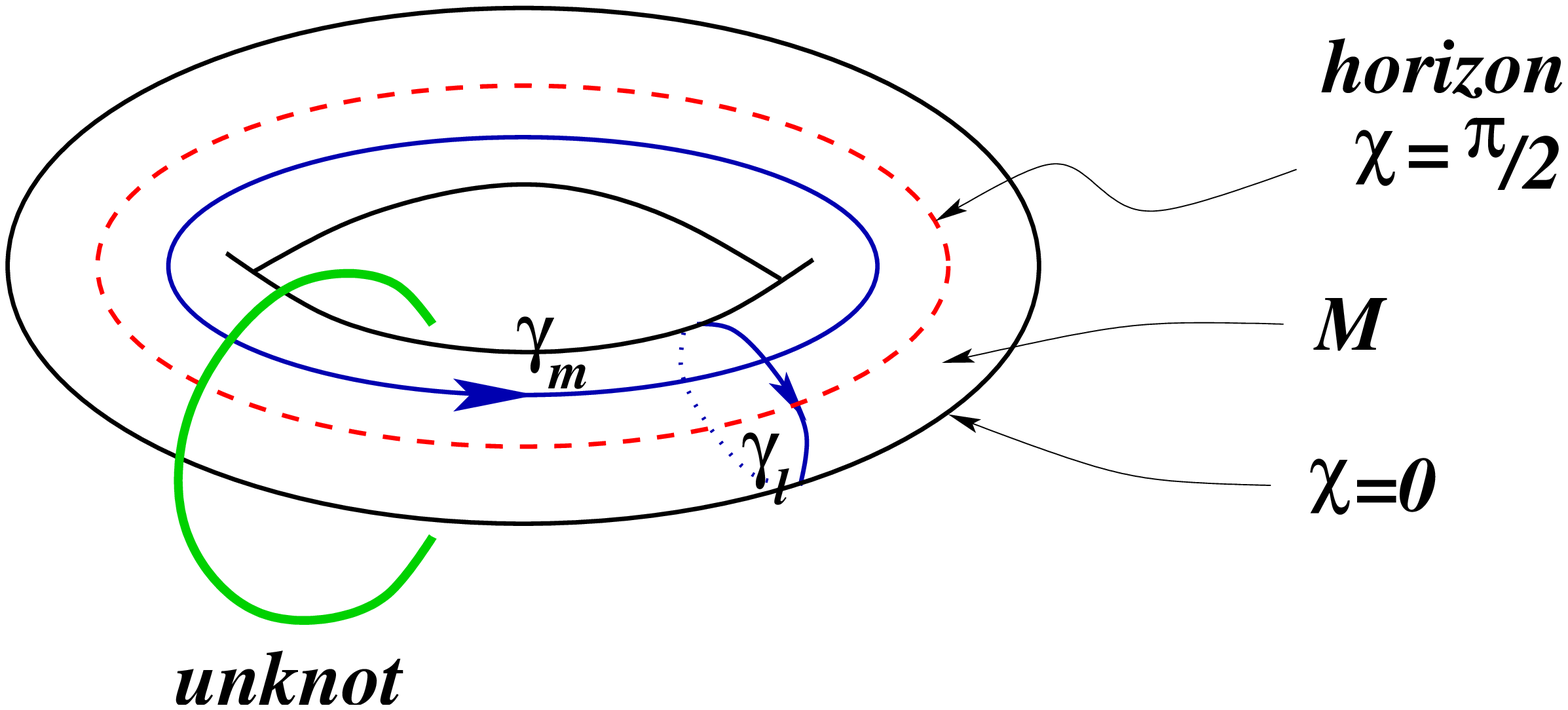}}

Using \xyzvia\ and \btzident, it is easy to see that the resulting
geometry is a solid torus, $M \cong {\bf D}^2 \times \S^1$,
shown on \euclbtzfig.
The core of the torus (the circle $\chi=\pi/2$) corresponds to
the position of the ``horizon''.
The solid torus $M$ can be also viewed as a complement of the unknot
(the trivial knot) in the 3-sphere, $M = \S^3 \setminus K$.
The tubular neighborhood of the trivial knot $K$
is itself homeomorphic to a solid torus, {\it cf.} \lmcycles.
Thus, a union of these two solid tori joined along a common
torus boundary gives back the 3-sphere.

Notice, that the longitude $\g_l$ of the unknot is contractible in $M$,
whereas the meridian $\g_m$ is a non-trivial cycle in $M$.
Therefore, it follows that the holonomy $l$ must be trivial,
whereas the value of $m$ can be arbitrary.
Hence, we conclude that the A-polynomial of the unknot
has the following simple form, {\it cf.} \aunknot,
\eqn\lminusone{  A_{{\rm unknot}} (l,m) = l-1 }
%
It turns out that this result has a nice physical interpretation.
Namely, the vanishing of the polynomial \lminusone\ can be
viewed as a ``mass shell'' condition for the so-called
off-shell BTZ black hole, which we are going to describe next.

The geometry of the off-shell BTZ black hole can be
obtained from \btzident\ by replacing the second
identification with a more general
condition \refs{\CToffshell, \CTaspects, \CarlipBH, \Carliplect}:
\eqn\btzprimeident{
(b'): \quad (R,\th,\chi) \sim (e^{\Si} \cdot R,\th + \Theta, \chi)
}
The resulting identifications \btzident\ - \btzprimeident\
can be represented by the elements of the $SL(2,\IC)$ group,
\eqn\btzhol{
\rho (a) = \pmatrix{e^{\pi (r_+ + i \vert r_- \vert)} & 0 \cr
0 & e^{- \pi (r_+ + i \vert r_- \vert)} }
\quad , \quad
\rho (b') = \pmatrix{ e^{(\Si + i \Theta)/2} & 0 \cr
0 & e^{-(\Si + i \Theta)/2} }
}
acting on the hyperbolic 3-space as in \gammaonh.
These two elements generate the holonomy group
$\Gamma \subset SL(2,\IC)$, such that the quotient space
$M=\IH^3 / \Gamma$ is a solid torus with a conical singularity
at the ``horizon'', which resembles the space-time geometry
around the spinning particle in 2+1 dimensions \DJT.
In this more general family of solutions, the usual
BTZ black hole metric without a conical singularity
is reproduced when
\eqn\btzonshell{\Si=0 \quad , \quad \Theta=2\pi}
Following the standard terminology,
we refer to this condition as the ``mass shell'' condition.

In what follows, our goal will be to construct a flat
$SL(2,\IC)$ connection corresponding to the more general
class of BTZ black home geometries
and, in particular, to demonstrate that the mass shell
condition \btzonshell\ is equivalent to the vanishing
of the A-polynomial \lminusone.
For this, we need to evaluate the holonomies
of the gauge connection,
\eqn\holfla{\rho (\g) = P \exp \oint_{\g} \CA,}
along the longitude, $\g_l$, and the meridian, $\g_m$.
One can choose to parameterize these cycles by a real
variable $s \in [0,1]$, such that
\eqn\btzlm{\eqalign{
\g_m  &:~~ s \mapsto
(R_0 e^{2 \pi r_+ s} , \th_0 + 2 \pi \vert r_- \vert s , \chi_0) \cr
\g_l  &:~~ s \mapsto
(R_0 e^{s \Si} , \th_0 + s \Theta , \chi_0) }}
Notice, that due to the identifications \btzident\ - \btzprimeident,
the curves $\g_m$ and $\g_l$ are manifestly closed.

In order to evaluate the holonomies \holfla, we also need to
construct the complex valued gauge connection,
$\CA^a = w^a + i e^a$, from the components of the vielbein and
the spin connection in the Euclidean BTZ black hole geometry.
The latter can be taken, for example, in the following
form \refs{\CTaspects,\CarlipBH}:
\eqn\btzee{
\eqalign{
e^1 & = {1 \over \sin \chi} {dR \over R} \cr
e^2 & = {1 \over \sin \chi} d \chi \cr
e^3 & = \cot \chi \cdot d \th }
\quad \quad \quad \quad
\eqalign{
w^1 & = - {1 \over \sin \chi} d \th \cr
w^2 & = 0 \cr
w^3 & = \cot \chi \cdot {dR \over R} }
}
It is easy to verify that the corresponding
$SL(2,\IC)$ gauge connection is indeed flat.
However, as we shall see in a moment,
this choice of the vielbein and the spin connection
leads to a singular connection\foot{
I am endebted to G.~Moore and A.~Strominger
for very helpful comments and suggestions on these points.}, $\CA$.
Indeed, even though \btzee\ defines a smooth metric \rchimetric,
it corresponds to a singular gauge field, $\CA^1 = w^1 + i e^1$,
which is not well defined near the center of the solid torus,
$\chi = \pi/2$, where the angular variable $\th$ is
ill-defined and $w^1 \approx - d \th$.

Another way to see that \btzee\ does not correspond to
a smooth $SL(2,\IC)$ connection over $M$ is to evaluate
the holonomies \holfla\ around the 1-cycles \btzlm.
The resulting holonomies turn out to be in the same
conjugacy class as the generating elements \btzhol\
of the group $\Gamma$.
Therefore, comparing \btzhol\ with \lmholonomies,
we conclude that in the present case
the holonomies $l$ and $m$ are given by
\eqn\btzlmviar{l = e^{(\Si + i \Theta)/2}
\quad , \quad
m = e^{\pi (r_+ + i \vert r_- \vert)}
}
In particular, we find that, with the choice \btzee,
the ``mass shell'' condition \btzonshell\ looks like
\eqn\lplusone{ l+1=0 }
and has the `wrong' sign compared to \lminusone.
This result does not agree with the fact that
the 1-cycle $\g_l$ becomes contractible in the on-shell
BTZ black hole geometry and, therefore,
the corresponding holonomy should be trivial.
Finally, let us remark that this problem can not be
fixed by applying a gauge transformation to \btzee\
since any (non-singular) gauge transformation can not
change the holonomy.

Therefore, we need to construct a flat $SL(2,\IC)$ gauge
connection that would be non-singular everywhere inside $M$.
It is convenient to introduce a new set of coordinates:
\eqn\newbtzcoord{\eqalign{
\varrho & = \log R \cr
\xi & = f(\chi) \cos \th \cr
\eta & = f(\chi) \sin \th
}}
where the function,
\eqn\fviachi{f(\chi) = {\cos \chi \over 1 + \sin \chi},}
is chosen such that the metric induced in the $(\xi,\eta)$-plane
is conformal to the usual Euclidean metric, $ds^2 = d \xi^2 + d \eta^2$.
Specifically, rewriting the BTZ black hole metric \rchimetric\
in the coordinates \newbtzcoord, we find
\eqn\torusmetric{
ds^2 = \left( {1 + f^2 \over 1-f^2} \right)^2
\Big[ d \varrho^2 + {4 \over (1+f^2)^2} (d\xi^2 + d\eta^2) \Big] }

The corresponding components of the vielbein
and the spin connection can be written as
\eqn\btzeee{
\eqalign{
e^1 & = {1+f^2 \over 1-f^2} d \varrho \cr
e^2 & = {2 \over 1-f^2} d \xi \cr
e^3 & = {2 \over 1-f^2} d \eta
}
\quad \quad \quad \quad
\eqalign{
w^1 & = \xi e^3 - \eta e^2 \cr
w^2 & = {2 \eta \over 1+f^2} e^1 \cr
w^3 & = - {2 \xi \over 1+f^2} e^1
}
}
Combining these together we find the explicit
expression for the components of the flat
$SL(2,\IC)$ gauge connection, $\CA^a = w^a + ie^a$,
\eqn\asmooth{\eqalign{
\CA^1 & = {2 \xi \over 1-f^2} d\eta - {2 \eta \over 1-f^2} d\xi
+ i {1+f^2 \over 1-f^2} d \varrho \cr
\CA^2 & = {2 \eta \over 1-f^2}d\varrho + i {2 \over 1-f^2} d \xi \cr
\CA^3 & = - {2 \xi \over 1-f^2}d\varrho + i {2 \over 1-f^2} d \eta
}}
which is non-singular everywhere in the interior of
the solid torus, $M = \S^1 \times {\bf D}^2$.

Now let us  evaluate the $SL(2,\IC)$ holonomies \holfla.
For simplicity, let us consider a non-rotating
BTZ black hole, with $J=0$.
Then, the second identification in \btzident\ is trivially realized
in the coordinates \newbtzcoord, whereas the first one reads:
\eqn\newbtzident{(a): \quad (\varrho,\xi,\eta) \sim
(\varrho + 2 \pi r_+,\xi,\eta) }
Therefore, we can choose the cycles $\g_l$ and $\g_m$ to be
parametrized in the following way, {\it cf.} \btzlm, 
\eqn\btznewlm{\eqalign{
\g_m  &:~~ s \mapsto
(\varrho_0 + 2 \pi r_+ s , ~\xi=0, ~\eta=0) \cr
\g_l  &:~~ s \mapsto
(\varrho_0, ~\xi=\xi_0 \cos 2 \pi s, ~\eta=\xi_0 \sin 2 \pi s) }}

Since only the $\CA^1$-component of the gauge connection
is non-trivial along the curve $\g_m$, we can easily
find the corresponding holonomy,
%
\eqn\btzlhol{
\rho (\g_m) = \pmatrix{e^{\pi r_+} & 0 \cr 0 & e^{-\pi r_+}}   }
Comparing this result with \btzlmviar, we find that the holonomy
of the gauge connection \asmooth\ around $\g_m$ is the same as
in the previous calculation, based on the singular gauge connection \btzee.
On the other hand, since the connection \asmooth\ is flat
and non-singular everywhere in the interior of the solid
torus, the holonomy around a contractible cycle $\g_l$
is guaranteed to vanish automatically,
$$
l=1
$$
However, since many components of the gauge connection \asmooth\
do not vanish along the curve $\g_l$, to verify this directly
would require some work. Perhaps the best way to approach this
problem would be to find a gauge transformation,
which removes non-commuting components of the gauge connection,
and allows to write the holonomy \holfla\ in a simple form.


\appendix{B}{Quantization for Torus Knots}

One simple way of classifying knots (in $\IR^3$) is to
associate to every knot $K$ a non-negative number, $g(K)$,
called the genus of the knot. Indeed, every knot can be embedded
in some Riemann surface of genus $g$, and, as the name suggests,
$g(K)$ is the minimal value of genus for which this can be done
(without self-crossing or breaking the knot).
For example, the only knot with $g=0$ is the unknotted circle.
The next simplest case corresponds to knots of genus one,
the so-called torus knots.

Since torus knots are special in a number of ways,
we can't really call them generic representatives.
In particular, since torus knots are not hyperbolic,
most of the motivation discussed in the introduction
does not apply here.
Nevertheless, torus knots provide a simple toy model,
which includes all the ingredients that one would find in
a more complicated non-linear quantum system,
say, corresponding to a figure-eight knot.
Therefore, in order to emphasize the general ideas,
here we will ignore the subtleties,
slightly oversimplifying our discussion.

\ifig\ttknot{Trefoil knot on a torus.}
{\epsfxsize5.0in\epsfbox{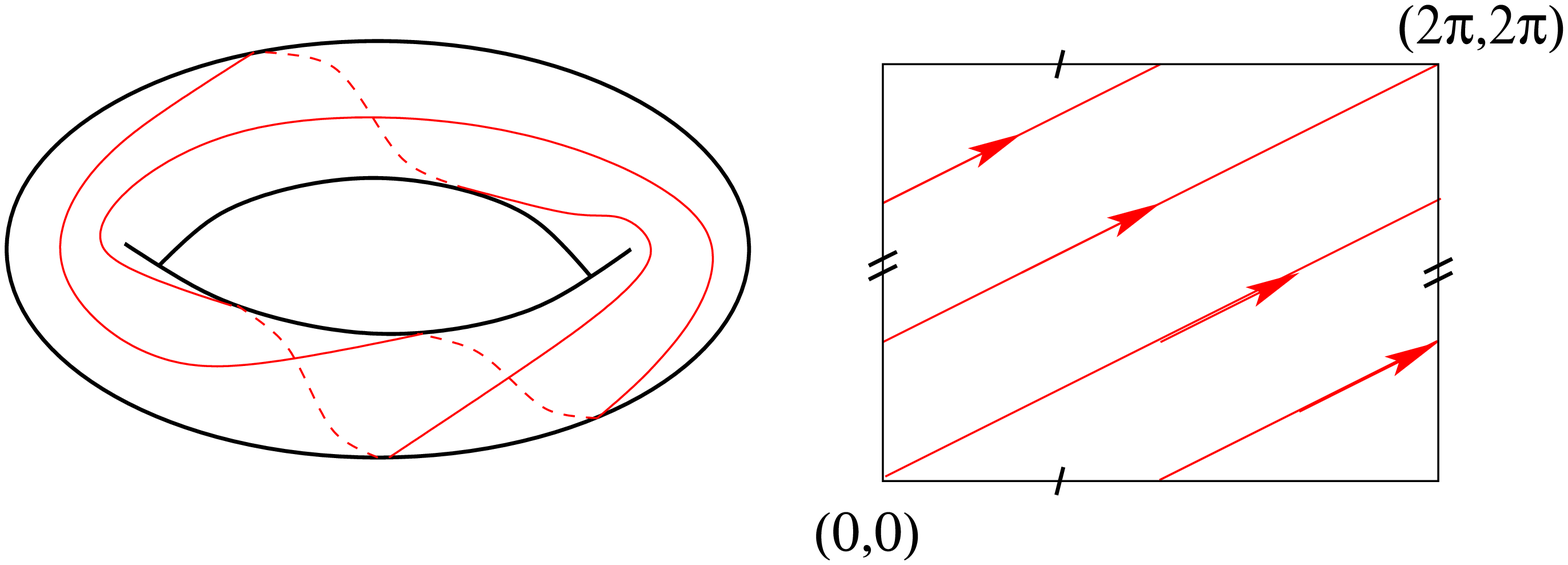}}

By definition, torus knots are knots which can be drawn
on the surface of a torus. They are labeled by an unordered
pair of relatively prime numbers $(r,s)$, which refers to
the number of times the image of the knot winds basic
cycles on the torus. For example, the trefoil knot is a torus
knot of type $(2,3)$, see \ttknot.

The A-polynomial of a non-trivial $(r,s)$-torus knot
has a very simple form \CCGLS:
\eqn\torusa{A(l,m) = lm^{rs}+1}
The zero locus of this polynomial defines a character
variety $L$, linearly embedded into $\CP = \IC^* \times \IC^*$.
In terms of the $(u,v)$-variables \uvdef,
the variety $L$ can be defined by the following two equations:
\eqn\torusuv{\eqalign{
& \r (u) + rs \cdot \r (v) =0 \cr
& \i (u) + rs \cdot \i (v) - \pi =0
} }
Our goal is to quantize a Hamiltonian system
associated with $L \hookrightarrow \CP$ and, in particular,
to find the quantum wave function $Z(M)$ supported on $L$.
The result can be interpreted as a partition
function in the $SL(2,\IC)$ Chern-Simons theory on a complement
$M = \S^3 \setminus K$ of the torus knot $K$.
Notice, that $M$ is not a hyperbolic 3-manifold.
Therefore, as suggested by the asymptotic expression \semiclassz,
it is natural to consider a special case\foot{In another special case,
$k=0$ and $\s \ne 0$, we find that the action integral
$S$ is identically zero.} corresponding to $\s=0$.

In this case, the canonical commutation relations
that follow from the Chern-Simons action \iaction\
look like:
\eqn\torusscomm{\eqalign{
& [\r (v) , \r (u) ] = [ \i (u) , \i (v) ] = {2 \pi i \over k} \cr
& [ \r (u) , \i (v) ] = - [ \r (v) , \i (u) ] = 0   }}
and the Liouville 1-form \tottheta\ is given by
%
\eqn\torusktheta{
\th = {k \over \pi}
\Big( \r (v) d \r (u) + \i (u) d \i (v) \Big) }
%
It can be written in the canonical form \canoneform\
provided that we define the coordinates $q_i$
and the conjugate momenta $p_j$ as
\eqn\pqtorusk{
\pmatrix{p_1 \cr p_2} = \pmatrix{\r (v) \cr \i (u)}
\quad {\rm and} \quad
\pmatrix{q_1 \cr q_2} = \pmatrix{\r (u) \cr \i (v)}
}
Here we omit the overall factor $k/4\pi$,
which can be treated as the inverse Planck constant.
In these variables, the Lagrangian submanifold $L$
can be written as:
%
\eqn\lkpq{\eqalign{
& q_1 + rs \cdot p_1 =0 \cr
& rs \cdot q_2 + p_2 - \pi =0
}}

Notice, that the Hamiltonian system splits into two
non-interacting subsystems described by the canonical
variables $(q_1,p_1)$ and $(q_2,p_2)$, respectively.
The phase space of the first system is non-compact, whereas
in the second system both the coordinate and the momentum are compact.
For now, let us ignore the compactness of $q_2$ and $p_2$.
Then, in the coordinate representation the action integral reads:
\eqn\storusk{\eqalign{
S & = {k \over \pi} \int p_1 d q_1 + p_2 d q_2 = \cr
& = - {k \over \pi}
\left( {1 \over 2rs} q_1^2 + {rs \over 2} q_2^2 - \pi q_2 \right)
}}
If $\psi$ is a constant half-density on $L$,
then the transformation rule for half-densities
implies that the Reidemeister-Ray-Singer torsion, $T$, is
also constant (independent on $q_i$) in this class of examples.
Therefore, in the semi-classical limit the partition
function of the $SL(2,\IC)$ Chern-Simons theory on
the complement of a torus knot looks like:
\eqn\torussemiclass{
Z(M) \sim
\exp \left(
- {i k \over 2 \pi rs} q_1^2 - {i krs \over 2 \pi} q_2^2
+ i k q_2 \right) + \ldots }
%
%
This result gives a prediction for
the Chern-Simons invariant of the torus knot complement $M$
(as a function of the holonomies $l$ and $m$).
Using \torusuv\ and \pqtorusk, we can write it as:
\eqn\toruscs{
CS(M)
= {1 \over 2 \pi^2 rs}
\Big( \log^2 \vert l \vert + \arg^2 (l) - \pi^2 \Big)
}

Here, we slightly oversimplified our discussion assuming
that the phase spaces of both dynamical systems are non-compact.
A nice way to incorporate the compactness of $q_2$ and $p_2$
is to notice that they parameterize a torus\foot{Once again, we should
remind that we tacitly omit the quotient by the Weyl group.},
which can be viewed as a phase space of the $SU(2)$ Chern-Simons
theory on a 3-manifold $M$ with boundary $\Si$, see \ssrep.
Hence, the exact partition function of the $SL(2,\IC)$ Chern-Simons
theory on $M$ can be written as a product,
\eqn\ztorusexact{
Z_{SL(2,{\bf C})} (M) = Z_{SU(2)} (M) \times
\exp \left( - {i k \over 2 \pi rs} q_1^2
+ {i \pi \over 4} {\rm sign}~ (rs) \right) }
where $Z_{SU(2)} (M)$ is the partition function of the $SU(2)$
Chern-Simons theory
(see {\it e.g.} \refs{\EMSS, \Weitsman, \BNair, \Jackiwsch, \Manoliu}),
and the rest represents the wave function in the first system,
with non-compact phase space parametrized by $p_1$ and $q_1$.
Notice, that the semi-classical approximation is exact in this case.
This has to be compared with the computation of
the Reshetikhin-Turaev-Witten invariants \refs{\WittenJones,\RT}
for torus knots in $\S^3$, where the stationary phase approximation
is also exact, see {\it e.g.} \LR.

\listrefs
\end